\documentstyle[aps,epsf,preprint,eqsecnum,vmargin]{revtex}

\newcommand{\fop}[1]{{\hat #1}}
\newcommand{\Ac}{{\mathcal A}}
\newcommand{\fopac}{\fop{{\mathcal A}}}
\newcommand{\Fc}{{\mathcal{F}}}
\newcommand{\fopfc}{\fop{{\mathcal F}}}
\newcommand{\Pc}{{\mathcal{P}}}
\newcommand{\foppc}{\fop{{\mathcal{P}}}}
\newcommand{\Vc}{{\mathcal{V}}}
\newcommand{\fopvc}{\fop{{\mathcal{V}}}}
\newcommand{\Sc}{{\mathcal{S}}}
\newcommand{\fopsc}{\fop{{\mathcal{S}}}}
\newcommand{\Lc}{{{\mathcal L}}}
\newcommand{\foplc}{\fop{{{\mathcal L}}}}
\newcommand{\cur}[1]{{\bf{\hat #1}}}

\newcommand{\fopa}{\fop{A}}
\newcommand{\fopf}{\fop{F}}

\newcommand{\fopw}{\fop{W}}
\newcommand{\fopx}{\fop{X}}

\newcommand{\D}{\displaystyle}

\renewcommand{\vec}{\bbox}

\begin{document}
\hyphenation{Lo-rentz
  in-fi-ni-te-si-mal}
\begin{center}
  {\large  Wigner functions in covariant and single-time formulations}
  \\
  Stefan Ochs and Ulrich Heinz
  \\
  Institut f\"ur theoretische Physik,
  93040 Universit\"at Regensburg
\end{center}

\begin{abstract}
  We will establish the connection between the Lorentz covariant and
  so-called single-time formulation for the quark Wigner operator. To
  this end we will discuss the initial value problem for the Wigner
  operator of a field theory and give a discussion of the
  gauge-covariant formulation for the Wigner operator including some
  new results concerning the chiral limit. We discuss the gradient
  or semi-classical expansion and the color and spinor decomposition
  of the equations of motion for the Wigner operator. The single-time
  formulation will be derived from the covariant formulation by taking
  energy moments of the equations for the Wigner operator. For
  external fields we prove that only the lowest energy moments of the
  quark Wigner operator contain dynamical information.
\end{abstract}
\hspace{20ex}

\noindent
number of pages: 92\\
number of tables: 2 \\
number of figures: 0
\newpage

\noindent running head: {\bf Single-time Wigner functions}

\vskip 1cm

\noindent {\em Send proofs to:}\\
Ulrich Heinz, \\
Institut f\"ur Theoretische Physik \\
93040 Regensburg\\
Germany\\
telephone: ++49\ 941\ 943\ 2004/2008\\
telefax: ++49\ 941\ 943\ 3887\\
email: Ulrich.Heinz@physik.uni-regensburg.de

\newpage

\section{Introduction}
\label{sec1}

Ultra-relativistic heavy ion experiments nowadays reach extreme beam
energies of up to 200 GeV/A at CERN. The compressed nuclear matter
becomes so hot and dense that a Quark-Gluon-Plasma (QGP) is expected
to form \cite{mueller}. The quest for this new state of matter
\cite{QM} motivates the following work. It will deal with the
description of ultra-relativistic heavy ion collisions in the
framework of a transport theory for the underlying degrees of freedom,
namely quarks and gluons. So far Quantum Chromodynamics (QCD) as the
gauge theory for quarks and gluons has been successful in describing
equilibrium properties (including linear response functions to
external perturbations) through (i) perturbation theory in the
framework of thermal field theory (TFD) \cite{lebellac,kapusta},
especially in connection with the Hard Thermal Loop resummation scheme
of Braaten and Pisarski \cite{HTL}, and (ii) (non-perturbatively)
through numerical Monte Carlo simulations of lattice gauge theory
\cite{laermann,lattice}. Since, however, the formation of a QGP is, at
least in the early dynamical stages, governed by non-equilibrium
processes we also need a kinetic theory which can deal with
non-equilibrium phase space dynamics. For this the proper framework is
relativistic transport theory.

{\em Classical} relativistic kinetic theory \cite{degroot,NIC} has been
used in practice for relativistic heavy ion collisions in the form of
numerical cascade codes, including mean field as well as quantum
statistical scattering effects for fermions (Pauli suppression)
\cite{BUU,daniel} and bosons (stimulated scattering) \cite{WELKE,BOSE}.
A classical relativistic transport theory for colored degrees of
freedom was developed in \cite{prl,KFT}. The basis for {\em quantum}
transport theory is the Wigner formalism
\cite{WIG,QCT,chou85,EH89,malfliet}; through a systematic gradient
expansion it can be reduced to classical kinetic theory \cite{WIG2}.

Quantum transport equations for gauge theories require a gauge
covariant definition of the Wigner operator \cite{prl,EH89,LINK,QTT}.
Furthermore, proper inclusion of field theoretic retardation effects
is facilitated by a Lorentz covariant formulation \cite{EH89}.

On the other hand, if one wants to solve the kinetic equations as an
initial value problem in time, a Lorentz covariant formulation leads
to conceptual difficulties. These will be shortly reviewed in
Sec.~\ref{sec2}. This led Bialynicki-Birula, Gornicki, and Rafelski
(BGR) \cite{BGR1} to propose a so-called single-time formulation for
relativistic quantum transport theory which breaks explicitly the
Lorentz covariance of the theory but can be solved as an initial value
problem \cite{BGR1,BGR2,BGR3}.

Our goal in this paper is to establish the connection between the
Lorentz covariant and single-time formulations of quantum transport
theory. The present work extends the previous study by Zhuang and
Heinz \cite{HIE} to the case of non-Abelian gauge theories. We
concentrate on the dynamics of fermion fields in the background of
a (classical or fluctuating) gauge field. We start from the Lorentz
covariant formulation of relativistic quantum transport theory,
discussing its equations of motion and their semiclassical expansion
as well as their color and spinor decomposition (Sec.~\ref{sec3}).
Most of this reviews (sometimes in more elegant notation) previous
work which is relevant for our analysis here. We do, however, point
out that quite generally transport and mass shell equations need to be
solved only for 8 of the 16 spinor components, either for the scalar,
pseudoscalar, and tensor densities or for the vector and axial vector
densities. The remaining 8 components can then be obtained from
constraints (Sec.~\ref{sec3.6}). This was apparently not realized
before. Also the color decomposition in Sec.~\ref{sec3.4} has not been
presented before. Some new results are also contained in
Sec.~\ref{sec3.7} on the chiral (massless) limit and in
Sec.~\ref{sec3.9} on the classical limit. In particular we show that
in the chiral limit the kinetic equations for the right-handed and
left-handed components decouple which considerably simplifies the
structure of the covariant transport equations.

In Sec.~\ref{sec4} we derive the single-time form of the transport
equations by taking energy moments of the covariant equations. This
method was established in Ref.~\cite{HIE} for simpler theories and is
here extended to the case of general (abelian or non-abelian) gauge
theories. The covariant transport equations are equivalent to an
infinite hierarchy of coupled single-time transport equations for the
energy moments of the covariant Wigner operator. We construct this
hierarchy and discuss its practical truncation for applications. The
general moment hierarchy is written down for the Wigner {\em operators}
and their equations of motion which, after taking ensemble expectation
values, would generate the whole coupled BBGKY hierarchy of kinetic
equations for 1-body, 2-body, etc. phase-space distribution {\em functions}.

In the present paper the truncation of the energy moment hierarchy is
discussed explicitly only for the simple limit of external gauge
fields (Hartree limit). This corresponds to Vlasov-type transport
equations without collision terms. In this limit we find a simple and
exact truncation scheme (``exact'' meaning that no additional
approximations like e.g. a semi-classical gradient expansion are
neccessary) in which only one dynamical single-time transport equation
for the lowest energy moment must be solved while all higher order
moments can be obtained from the solution via constraints. This
explains the success of the BGR approach \cite{BGR1} (who only studied
the lowest energy moment and never looked at the full moment
hierarchy) for external field problems and at the same time provides a
basis for generalizing it to include back-reaction and collision
effects. In the last section of the main text, Sec.~\ref{sec4.4}, we
write these equations out explicitly for QED to leading order of a
semiclassical gradient expansion to recover the well-known Vlasov
equations supplemented by spin precession terms.

A short summary of the achievements and an outlook on the remaining open
problems is given in Sec.~\ref{sec5}. The Appendix contains helpful
formulae and other technical details.

\section{The initial value problem}
\label{sec2}

The Wigner operator for a field theory with second quantized fields
$\Psi$, $\Psi^\dag$ is given by the Fourier transformation of the
density matrix
 \begin{equation}
   \fop{\varrho}(x_+,x_-) = \Psi (x_+) \Psi^\dag (x_-),\quad x_\pm = x\pm y/2
 \end{equation}
over the relative coordinate $y = x_+ - x_-$
 \begin{equation}
 \label{wigop}
   \fopw (x,p) \equiv \int \frac{d^4y}{(2 \pi)^4}\ e^{-ip\cdot y}
   \fop{\varrho}\left(x+{\textstyle{y\over 2}},x-{\textstyle{y\over 2}}
   \right).
 \end{equation}
$x = (x_+ + x_-)/2$ is the center of mass coordinate, and $p$ is the
canonically conjugate momentum to $y$.

Using the translation operators $\exp(-y\cdot\partial_x/2)$ (acting to
the right) and $\exp(y\cdot\partial_x^\dag/2)$ (acting to the left)
to rewrite the density matrix,
 \begin{eqnarray}
 \label{dichteopqm}
  \fop{\varrho} \left(x+{\textstyle{y\over 2}},
                      x-{\textstyle{y\over 2}}\right)
  &=& \Psi\left(x+{\textstyle{y\over 2}}\right)
      \Psi^\dag(x-{\textstyle{y\over 2}})
  = \Psi(x) e^{y\cdot(\partial^\dag_x - \partial_x/2)} \Psi^\dag(x)
 \nonumber\\
  &=& \Psi (x) e^{i\fop{\pi}\cdot y} \Psi^\dag (x),
  \qquad \fop{\pi} = \frac{i}{2}(\partial_x - \partial^\dag_x),
\end{eqnarray}
the connection with the classical phase space density becomes apparent:
 \begin{eqnarray}
 \label{impulsqm}
   \int\frac{d^4y}{(2 \pi)^4}\ e^{-ip\cdot y}
   \fop{\varrho}(x+y/2,x-y/2)
   &=& \int\frac{d^4y}{(2 \pi)^4}\ \Psi(x)
   e^{-i(p -\fop{\pi})\cdot y}\Psi^\dag(x)
 \nonumber\\
   &=& \Psi(x) \delta(p - \fop{\pi}) \Psi^\dag(x).
 \end{eqnarray}
$\fop{\pi} = i(\partial_x - \partial^\dag_x)/2 $ is the momentum
operator. The Wigner operator (more exactly: its ensemble expectation
value) thus is a quantum mechanical and off-shell generalization of
the classical phase space density
 \begin{equation}
 \label{klasspsdichte}
   f(\vec x, \vec p, t) = \sum_i
   \delta(\vec x - {\vec x}_i(t))\,
   \delta(\vec p - {\vec p}_i(t)),
 \end{equation}
which gives the probability density to have a particle with (on-shell)
momentum $\vec p$ at space-time coordinate $(\vec x,t)$
(where $x=(x_+ + x_-)/2$). In contrast to the classical phase-space
density $f$, $W(x,p) \equiv \langle \fopw(x,p) \rangle$ and its energy
average $W(\vec x,\vec p, t) = \int dp_0\, W(x,p)$ are in general not
positive definite, but can become negative at small phase-space
distances. Only if averaged over sufficiently large phase-space
volumes $\Delta V = \Delta^3x \, \Delta^3p \gg (2 \pi)^3$ with
some positive definite weight function $g(\vec x,\vec p)$,
\begin{equation}
  f(\vec x,\vec p,t) = \int_{\Delta V} {d^3x'\,d^3p' \over (2 \pi)^3}\,
  W(\vec x-\vec x',\vec p - \vec p',t)\cdot g(\vec x',\vec p'),
\end{equation}
the resulting $f(\vec x, \vec p,t)$ is positive definite and can be
interpreted as a classical phase-space distribution \cite{WIG}.

\subsection{Initial value problem and energy moments}
\label{sec2.1}

For a covariant quantum field theory the density operator
$\fop{\varrho} (x_+,x_-)$ and, in general, all 2-point functions
contain two time para\-meters, $t_1 = x_0 + y_0/2$ and $t_2 = x_0 -
y_0/2$. The occurrence of two time parameters is necessary for a
proper description of signal propagation and retardation effects. It
poses, however, conceptual problems for a transport theoretic approach
because the corresponding covariant Wigner density cannot be
initialized with the help of asymptotic field configurations at $t =
-\infty$ \cite{INI}. To see this let us rewrite the covariant Wigner
function (see (\ref{wigop})) as follows:
 \begin{equation}
 \label{def_Wig4}
  W_4(x,p) = \int_{-\infty}^\infty\! \frac{dy_0}{2\pi} \, e^{-ip_0y_0}
  \int \frac{d^3y}{(2 \pi)^3} \, e^{i\vec p \cdot \vec y}
  \left\langle
  \Psi \left(\vec x + {\textstyle{\vec y\over 2}},x_0 +
    {\textstyle{y_0\over2}}\right)\,
  \Psi^\dag \left(\vec x - {\textstyle{\vec y\over 2}},x_0 -
    {\textstyle{y_0\over2}}\right) \right\rangle
  \, .
 \end{equation}
Obviously, due to the Fourier transformation over relative times
$y_0$, the covariant Wigner function $W_4$ at any fixed time $x_0$
requires knowledge of the fields $\Psi$, $\Psi^\dag$ at {\em all}
times. $W_4$ at $x_0\to -\infty$ can thus not be initialized by giving
the fields only at $t=-\infty$.

The equal-time (or single-time) Wigner function
 \begin{equation}
 \label{def_Wig3}
  W_3(x,\vec p) = \int \frac{d^3y}{(2 \pi)^3}\, e^{i\vec p \cdot \vec y}
  \left\langle
  \Psi \left(\vec x + {\textstyle{\vec y\over 2}},x_0\right)\,
  \Psi^\dag \left(\vec x - {\textstyle{\vec y\over 2}},x_0 \right)
  \right\rangle \, ,
 \end{equation}
which uses the fields at equal times and thus depends only on a single
time parameter, has clearly no such problem. Initializing $W_3$ at
$x_0=-\infty$ by vacuum solutions for the fields, it has been used in
\cite{BGR2} to calculate the pair production rate in QED in a constant
external electric field, reproducing the correct Schwinger rate
\cite{PAAR}. On the other hand, by comparing (\ref{def_Wig4}) and
(\ref{def_Wig3}) one sees that the equal-time Wigner function $W_3$ is
simply the energy integral of the covariant Wigner function $W_4$:
 \begin{equation}
 \label{enav}
   W_3(x,\vec p) = \int_{-\infty}^\infty dp_0\, W_4(x,\vec p,p_0)\, .
 \end{equation}
As such it contains less dynamical information than $W_4$ since the
complete off-shell structure of $W_4$ is averaged over.

While the formulation of the transport theory as an initial value
problem thus seems to require a single-time language, it can clearly
not be based on $W_3$ alone. Retardation and memory effects in the
covariant approach are reflected in the off-shell behaviour of the
covariant Wigner function $W_4$ which should not be thrown away by
simply averaging it over $p_0$. Instead, one must study, in addition
to $W_3$ (which is the zeroth $p_0$-moment of $W_4$) also the higher
$p_0$-moments of the covariant Wigner function:
 \begin{eqnarray}
 \label{moments}
  \int_{-\infty}^\infty dp_0\ p_0^n\ W_4(x,p) &=&
  \biggl(\frac{i}{2}\biggr)^n \sum_{k=0}^n (-1)^k
  \left(\!{n \atop k} \!\right)
  \int \frac{d^3y}{(2\pi)^3}\ e^{i\vec p\cdot\vec y}
  \nonumber\\
  && \times\left\langle \left[ \partial_{x_0}^{(k)}
  \Psi \left(\vec x + {\textstyle{\vec y\over 2}},x_0\right)\right]
  \left[ \partial_{x_0}^{(n-k)}
  \Psi^\dag \left(\vec x - {\textstyle{\vec y\over 2}},x_0\right)
  \right]\right\rangle\, .
 \end{eqnarray}
The full information on $W_4$ is equivalent to knowing {\em all} (and
not just the lowest)of its $p_0$ moments, and initializing them at
$t=-\infty$ requires the knowledge of {\em all} time derivatives of
the fields $\Psi$, $\Psi^{\dag}$ at $t=-\infty$ (or, equivalently, to
knowing the fields for all times).

This is clearly impractical. It follows from these considerations that
a  formulation of covariant transport theory as an initial value
problem for equal-time energy moments of the covariant Wigner function
must be accompanied by a truncation prescription which allows to
restrict the description to a finite number of such energy moments.

The strategy of the present paper will thus be to first derive from
the underlying field theoretic Lagrangian a set of covariant equations
of motion for the covariant Wigner operator $\hat W_4$ and then, by
taking $p_0$-moments of these covariant equations, set up the coupled
hierarchy of equal-time equations for the energy moments of $\hat W_4$
\cite{HIE,OCHS}. We will then study the truncation of this hierarchy.

In general, truncating the moment hierarchy will entail nontrivial
approximations. However, if the fermion fields evolve under the
influence of an {\em external} gauge field, i.e. they do not interact
with each other, neither directly via scattering nor indirectly via
backreaction of their electric current on the external field, their
time evolution is uniquely fixed once their values at $t=-\infty$ have
been specified. (The Dirac equation is a first order differential
equation in time.) It should thus be sufficient to specify initial
conditions for $W_3$ only, but not for any higher energy moments. This
can only be true if the time evolution of the system in the equal-time
framework can be described completely in terms of $W_3$ only. This
turns out to be correct in the sense that, in the external field
limit, only for $W_3$ a genuine equal-time transport equation must be
solved while all higher energy moments of $W_4$ can be obtained from
the solution of this equation via constraints. This will be discussed
in more detail in Sec.~\ref{sec4}.

\subsection{Formulation with orthogonal polynomials}
\label{sec2.2}

In Ref.~\cite{HIE} it was suggested to construct the equal-time
moments of the covariant Wigner density $W_4$ from orthogonal
polynomials of $p_0$ rather than from the powers of $p_0$ used in
Eq.~(\ref{moments}). Orthogonal polynomials have the advantage that
the complete function $W_4$ is easily reconstructed from the set of
moments using the orthogonality relations. Their disadvantage is
that, if they are defined over the full $p_0$ interval
$(-\infty,\infty)$, they require additional weight functions which
fall off sufficiently rapidly at $\pm \infty$. Only if one restricts
$p_0$ to a finite interval $[-a,a]$ one can use orthogonal polynomials
without weight function, the Legendre polynomials. This was done in
Ref.~\cite{HIE}. The price for doing so was the technical requirement
that the Wigner density $W_4$ falls off as a function of $p_0$
sufficiently fast near $\pm \infty$ that its contributions outside a
large but finite interval can be neglected.

If one wants to avoid this technicality and rather work with
orthogonal polynomials on the infinite $p_0$ interval, one must use
additional weight functions $g(p_0)$ in order to obtain suitable
orthogonality relations. We will now show that this does not really
lead to a proper single-time formulation. Using the reduced energy
variable $z=p_0/E$ with $E=\sqrt{{\vec p}^{\, 2} + m^2}$, the general
moment expansion then reads
 \begin{eqnarray}
 \label{orthpol}
  \fopw(x,\vec p, p_0) &=& \sum_{k=0}^\infty \frac{1}{N_k}
  \hat w_k(x,\vec p) \,  u_k(z) \, ,
 \\
 \label{orthpol1}
  \hat w_k(x,\vec p) &=& \int_\infty^\infty dz\, g(z)\, u_k(z)\,
  \fopw(x,\vec p, Ez) \, ,
 \\
 \label{orthpol2}
  \delta_{kl} N_k &=& \int_\infty^\infty dz\, g(z)\, u_k(z)\, u_l(z)\,  .
 \end{eqnarray}
Inserting the definition (\ref{wigop}) of the Wigner operator we get
 \begin{eqnarray}
 \label{orthwig}
    \hat w_k(x,\vec p) &=& \int dz\, g(z)
    \int \frac{d^4y}{(2\pi)^4}\,
    \left[
      u_k(i \partial_{y_0}/E) e^{-ip\cdot y}
    \right]
    \fop{\varrho}(x_+,x_-)
 \nonumber\\
    &=& \int \frac{d^4y}{(2\pi)^4}\ e^{i\vec p \cdot\vec y}
    \left[
      u_k(i\partial_{y_0}/E) \int dz\, g(z)\, e^{-iEzy_0}
    \right]
    \fop{\varrho} (x_+,x_-)
 \nonumber\\
    &=& \int \frac{d^4y}{(2\pi)^4}\ e^{i\vec p \cdot\vec y}
    \left[
      \int dz\ g(z) e^{-iEzy_0}
    \right]
    u_k(-i\partial_{y_0}/E)\fop{\varrho} (x_+,x_-)\, ,
 \end{eqnarray}
with partial integration in the last line, assuming that
$\fop{\varrho} (x_+, x_-)$ vanishes identically at
$\vert{x_0}_+\vert$, $\vert{x_0}_-\vert \to \infty$. However,
Eq.~(\ref{orthwig}) only then strictly corresponds to a formulation in
terms of only a single time parameter if
 \begin{equation}
  \label{orthdelt}
  \int_\infty^\infty dz\, g(z)\, e^{-iEzy_0} \stackrel{!}{=}
  \delta(Ez) = \delta (y_0).
\end{equation}
Since for orthogonal polynomials on the full $p_0$ interval $g(z)
 {\not\equiv} 1$ this is not possible, a strict single-time formulation
cannot be achieved in this way.

If we insist on a true single-time formulation there are thus only
two possibilities: (1) One tries to keep the computational
advantages of orthogonal polynomials. In this case one must restrict
$p_0$ to a finite interval and assume that outside that interval the
Wigner function vanishes \cite{HIE}. (2) One does not want to place
such a drastic restriction on $W_4$ but allows for, say, exponential
tails in the Wigner function at large $p_0$. In that case one should
use the moments (\ref{moments}) with simple powers of $p_0$ which are
not orthogonal on each other. We will here take the second choice.

Let us remark that the moment expansion of the covariant Wigner
operator needs not be restricted to energy moments. In \cite{MOM2} a
systematic expansion of the Wigner operator into moments of all four
components of the momentum $p$ was given. One can show that such an
expansion corresponds to the semiclassical $\hbar$ expansion or,
equivalently, to a gradient expansion around $x$. We will explain this
in more detail in Sec.~\ref{sec4} when we discuss the connection
between the expansion in energy  moments and a temporal gradient
expansion around $x_0$.

\subsection{Green functions and the Wigner operator}
\label{sec2.3}

Related to the initial value problem of non-equilibrium dynamics in
field theory is the doubling of the Hilbert space between zero and
finite temperature field theory \cite{SPEC}.

The connection between the non-equilibrium Wigner formalism and the
two-point Green functions of a field theory in thermodynamic
equilibrium is given via the density operator \cite{GRE}:
 \begin{equation}
 \label{green}
  i \,G(x_+, x_-) =
  \langle 0\vert T \, \fop{\varrho} (x_+,x_-)\vert 0 \rangle.
\end{equation}
$T$ is the path ordering operator for a given time contour $C$.
The most perfect formal agreement \cite{chou85,malfliet,SPEC} between
vacuum field theory, finite temperature equilibrium thermodynamics and
non-equilibrium dynamics is obtained if one chooses as the time
contour $C$ the ``Closed Time Path'' (CTP) \cite{KEL} which goes from
$t=-\infty$ to $t=+\infty$ infinitesimally above the real time axis,
then returns to $t=-\infty$ infinitesimally below the real time axis,
and (for equilibrium thermodynamics) finally proceeds vertically to
$-\infty-i\beta$ (where $\beta= 1/T$) in order to ensure the KMS
condition, i.e. the periodicity of the thermal Green functions in
imaginary time. The last piece is missing in non-equilibrium
situations in which case the Green functions do not satisfy the
KMS condition.

In the CTP formalism the 2-point Green functions can be split into
four pieces according to the four possibilities to distribute the two
time arguments on the upper and lower parts of the real-time
contour. In matrix notation one writes \cite{chou85,malfliet,SPEC}
 \begin{equation}
 \label{green2dim}
  G(x,y) =
  \left(
    \begin{array}{cc}
      G^c(x,y) & G^<(x,y)\\
      G^>(x,y) & G^a(x,y)
    \end{array}
  \right).
 \end{equation}
The covariant Wigner function $W_4$ is the 4-dimensional Wigner
transform (Fourier transform with respect to $y$) of
$G^< (x+y/2,x-y/2)$.

In thermodynamic equilibrium formal identities, in particular the KMS
condition, allow to express all four components of the $2\times 2$
matrix (\ref{green2dim}) in terms of a single real spectral density
\cite{SPEC}. In non-equilibrium situations, the absence of the KMS
condition implies that at least two components of (\ref{green2dim}) are
independent and probably should be kept for a complete dynamical
description. While $G^c$ and $G^a$ contain also information on the
virtual excitations of the vacuum, $G^<$ and $G^>$ (which are again
related by the spectral density) describe exclusively the dynamics of
the real excitations in the medium \cite{malfliet,malfliet1}. Here we
will be mostly interested in the case of the external fields where
knowledge of the dynamics of $G^<$ resp. $W_4$ is sufficient.

\section{Lorentz-covariant formulation}
\label{sec3}

In the following we will discuss the gauge theories QED and QCD
\cite{LINK,QTT}. We will consider only the fermion Wigner
operator and treat the photon or gluon degrees of freedom through
their classical gauge fields. Problems connected with a suitable gauge
choice will not be discussed here, the reader is referred to
\cite{LINK,elze88,RAD}.

\subsection{Gauge-covariant Wigner operator}
\label{sec3.1}

For a gauge covariant theory the Wigner operator must transform
covariantly under gauge transformations. This is achieved through a
gauge covariant definition of the density operator \cite{prl,LINK}
 \begin{equation}
 \label{dichteop1}
   \fop{\varrho} \left(x+{\textstyle{y\over 2}},
                       x-{\textstyle{y\over 2}} \right)
   = \overline{\psi}(x)
   e^{y \cdot D^{\dag}(x)/2} \otimes e^{-y \cdot D(x)/2} \psi(x),
\end{equation}
with the covariant derivative
 \begin{equation}
 \label{kov_abl}
   D_\mu(x) = \partial_\mu - ig \fopa_\mu (x)
 \end{equation}
instead of the partial derivative in the translation operator of
Eq.~(\ref{dichteopqm}). The direct product in Eq.~(\ref{dichteop1}) is
over spinor and (in the case of QCD) color indices. Both the fields
${\bar \psi}_\beta^a$, $\psi_\alpha^a$, $a = 1,2,\dots,N_c$, and the
covariant translation operators are group elements of $SU(N_c)$ albeit
in different representations.

With (\ref{dichteop1}) the Wigner operator reads
 \begin{equation}
 \label{wigopkov}
  \fopw (x,p) = \int \frac{d^4 y}{(2\pi)^4}\, e^{-ip\cdot y}
  \, \bar\psi(x) e^{y \cdot D^\dag (x)/2}
        \otimes  e^{-y \cdot D(x)/2} \psi(x).
 \end{equation}
Similar manipulations as in Eq.~(\ref{dichteopqm}) show that using the
covariant instead of the partial derivative results in $p$ being the
physical or kinetic momentum rather than the canonical one
\cite{prl,LINK}. The gauge covariant translation operators in
(\ref{dichteop1},\ref{wigopkov}) can be rewritten as link operators
\cite{LINK}
 \begin{equation}
 \label{link}
   \fop{\varrho} \left(x+{\textstyle{y\over 2}},
                       x-{\textstyle{y\over 2}} \right)
   = \overline{\psi}\left(x+{\textstyle{y\over 2}}\right)
     U\left(x+{\textstyle{y\over 2}},x\right) \otimes
     U\left(x,x-{\textstyle{y\over 2}},x\right)
     \psi\left(x-{\textstyle{y\over 2}}\right)
 \end{equation}
with
 \begin{equation}
 \label{linkdef}
     U(a,b) = P \exp\left( {ig\over\hbar c} \int_b^a dz^\mu \hat
         A_\mu(z) \right)\, ,
 \end{equation}
where the integral in the exponent is taken along the straight path
connecting $b$ and $a$, and $P$ denotes path ordering of the operators.
The link operators are unitary and fulfill
 \begin{eqnarray}
  U^\dag(a,b) &=& U^{-1}(a,b) = U(b,a)
 \nonumber\\
  U(a,b) \, U(b,a) &=& U(a,a) = {\bf 1}.
 \end{eqnarray}
The density operator can then be rewritten as
 \begin{equation}
 \label{defdichteop}
  \fop{\varrho}\left(x+{\textstyle{y\over 2}},
                     x-{\textstyle{y\over 2}}\right)
  = \overline{\psi}\left(x+{\textstyle{y\over 2}}\right)
        U\left(x+{\textstyle{y\over 2}},x\right) \otimes
        U\left(x,x-{\textstyle{y\over 2}}\right)
        \psi\left(x-{\textstyle{y\over 2}}\right) \, .
 \end{equation}

In the following we will use the conventions from \cite{LINK}
for an operator $\fop{O}$ operating on the color and spinor indices of
$\fop{\varrho}$:
 \begin{eqnarray}
 \label{konvention1}
  \fop{O}\, \fop{\varrho}\left(x+{\textstyle{y\over 2}},
                               x-{\textstyle{y\over 2}}\right)
  &{=}& \overline{\psi}\left(x+{\textstyle{y\over 2}}\right)
        U\left(x+{\textstyle{y\over 2}},x\right) \otimes
        \fop{O} U\left(x,x-{\textstyle{y\over 2}}\right)
        \psi\left(x-{\textstyle{y\over 2}}\right)
 \nonumber\\
 \label{konvention2}
  \fop{\varrho}\left(x+{\textstyle{y\over 2}},
                     x-{\textstyle{y\over 2}}\right) \,\fop{O}
  &{=}& \overline{\psi}\left(x+{\textstyle{y\over 2}}\right)
        U\left(x+{\textstyle{y\over 2}},x\right) \fop{O} \otimes
        U\left(x,x-{\textstyle{y\over 2}}\right)
        \psi\left(x-{\textstyle{y\over 2}}\right). \quad
\end{eqnarray}
Color octet objects in QCD will be denoted as matrices, $O = O_a t_a$,
with $t_a$ being the generators in the fundamental representation of
the gauge group. They are given in Appendix~\ref{appa} together with
useful color trace formulae. Occasionally we will also use their
adjoint representation $T_a$ with
 \begin{equation}
 \label{adjoint}
  (T_a)_{bc} = - i\hbar f_{abc}
 \end{equation}
and norm
 \begin{equation}
  {\rm tr}(T_a T_b) = 3\hbar^2\delta_{ab}.
 \end{equation}
With our normalization the $t_a$ have eigenvalues $\sim {\hbar\over 2}$
proportional to the physical color-spin of quarks, while the $T_a$
have eigenvalues $\sim\hbar$ corresponding to the color spin of gluons.

>From the Wigner operator one obtains the operators for the fermionic
baryon (charge) and color currents via
\begin{eqnarray}
  \label{stroeme}
  j_\nu^{\mbox{\tiny baryon}}(x) &=&
  -{\rm tr} \left(\gamma_\nu \bar \psi (x) \psi(x)\right) =
  \int d^4 p\ {\rm tr}
  \left({\bf 1}_{3\times3} \gamma_\nu \fopw(x,p)\right), \\
  \label{stroeme2}
  j_{\nu a}^{\mbox{\tiny color}}(x) &=& -{\rm tr}
  \left(\gamma_\nu \frac{\lambda_a}{2}\bar \psi (x) \psi(x)\right) =
  \int d^4 p\ {\rm tr}
  \left(\frac{\lambda_a}{2} \gamma_\nu \fopw(x,p)\right).
\end{eqnarray}
The baryon current projects onto the color singlet, the color current
onto the color octet contributions of the Wigner operator. In analogy
we will later on perform a color decomposition for the equations of
motion of the Wigner operator.

As spinor matrices $\fop{\varrho}$ and $\fopw$ transform according to
 \begin{equation}
   \fop{\varrho}^\dag \left(x+{\textstyle{y\over 2}},
                            x-{\textstyle{y\over 2}}\right)
   = \gamma^0  \fop{\varrho}\left(x-{\textstyle{y\over 2}},
                                  x+{\textstyle{y\over 2}}\right)\gamma^0
 \end{equation}
with exchanged arguments $x_\pm$ and
 \begin{equation}
   \fopw^\dag (x,p) = \gamma^0 \fopw (x,p) \gamma^0.
\end{equation}

\subsection{Equations of motion for the Wigner operator}
\label{sec3.2}

The basis for the dynamics of strong interactions is the Lagrange
density of QCD (since flavor quantum numbers don't matter for our
purpose we consider only one quark flavor):
\begin{equation}
  \label{lagrangeQCD}
  {\mathcal L}_{\mbox{\scriptsize QCD}} =
  i\hbar c \overline{\psi} (x) \gamma^\mu D_\mu(x) \psi(x)
  -m c^2 \overline{\psi} \psi - \frac{1}{4\hbar c} \fopf_{\mu\nu}(x)
  \fopf^{\mu\nu}(x).
\end{equation}
$\psi(x)$, $\overline{\psi}(x) = \psi^\dag \gamma^0$ are the spinor
fields,
 \begin{equation}
 \label{kov_abl_hbar}
    D_\mu(x) = \partial_\mu - \frac{ig}{\hbar c} \fopa_\mu(x),
 \end{equation}
is the covariant derivative (see Eq.~(\ref{kov_abl})), $m$ is the bare
mass, and $\fopf_{\mu\nu}(x)$ is the field strength tensor. The latter
is defined through the commutator of the covariant derivative,
 \begin{equation}
 \label{deffmunu}
  \fopf_{\mu\nu}(x) \equiv - \frac{\hbar c}{ig}[D_\mu , D_\nu] =
  (\partial_\mu \fopa_\nu (x)) - (\partial_\nu \fopa_\mu (x))
  - \frac{ig}{\hbar c} [\fopa_\mu (x) , \fopa_\nu (x)],
\end{equation}
and thus satisfies the Jacobi identity
 \begin{equation}
 \label{jacobi1}
  [D_\alpha(x), \fopf_{\mu\nu}(x)] + [D_\nu(x), \fopf_{\alpha\mu}(x)] +
  [D_\mu(x), \fopf_{\nu\alpha}(x)] = 0.
\end{equation}
Since for the semiclassical and gradient expansions we will later need
to count powers of $\hbar$ we displayed them here once explicitly.
Until further notice we will, however, from now on return again to
natural units with  $\hbar=c=1$.

Inserting the Dirac equation and its adjoint,
\begin{eqnarray}\label{dirac}
  (i\gamma^\mu D_\mu(x) -m)\psi(x) &=& 0,\\
  \label{diracad}
  \overline{\psi}(x)(i\gamma^\mu D^{\dag}_\mu(x) -m) &=& 0,
\end{eqnarray}
(with the hermitean adjoint of the partial derivative defined as before
as acting to the left) into the definition (\ref{defdichteop}) of the
density operator and, following \cite{LINK,QTT}, pulling the
derivatives out in front by using the derivative rules for link
operators as given in Appendix~\ref{appc}, one finds the following
equations of motion for the covariant fermion Wigner operator
\cite{LINK}:
 \begin{eqnarray}
 \label{bewegwig1}
  2m \fopw (x,p) \!&{=}&\!\gamma^\mu
  \Bigl[
    \Bigl(
      i D_\mu (x)+p_\mu-g\int_{-1/2}^0ds\,(1-2s)\,
      {^{[x]}\fopf_{\nu\mu}}(x+is\partial_p)i\partial_p^\nu
    \Bigr)\fopw(x,p)
 \nonumber\\
  && + \fopw(x,p)
    \Bigl(
      -i D_{\mu}(x)+p_\mu+g\int_{-1/2}^{0}\!\!\! ds\,(1+2s)\,
      {^{[x]}\fopf_{\nu\mu}}(x+is\partial_p)i\partial_p^\nu
    \Bigr)
  \Bigr],
 \nonumber\\ \quad
 \\
  \label{bewegwig2}
  2m \fopw(x,p)\!&{=}&\!\Bigl[
    \Bigl(
      -iD_\mu(x) + p_\mu +
      g \int_{-1/2}^0 ds\,(1 + 2s)\,
      {^{[x]}\fopf_{\nu\mu}}(x+is\partial_p)i\partial_p^\nu
    \Bigr)\fopw(x,p)
  \nonumber\\
  &&+\fopw (x,p)\Bigl(
      iD_\mu(x)
      + p_\mu - g\int_{-1/2}^{0} ds\,(1 - 2s)\,
      {^{[x]}\fopf_{\nu\mu}}(x+is\partial_p)i\partial_p^\nu
    \Bigr)
  \Bigr]\gamma^\mu.
 \nonumber\\
 \end{eqnarray}
Here, for notational convenience, momentum derivatives standing to the
right of the Wigner operator are defined in the sense of partial
integration as
 \begin{equation}
 \label{padj}
   \hat W(x,p) \partial_p^{\nu_1} \dots \partial_p^{\nu_k} =
   (-1)^k \partial_p^{\nu_k} \dots \partial_p^{\nu_1} \hat W(x,p)\, .
 \end{equation}
The operator
 \begin{equation}
  \label{schwingerstringdef}
  {^{[x]}\fopf}_{\nu\mu}(z(s)) \equiv
  U(x,z(s))\fopf_{\nu\mu}(z(s))U(z(s),x)
  \quad \mbox{with}\ z(s) = x + sy
 \end{equation}
is called Schwinger string and connects the field strength tensor at point
$z$ gauge covariantly with the point $x$ along a straight line path.
The Schwinger string is the essential quantity to describe the gluonic
degrees of freedom as non-local operators.

To get a more compact notation we define generalized non-local
momentum and derivative operators $\Pi_\mu$ and $\Delta_\mu$,
respectively\footnote{Our notation here differs somewhat from that of
  Ref.~\cite{MOM}; it is optimized for a simultaneous description of
  non-abelian and abelian gauge interactions, while in \cite{MOM} only
  the abelian case of QED with classical external fields was
  studied. In that limit $\Delta$ and $\Pi$ as defined here reduce to
  the definitions given in \cite{MOM}, see
  Eqs.~(\ref{qedopa},\ref{qedopb}) below.}
(here we display $\hbar$ and $c$ explicitly for later use):
 \begin{eqnarray}
 \label{allgop}
  \Pi_\mu &=& p_\mu + {2g\over c}\int_{-1/2}^0 ds\,
      is\hbar \partial_p^\nu\
      {^{[x]}\fopf_{\nu\mu}}(x+is\hbar\partial_p) \, ,
 \\
 \label{allgop2}
  \Delta_\mu & = & \hbar D_\mu(x) +
  {ig\over c}\int_{-1/2}^0 ds\, i\hbar\partial_p^\nu\
  {^{[x]}\fopf_{\nu\mu}}(x+is\partial_p) \, .
 \end{eqnarray}
The equations of motion for the Wigner operator are then
\begin{eqnarray}
  \label{bewegkom1}
  2 m \fopw(x,p) &=&
  \gamma^\mu
  \left(
    \{\Pi_\mu,\fopw(x,p)\} + i[\Delta_\mu,\hat W(x,p)]
  \right),
 \\
  \label{bewegkom2}
  2 m \fopw(x,p) &=&
  \left(
    \{\Pi_\mu,\fopw(x,p)\} - i[\Delta_\mu,\fopw(x,p)]
  \right) \gamma^\mu.
\end{eqnarray}
With $\Pi_\mu^\dag = \Pi_\mu, (i\Delta_\mu)^\dag = i \Delta_\mu$ and
$[\Pi_\mu,\gamma^0] = [ \Delta_\mu,\gamma^0] = 0$ we find
 \begin{equation}
  \gamma^0(\Pi_\mu {\pm} i \Delta_\mu)^\dag \gamma^0
  = \Pi_\mu {\pm} i \Delta_\mu,
 \end{equation}
which means that Eqs.~(\ref{bewegkom1}) and (\ref{bewegkom2}) are
adjoint to each other: $\gamma^0 (\ref{bewegkom1}) \gamma^0 =
(\ref{bewegkom2})^\dag$. Adding and subtracting the two equations
gives
 \begin{eqnarray}
 \label{bewwig1}
  4m\fopw(x,p) &=&\{\gamma^\mu,\{\Pi_\mu,\fopw(x,p)\}\} +
  i [\gamma^\mu,[\Delta_\mu,\fopw(x,p)]]\, ,
 \\
 \label{bewwig2}
  0 &=& [\gamma^\mu,\{\Pi_\mu,\fopw(x,p)\}] +
  i \{\gamma^\mu,[\Delta_\mu,\fopw(x,p)]\}\, .
 \end{eqnarray}
This corresponds to a separation into real and imaginary parts. The
generalized momentum operator is always working as an anti-commutator,
the generalized derivative operator always as a commutator on
$\fopw$. The only difference between the two equations is that for
the spinor structure commutators are replaced by anti-commutators.
Defining therefore
 \begin{equation}
  \label{kommutatorplusminus}
  [ \hat A, \hat B]_- \equiv [ \hat A, \hat B]\, , \qquad
  [ \hat A, \hat B]_+ \equiv \{ \hat A, \hat B \}\, ,
 \end{equation}
we can combine the two equations as
\begin{equation}
  \label{bewwigtot}
    [2m {\bf 1}, \fopw(x,p)]_{\pm} =
    [\gamma^\mu,\{\Pi_\mu, \fopw
    (x,p)\}]_{\pm} +
    i[\gamma^\mu,[\Delta_\mu,\fopw (x,p)]]_\mp.
\end{equation}
These are the equations of motion for the Wigner operator for
particles with spin $1/2$.

The equations are closed by the Yang-Mills-equations for the field
strength tensor. These couple to the Wigner operator for the fermion
fields via the current $j_\nu$:
 \begin{equation}\label{bewegfmunu}
  [D^\mu(x), \fopf_{\mu\nu}(x)] =
  j_\mu(x) = t_a \mbox{tr}\ t_a \gamma_\nu \fopw (x,p).
 \end{equation}
The trace sums over spinor and color degrees of freedom and contains an
integral over momentum space.

Eqs.~(\ref{bewwigtot}) and (\ref{bewegfmunu}) contain the full
dynamical information of the system on the operator level.

\subsection{Gradient expansion}
\label{sec3.3}

The classical limit of these equations is obtained via a gradient
expansion of the non-local operators $\Pi$ and $\Delta$ in
Eqs.~(\ref{allgop},\ref{allgop2}). It is generated by a Taylor
expansion for the Schwinger string around $x$ in powers of $is\hbar
\partial_p$. As such it is automatically also an expansion in powers
of $\hbar$; since, however, the color decomposition discussed in the
following subsection introduces additional powers of $\hbar$, we will
refer to the semiclassical expansion in the sense of Wigner and
Kirkwood only as the ``gradient expansion''. As we will see it is
actually an expansion in {\em covariant} derivatives. It thus
preserves gauge covariance.

The Taylor expansion of the Schwinger string is given by
 \begin{equation}
  \label{tayl}
  {^{[x]}\fopf_{\nu\mu}}(x+is\hbar \partial_p) =
  \sum_{k=0}^\infty \frac{1}{k!} (is\hbar)^k
  \left[
    (\partial_{p\alpha}\partial_y^\alpha)^k \
    {^{[x]\!}\fopf_{\nu\mu}}(x+y)
  \right]_{y=0}.
 \end{equation}
The connection between the expansion of the Schwinger string into
local operators $[(\partial_p\cdot \partial_y)^n\
{^{[x]\!}F_{\nu\mu}(x+y)}]_{y=0}$ and a gradient expansion in
covariant derivatives $D_\mu(x)$ can be shown from the expansion of
the Schwinger string in an exponential as derived in \cite{LINK}:
 \begin{eqnarray}
 \label{sstrdef1}
  {^{[x]}\fopf_{\mu\nu}(x + is \hbar\partial_p)}
  &=& \exp[is\hbar\partial_p^\alpha \tilde{\mathcal D}_\alpha(x)]
  \fopf_{\mu\nu}(x)
\\
\label{sstrdef1b}
  &=& \sum_{k=0}^{\infty} \frac{1}{k!}
(is\hbar)^k \left(\partial_p^{\alpha}
  \tilde{\mathcal D}_{\alpha}(x)\right)^k \fopf_{\mu \nu} (x).
 \end{eqnarray}
$\tilde{\mathcal D}_\alpha(x)$ is defined as the covariant derivative
in the adjoint representation \cite{LINK}:
 \begin{equation}\label{deriv}
  \tilde{\mathcal D}_{\alpha}(x) \fopf_{\mu\nu}(x) \equiv
  [ D_\alpha(x) , \fopf_{\mu\nu}(x)].
 \end{equation}
Comparison between Eqs.~(\ref{tayl}) and (\ref{sstrdef1b}) gives the
following identity (see also Appendix~\ref{appc} for a direct
calculation from of the l.h.s. of Eq.~(\ref{deriv}) for $n=1$):
 \begin{equation}
  \label{schwinger_abl}
  \left[(\partial_p\cdot \partial_y)^n \
    {^{[x]\!}\fopf_{\nu\mu}(x+y)}\right]_{y=0} =
    \lbrack{\!}\lbrack (\partial_p \cdot D(x))^n
    ,\fopf_{\nu\mu}(x) \rbrack{\!}\rbrack \, .
 \end{equation}
Here we defined a generalized commutator through
 \begin{eqnarray}
 \label{doppelkommutator}
  \lbrack{\!}\lbrack A^n , B \rbrack{\!}\rbrack &=&
  \Bigl[ A , \lbrack{\!}\lbrack A^{n-1} , B \rbrack{\!}\rbrack
  \Bigr],
 \nonumber\\
  \lbrack{\!}\lbrack A^{0} , B \rbrack{\!}\rbrack &=& B,
 \end{eqnarray}
or explicitly
 \begin{equation}
  \label{doppelkommutatorexpl}
  \lbrack{\!}\lbrack A^n , B \rbrack{\!}\rbrack =
  \underbrace{[A ,\, [A ,\dots ,\, [A}_{\D n \ \mbox{times}},B]\dots]]
  =  \sum_{k=0}^n {n \choose k} (-1)^k A^{n-k} B A^k.
 \end{equation}
Using Eqs.~(\ref{tayl},\ref{schwinger_abl}) in
Eqs.~(\ref{allgop},\ref{allgop2}) and performing the $s$-integration
we find for the non-local operators
\begin{eqnarray}
  \label{D_op1}
  \Delta_\mu &=& \hbar D_\mu - \hbar \frac{g}{2c}
  \sum_{k=0}^\infty \left(-\frac{i\hbar}{2}\right)^k
  \frac{1}{(k+1)!}
  \lbrack{\!}\lbrack (\partial_p \cdot D(x))^k , \fopf_{\nu\mu}(x)
  \rbrack{\!}\rbrack \partial_p^\nu,
  \\
  \label{P_op1}
  \Pi_\mu &=& p_\mu - \hbar \frac{ig}{2c}
  \sum_{k=0}^\infty \left(-\frac{i\hbar}{2}\right)^{k}
  \frac{k+1}{(k+2)!}
  \lbrack{\!}\lbrack (\partial_p \cdot D(x))^{k} , \fopf_{\nu\mu}(x)
  \rbrack{\!}\rbrack \partial_p^\nu.
\end{eqnarray}
Using the definition (\ref{deffmunu}) for the field strength tensor
and in (\ref{P_op1}) the identity
 \begin{equation}
  \label{P_id1}
  D_\mu = [\partial_p \cdot D(x), p_\mu]
 \end{equation}
we finally get
 \begin{eqnarray}
 \label{D_op2}
  \Delta_\mu &=& D_{\mu}(x) + \sum_{k=1}^\infty \frac{1}{k!}
  \lbrack{\!}\lbrack (-{\textstyle{i\hbar\over 2}}\partial_p \cdot
  D(x))^k , \hbar D_\mu(x) \rbrack{\!}\rbrack
 \\
 \label{P_op2}
  \Pi_\mu &=& p_{\mu} + 2 \sum_{k=1}^\infty \frac{1-k}{k!}
  \lbrack{\!}\lbrack (-{\textstyle{i\hbar\over 2}}\partial_p \cdot
  D(x))^{k} , p_\mu \rbrack\!\rbrack .
 \end{eqnarray}
Eqs.~(\ref{D_op2},\ref{P_op2}) identify $\Delta_\mu$ and $\Pi_\mu$ as
non-local generalizations of the covariant derivative $\hbar D_\mu$
and the momentum $p_\mu$.

The (hermitean) operator for the expansion about the classical limit
is thus given by
 \begin{equation}
 \label{delta-op}
   \triangle = \partial_p^W \cdot iD^F(x),
 \end{equation}
with $\hbar$ as the expansion parameter. The superscript $W$ means
that the momentum derivative acts on the momentum dependence of the
Wigner operator. The superscript $F$ on the covariant derivative
reminds us that it acts only on the field strength operator following
it, see Eqs.~(\ref{D_op1},\ref{P_op1}). From
Eqs.~(\ref{D_op2},\ref{P_op2}) or (\ref{P_id1}) we find that
$\Delta_\mu$ is one order higher in $\triangle$ and $\hbar$ than
$\Pi_\mu$.

Since, at least for QED, the measurable classical fields are
the field strengths $F_{\mu\nu}(x)$ we will use for the gradient
expansion Eqs.~(\ref{D_op1},\ref{P_op1}) rather than
Eqs.~(\ref{D_op2},\ref{P_op2}).

The validity of the gradient expansion requires
 \begin{equation}
  \hbar \triangle =  \hbar \partial_p^W \cdot iD^F(x) \ll 1,
 \end{equation}
i.e. that the Wigner function is sufficiently smooth in momentum space
and the field strengths vary sufficiently slowly (in a covariant
sense) in coordinate space. The corresponding length scales on
which these functions vary must satisfy
\begin{equation}
  (\Delta p)_W \cdot (\Delta x)_F \gg \hbar.
\end{equation}

Inserting the expansions (\ref{D_op1},\ref{P_op1}) together with the
Ansatz
 \begin{equation}
  \label{hbarx}
  \fopw (x,p) = \sum_{k=0}^\infty \hbar^k \fopw^{(k)} (x,p)
 \end{equation}
for the Wigner operator into the equations of motion (\ref{bewwigtot})
we obtain
 \begin{eqnarray}
  \lefteqn{[2m {\bf 1},  \fopw^{(n)}]_{\pm}\ =\qquad \qquad \qquad }
  \\&&
  \!\!\!\!\biggl[
  \gamma^\mu, 2p_\mu \fopw^{(n)} -
  \frac{ig}{2c} \sum_{k=0}^{n-1}
  \frac{k+1}{(k+2)!}
  \Bigl\{
  \lbrack\!\lbrack
  (\frac{-i}{2}
  \Delta)^{k}, \fopf_{\nu\mu}(x)
  \rbrack\!\rbrack
  \partial_p^\nu,
  \fopw^{(n-k-1)}
  \Bigr\}
  \biggr]_{\pm}
  \nonumber\\
  &&
  \!\!\!\!+ i\biggl[
  \gamma_\mu, \left[D_\mu(x),\fopw^{(n-1)} \right]
  - \frac{g}{2c} \sum_{k=0}^{n-1}
  \frac{1}{(k+1)!}
  \Bigl[
  \lbrack\!\lbrack
  (\frac{-i}{2}
  \Delta)^{k},
  \fopf_{\nu\mu}(x)
  \rbrack\!\rbrack
  \partial_p^\nu,
  \fopw^{(n-k-1)}
  \Bigr]
  \biggr]_{\mp}\!\!\!.
  \nonumber
\end{eqnarray}
With Table~\ref{tabhbarord} for the zeroth and first order terms in
the non-local operators we get
\begin{equation}
  \label{wig0ord}
  [2m {\bf 1},  \fopw^{(0)}]_{\pm} =
  [\gamma^\mu, 2p_\mu \fopw^{(0)}]_{\pm}
\end{equation}
to zeroth order and
\begin{eqnarray}
  \label{wig1ord}
  [2m {\bf 1},  \fopw^{(1)}]_{\pm} &=&
  \biggl[
  \gamma^\mu, 2p_\mu \fopw^{(1)} -
  \frac{ig}{4c}
  \Bigl\{
  \fopf_{\nu\mu}(x)
  \partial_p^\nu,
  \fopw^{(0)}
  \Bigr\}
  \biggr]_{\pm}
  \nonumber\\ &&
  + i\biggl[
  \gamma_\mu,
  \Bigl[
    D_\mu(x) - \frac{g}{2c} \fopf_{\nu\mu}(x) \partial_p^\nu,\fopw^{(0)}
  \Bigr]
  \biggr]_{\mp}
\end{eqnarray}
to first order in $\hbar$.

We will examine the resulting classical limit more closely in the
context of the spinor decomposition for the Wigner operator since
there we will find an easy connection to the classical mass shell
condition and transport equation.

\subsection{Color decomposition}
\label{sec3.4}

For QCD it is useful to decompose the Wigner operator into its
(observable) color singlet and its (unobservable) color octet
contributions \cite{prl}:
\begin{eqnarray}
  \label{wigcol}
  \fopw (x,p) &=& \fopw_s(x,p) {\bf 1} + \fopw^a(x,p) \ t^a
  \\
  \fopw_s(x,p) &=& \frac{1}{3} {\rm tr} \, \fopw(x,p) ,
  \\
  \fopw^a(x,p) &=&  \frac{2}{\hbar^2} {\rm tr} \, t^a\ \fopw(x,p)
\end{eqnarray}
The color decomposition of the Schwinger string \cite{COL} can be
calculated to any fixed order of the gradient expansion from the
explicit form
\begin{eqnarray}
  \label{schwingerentw}
    {^{[x]}\fopf_{\mu\nu}}(x + is \hbar \partial_p) &=&
    \sum_{n=0}^\infty \frac{(is\hbar)^n}{n!}
    \lbrack\!\lbrack (\frac{1}{2}
    \partial_p^\alpha \partial_\alpha {\bf 1} + \frac{ig}{\hbar c}
    \partial_p^\alpha \fopa_\alpha^a(x)t^a)^n
    \ ,\ \fopf_{\nu\mu}^b(x) t^b \rbrack\!\rbrack.
\end{eqnarray}
Combining the two decompositions we find to zeroth order in the
gradient expansion
\begin{eqnarray}
  \label{col0hbar}
  \mbox{singlet:} \quad [ m {\bf 1}   - \gamma^\mu p_\mu  ,
  \fopw_s^{(0)}]_{\pm} &=& 0,
  \\
  \label{col0hbarb}
  \mbox{octet:} \quad  [ m {\bf 1}   - \gamma^\mu p_\mu  ,
  \fopw^{(0)a}]_{\pm} &=& 0,
\end{eqnarray}
and to first order
\begin{eqnarray}
  \label{col1hbar}
  \lefteqn{
    \mbox{singlet:} \quad [ m {\bf 1} - \gamma^\mu p_\mu  ,
    \fopw_s^{(1)}]_{\pm} +
    \frac{ig }{4c} [\gamma^\mu,\partial_\mu \fopw_s^{(0)}]_{\mp} =
      \qquad\qquad\qquad}
  \nonumber\\
  &&
  -\frac{ig\hbar^2}{24c}
  \left(
    \frac{1}{2}
    [\gamma^\mu,\partial_p^\nu[\fopf_{\nu\mu}^a(x),\fopw^{(0)a}]]_{\pm}
    + [\gamma^\mu,\partial_p^\nu \{\fopf_{\nu\mu}^a(x),\fopw^{(0)a}\}]_{\mp}
  \right.
  \nonumber\\ &&
  \left.\qquad\quad
    - \frac{ig}{\hbar c}
    [\gamma^\mu,[\fopa_{\mu}^a(x),\fopw^{(0)a}] ]_{\mp}
  \right)
  \\
  \label{col1hbarb}
  \lefteqn{
    \mbox{octet:} \quad [ m {\bf 1} - \gamma^\mu p_\mu  ,
    \fopw^{(1)a}]_{\pm} +
    \frac{ig }{4c} [\gamma^\mu,\partial_\mu \fopw^{(0)a}]_{\mp} =
    \qquad\qquad\qquad}
  \nonumber\\
  &&
  -\frac{ig}{4c}\left(
    \frac{1}{2}
    [\gamma^\mu,\partial_p^\nu[\fopf_{\nu\mu}^a(x),\fopw_s^{(0)}]]_{\pm}
    + [\gamma^\mu,\partial_p^\nu \{\fopf_{\nu\mu}^a(x),\fopw_s^{(0)}\}]_{\mp}
  \right.
  \nonumber\\ &&
  \left.\qquad
    - \frac{ig}{\hbar c}
    [\gamma^\mu,[\fopa_{\mu}^a(x),\fopw_s^{(0)}] ]_{\mp}
  \right)
  \nonumber\\
  &&
  - \frac{ig\hbar}{8c}d^{abc}\left(
  \frac{1}{2}
    [\gamma^\mu,\partial_p^\nu [\fopf_{\nu\mu}^b(x),\fopw^{(0)c}]]_{\pm}
    + [\gamma^\mu,\partial_p^\nu \{\fopf_{\nu\mu}^b(x),\fopw^{(0)c}\}]_{\mp}
  \right.
  \nonumber\\ &&
  \left.\qquad\qquad\quad
    - \frac{ig}{\hbar c}
    [\gamma^\mu,[\fopa_{\mu}^b(x),\fopw^{(0)c}] ]_{\mp}
  \right)
  \nonumber\\
  &&
  + \frac{g\hbar}{8c}f^{abc}\left(
    \frac{1}{2}
    [\gamma^\mu,\partial_p^\nu \{\fopf_{\nu\mu}^b(x),\fopw^{(0)c}\}]_{\pm}
    + [\gamma^\mu, \partial_p^\nu [\fopf_{\nu\mu}^b(x),\fopw^{(0)c}]]_{\mp}
  \right.
  \nonumber\\ &&
  \left.\qquad\qquad\quad
    - \frac{ig}{\hbar c}
    [\gamma^\mu,\{\fopa_{\mu}^b(x),\fopw^{(0)c}\} ]_{\mp}
  \right).
\end{eqnarray}
Here we used the trace formulae given in Appendix~\ref{appa} and the
relation $[AB,C] = A[B,C] + [A,C]B$.

These equations exemplify a general result which to prove we will
leave to the reader: At any given order $n$ of the gradient expansion,
the ``leading'' color singlet and octet contributions
$\fopw^{(n)}_s,\fopw^{(n)}_a$ decouple from each other; the color
singlet and octet sectors couple only via the lower order components
$\fopw^{(n-1)}$.

Note that the color decomposition brings in additional powers of
$\hbar$ which are not connected to the gradient expansion, but are
related to the question to what extent color can be treated
classically \cite{KFT}. One should carefully differentiate between the
limit of classical color and the classical limit in the kinetic sense
which is defined through the gradient expansion.

The transition back from QCD to QED is made by formally setting the
color matrices equal to {\bf 1}. Nonetheless the fields $\fopa_\mu(x)$
are still operators in Fock space and neither commute with the Wigner
operator nor with each other. The equations of motion for the Wigner
operator are thus formally identical in QED and QCD, only that in QCD
the (anti-)commutators refer additionally to the color indices.

\subsection{External field limit and Wigner function}
\label{sec3.5}

Equations of motion for the Wigner {\em function} (rather than the
Wigner {\em operator}) are obtained by taking ensemble expectation
values of the operator equations. Writing for operators $\fop{O}$
working on $\fopw$
\begin{eqnarray}
  \langle [ \fop{O} , \fopw]_\pm\rangle &=&
  [\langle \fop{O}\rangle , \langle \fopw\rangle ]_\pm
  +   \langle [ \fop{O} , \fopw]_\pm\rangle -
  [\langle \fop{O}\rangle , \langle \fopw\rangle ]_\pm
  \nonumber\\
  &\equiv& [\langle \fop{O}\rangle , \langle \fopw\rangle ]_\pm
  + C(\fop{O},\fopw),
\end{eqnarray}
we can split the resulting equations into two parts: One part
contains only one particle Wigner functions $\langle \fopw \rangle$
and mean (gluon) fields (Vlasov term), while the other part describes
the correlations $C(\fop{O},\fopx)$ (collision term). (For a
discussion of the latter see the formulation with Green functions in
\cite{KOR}.) In this subsection we will study the structure of the
equations for the Wigner {\em function} under the assumption that the
correlations can be neglected and that the fields $\fopf_{\mu\nu}$ can
be approximated by external fields $\langle \fopf_{\mu\nu} \rangle =$
$F_{\mu\nu}$. Then
\begin{equation}
    \langle \fop{\varrho} \fopf_{\mu\nu}\rangle = \langle \fop{\varrho}
    \rangle F_{\mu\nu},
\end{equation}
and similarly for higher order operator products. This mean field
approximation will then lead to a generalized Vlasov equation
where collision terms, i.e. multi particle correlations, are
neglected.

For QCD the equations of motion for the Wigner function in an external
gauge field are formally identical to those for the Wigner operator
interacting with arbitrary gauge field operators, due to the
non-commuting color structure. For QED the equations simplify somewhat
since now the commutators of the gauge field with the Wigner function
vanish, and only the non-trivial commutators with the Dirac matrices
survive.

In the external field limit the equations of motion for the electron
Wigner function in QED read
 \begin{equation}
  \label{qed_gl}
  2m [{\bf 1}, W(x,p)]_\pm  =
  2 {\mathcal P}_\mu [ \gamma^\mu , W(x,p)]_\pm
  + i{\mathcal D}_\mu [\gamma^\mu , W(x,p)]_\mp ,
 \end{equation}
with (c.f. \cite{MOM})
 \begin{eqnarray}
  \label{qedopa}
  {\mathcal P}_\mu &=& p_\mu + e \int_{-1/2}^{1/2} ds\ s
  F_{\mu\nu}(x+is\partial_p) \ i\partial_p^\nu,
  \\
  \label{qedopb}
  {\mathcal D}_\mu &=&
  \partial_\mu + ie \int_{-1/2}^{1/2} ds
  F_{\mu\nu}(x+is\partial_p) \ i \partial_p^\nu,
 \end{eqnarray}
where $e = -g$ is the electric coupling constant. Using
Table~\ref{tabhbarordQED}, the two leading terms of the gradient
expansion yield the semiclassical equations
\begin{equation}
  \label{qed0hbar}
   [m {\bf 1} - p_\mu \gamma^\mu, W^{(0)}]_{\pm} = 0
\end{equation}
in zeroth order and
 \begin{equation}
  \label{qed1hbar}
  [m {\bf 1} - p_\mu \gamma^\mu, W^{(1)}]_{\pm}
  + \frac{ig}{4c} \partial_\mu [\gamma^\mu,W^{(0)}]_{\mp}
  =  \frac{ig}{2 c}F_{\mu\nu}(x)\partial_p^\nu
  [\gamma^\mu,W^{(0)}]_{\mp}
 \end{equation}
in first order of the expansion.

For QCD the zeroth order equation is again (\ref{qed0hbar}), while the
first order equation becomes after color decomposition
 \begin{eqnarray}
  \label{col1aeussere}
    \mbox{singlet:} \quad &{[}&
    m {\bf 1} - \gamma^\mu p_\mu,W_s^{(1)}]_\pm
    + \frac{ig}{4c} [\gamma^\mu, \partial_\mu W_s^{(0)}]_\mp
 \nonumber\\
    &&  =  - \frac{ig\hbar^2}{12c} F_{\nu\mu}^{a}\partial_p^\nu
    [\gamma^\mu , W^{(0)a}]_\mp,
 \\
    \mbox{octet:} \quad &{[}&
    m {\bf 1} - \gamma^\mu p_\mu , W^{(1)a}]_{\pm}
    + \frac{ig}{4c} [\gamma^\mu, \partial_\mu W^{(0)a}]_\mp
 \nonumber\\
    &&  =  - \frac{i g}{2c} F_{\nu\mu}^a \partial_p^\nu
    [\gamma^\mu, W_s^{(0)}]_\mp
    - \frac{ig\hbar}{4c} d^{abc} F_{\nu\mu}^b \partial_p^\nu
    [\gamma^\mu, W^{(0)c}]_\mp
 \nonumber\\
    && \phantom{=}
    + \frac{g\hbar}{8c} f^{abc} F_{\nu\mu}^b \partial_p^\nu
    [\gamma^\mu, W^{(0)c}]_\pm
    - \frac{ig^2}{4c^2} f^{abc} A_\mu^b [\gamma^\mu, W^{(0)c}]_\mp.
 \end{eqnarray}
The equations for the color singlet distributions become identical
to those of QED if we replace $W_s^{(n)} \to W^{(n)}$ and
$W^{(n)a} \to N_c(N_c-1)\,W^{(n)}$ with $N_c(N_c-1)=6$ for QCD.

The above mean field equations will be used in Sec.~\ref{sec4}.

\subsection{Spinor decomposition}
\label{sec3.6}

We now return to the full operator equations and perform a spinor
decomposition, using the following basis for the Clifford algebra:
 \begin{equation}
  \Gamma_i = 1,i\gamma^5,\gamma^\mu,\gamma^\mu \gamma^5,\sigma^{\mu\nu}.
 \end{equation}
In this basis the Wigner operator $\fopw$ is expanded as
 \begin{equation}
 \label{spinorexp}
  \fopw (x,p) = \fopfc + i \gamma^5 \foppc + \gamma^\mu \fopvc_\mu +
  \gamma^\mu\gamma^5 \fopac_\mu + \frac{1}{2} \sigma^{\mu\nu}
  \fopsc_{\mu\nu},
 \end{equation}
with the hermitian spinor components $\fopfc$ (scalar), $\foppc$
(pseudo-scalar), $\fopvc_\mu$ (vector), $\fopac_\mu$ (axial vector)
and $\fopsc_{\mu\nu}$ (tensor). These components are projected out
from $\hat W$ by taking traces with the corresponding basis elements
$\Gamma_i$ \cite{QTT}. Inserting the expansion (\ref{spinorexp}) into
Eqs.~(\ref{bewwigtot}) and projecting onto the various spinor channels
as described in \cite{QTT} one finds (after separating hermitean and
antihermitean parts)
 \begin{mathletters}
 \label{spinorgl1}
 \begin{eqnarray}
  2m\fopfc &=& \{\Pi_\mu,\fopvc^\mu\}
  \\ 2m\foppc &=&
  [\Delta_\mu,\fopac^\mu]
  \\ 2m\fopsc_{\mu\nu} &=& [\Delta_\mu,\fopvc_\nu] -
  [\Delta_\nu,\fopvc_\mu] + \varepsilon_{\alpha\beta\mu\nu}
  \{\Pi^\alpha,\fopac^\beta\}
  \\ 2m\fopvc_\mu &=& \{\Pi_\mu,\fopfc\} +
  [\Delta^\nu,\fopsc_{\mu\nu}] \\ 2m\fopac_\mu &=& - [\Delta_\mu,\foppc] +
  \{ \Pi^\nu,\foplc_{\mu\nu} \}
 \end{eqnarray}
 \end{mathletters}%
and
 \begin{mathletters}
 \label{spinorgl2}
 \begin{eqnarray}
  0 &=& [\Delta_\mu,\fopvc^\mu]\\
  0 &=& \{\Pi_\mu,\fopac^\mu\}\\
  0 &=& \{\Pi_\mu,\fopvc_\nu\} - \{\Pi_\nu,\fopvc_\mu\} -
  \varepsilon_{\alpha\beta\mu\nu} [\Delta^\alpha,\fopac^\beta]\\
  0 &=& [\Delta_\mu,\fopfc] -\{\Pi^\nu,\fopsc_{\mu\nu}\}\\
  0 &=& \{\Pi_\mu,\foppc\} + [\Delta^\nu,\foplc_{\mu\nu}].
 \end{eqnarray}%
 \end{mathletters}
Here we defined the adjoint tensor $\foplc_{\mu\nu}$ to
$\fopsc_{\mu\nu}$ by
 \begin{equation}
  \foplc_{\mu\nu} \equiv
  \frac{1}{2}\varepsilon_{\mu\nu\alpha\beta} \fopsc^{\alpha\beta}.
 \end{equation}
Up to the (anti-)commutator structure, these equations are identical
to those derived in Ref.~\cite{QTT} for QED in the external field
limit. Note that by contraction with the Levi-Civita tensor
Eqs.~(\ref{spinorgl1}iii,\ref{spinorgl2}iii) can also be written as
\begin{eqnarray}
  2m\foplc_{\mu\nu} &=&
  \varepsilon_{\alpha\beta\mu\nu}[\Delta^\alpha,\fopvc^\beta] -
  \{\Pi_\nu,\fopac_\mu\} + \{\Pi_\mu,\fopac_\nu\}\nonumber\\ 0 &=&
  [\Delta_\mu, \fopac_\nu] - [\Delta_\nu, \fopac_\mu] +
  \varepsilon_{\alpha\beta\mu\nu} \{ \Pi^\alpha, \fopvc^\beta\}. \nonumber
\end{eqnarray}

Eqs.~(\ref{spinorgl1},\ref{spinorgl2}) can be rewritten in the form of
generalized transport equations and generalized mass shell conditions.
To this end we introduce \cite{OCHS} a generalization of the
Lorentz-covariant total time derivative (also known as the ``drift
operator'') $m \ d\fopx(x,p) / d\tau = p_\mu \partial^\mu \fopx(x,p)$
via
\begin{equation}
  \{\Pi_\mu,[ \Delta^\mu,\fopx]\}
  \quad \mbox{or } \quad
  [\Delta^\mu,\{\Pi_\mu,\fopx \}]
\end{equation}
and the generalized mass shell operator
\begin{equation}
  \label{mshell}
  M^2 \fopx \equiv
  4 m^2 \fopx - \{\Pi_\mu,\{\Pi^\mu,\fopx \}\}
  + [\Delta_\mu,[\Delta^\mu,\fopx]]
\end{equation}
which will be used to generalize the classical mass shell conditions
$(m^2 - p^2)\fopx(x,p) = 0$. In the semiclassical limit and for
Abelian gauge theories the generalized drift operator becomes
the well-known Vlasov-operator $p_\mu\partial^\mu + e p^\mu
\partial_p^\nu \fopf_{\mu\nu}(x)$ while the generalized mass shell
operator simply turns into $m^2-p^2$.

The details of the derivation of the generalized transport and mass
shell equations are given in Appendix~\ref{appd}. One finds that only
8 of the 16 operators $\fopfc, \foppc, \fopvc_\mu, \fopac_\mu,
\fopsc_{\mu\nu}$ are independent and satisfy generalized transport and
mass shell equations while the other 8 functions are obtained from the
solutions via simple constraints. For the set of independent functions
one can choose either $\fopfc, \foppc,$ and $\fopsc_{\mu\nu}$ or
$\fopvc_\mu$ and $\fopac_\mu$. We will give here the equations
corresponding to the first choice; those for the second choice are
found in Appendix~\ref{appd3}.

The complete set of covariant equations of motion for the spinor
components of $\hat W(x,p)$ is:

\noindent (a) Transport equations:
 \begin{mathletters}
 \label{trans}
 \begin{eqnarray}
 \label{trans1}
  [\Delta_\mu,\{ \Pi^\mu,\fopfc \}] &{=}&
  [\Delta_\nu \Delta_\mu,\fopsc^{\mu\nu}]
 \\
 \label{trans2}
  [\Delta_\mu,\{ \Pi^\mu,\foppc \} ] &{=}&
  [\Delta_\nu \Delta_\mu,\foplc^{\mu\nu}]
 \\
 \label{trans3}
  [\Delta_\alpha ,\{\Pi^\alpha,\fopsc_{\mu\nu}\}] &{=}&
  [[p_\mu+\Pi_\mu,p_\nu+\Pi_\nu],\fopfc]
  + \varepsilon_{\alpha\beta\mu\nu} [\Delta^\alpha \Delta^\beta,\foppc]
 \nonumber\\
  && - \left( \{\fopsc_{\mu\alpha},[\Delta^\alpha,p_\nu+\Pi_\nu]\}
     + \{[p_\mu+\Pi_\mu,\Delta^\alpha],\fopsc_{\alpha\nu}\}
    \right)
 \end{eqnarray}
 \end{mathletters}
\noindent (b) Mass shell equations:
 \begin{mathletters}
 \label{mass}
 \begin{eqnarray}
  M^2\fopfc &{=}& \{[p^\mu+\Pi^\mu,\Delta^\nu],\fopsc_{\mu\nu}\}
 \\
  M^2\foppc &{=}& \{[p^\mu+\Pi^\mu,\Delta^\mu],\foplc_{\mu\nu}\}
 \\
  M^2\fopsc_{\mu\nu} &{=}& \{[p_\mu+\Pi_\mu,\Delta_\nu],\fopfc\}
                         - \{[p_\nu+\Pi_\nu,\Delta_\mu],\fopfc\}
 \nonumber\\
  &&  - [\fopsc_{\mu\alpha},[\Delta^\alpha,\Delta_\nu]]
      - [[\Delta_\mu,\Delta^\alpha],\fopsc_{\alpha\nu}]
 \nonumber\\
  && + [\fopsc_{\mu\alpha},[p^\alpha+\Pi^\alpha,p_\nu+\Pi_\nu]]
     + [[p_\mu+\Pi_\mu,p^\alpha+\Pi^\alpha],\fopsc_{\alpha\nu}]
 \nonumber\\
  && - \varepsilon_{\alpha\beta\mu\nu}
       \{[p^\alpha+\Pi^\alpha,\Delta^\beta],\foppc\}
 \end{eqnarray}
 \end{mathletters}
\noindent (c) Constraints:
 \begin{mathletters}
 \label{constraint}
 \begin{eqnarray}
  0 &=& [\Delta_\mu,\fopfc] - \{\Pi^\nu,\fopsc_{\mu\nu}\}
 \\
  0 &=& \{\Pi_\mu,\foppc\} + [\Delta^\nu,\foplc_{\mu\nu}].
 \end{eqnarray}
 \end{mathletters}
\noindent (d) Defining equations for the dependent spinor components:
 \begin{mathletters}
 \label{connection}
 \begin{eqnarray}
  2m\fopac_\mu &=& - [\Delta_\mu,\foppc] + \{\Pi^\nu,
  \foplc_{\mu\nu}\}
 \\
  2 m \fopvc_\mu &=& \{\Pi_\mu,\fopfc\} +
  [\Delta^\nu,\fopsc_{\mu\nu}].
 \end{eqnarray}
 \end{mathletters}

The general structure that transport and mass shell equations must be
solved for only 8 of the 16 spinor components (either for $\fopfc,
\foppc, \fopsc_{\mu\nu}$ or for $\fopvc_\mu, \fopac_\mu$) while the
other 8 components are obtained from constraints has apparently not
been noticed before. It was not made manifest in the equations given
in Ref.~\cite{QTT} where transport and mass shell conditions were
derived for the full Wigner operator but were not given explicitly for
each spinor component. Compared to previous work
Eqs.~(\ref{trans}-\ref{connection}) thus provide a considerable
structural simplification of the covariant transport theory for spinor
fields.

\subsection{Chiral Limit}
\label{sec3.7}

In this subsection we study  the covariant kinetic equations in the
chiral limit of vanishing fermion mass. The resulting equations of
motion for the Wigner operator and its spinor components can be easily
obtained by setting $m=0$ in Eq.~(\ref{bewwigtot}) and in
Eqs.~(\ref{spinorgl1},\ref{mass},\ref{connection}). In particular one
sees that now in Eqs.~(\ref{spinorgl1},\ref{spinorgl2}) $\fopac_\mu,
\fopvc_\mu$ completely decouple from $\fopfc, \foppc$, and
$\fopsc_{\mu\nu}$. A closer look at the resulting ``transport'' and
``mass shell'' equations reveals, however, that they are no longer
linearly independent but become linear combinations of each other. For
$m=0$ it no longer makes sense to distinguish between transport and
mass shell equations (which both contain the non-linear operators
$\Pi$ and $\Delta$ in second order), and the transport theory is
better formulated on the level of Eqs.~(\ref{spinorgl1},\ref{spinorgl2})
which are linear in the non-local operators.

These equations can be further diagonalized by introducing a
``helicity basis'' via
 \begin{mathletters}
 \label{helicity}
 \begin{eqnarray}
 \label{helicity1}
   \fop{a}_\mu^{(\pm)} &{=}& \fopvc_\mu\pm \fopac_\mu  \, ,
 \\
 \label{helicity2}
    \fop{f}^{(\pm)} &{=}& \fopfc \pm \foppc\, ,
 \\
 \label{helicity3}
    \fop{s}_{\mu\nu}^{(\pm)} &{=}& \fopsc_{\mu\nu}
    \pm  {\textstyle{1\over 2}} \varepsilon_{\alpha\beta\mu\nu}
    \fopsc^{\alpha\beta}\, ,
 \end{eqnarray}
 \end{mathletters}
with ${1\over 2} \varepsilon_{0ijk}\,\fop{s}^{jk(\pm)} =
\fop{s}_{0i}^{(\pm)}$. We find
 \begin{mathletters}
 \label{chiralAV1}
 \begin{eqnarray}
 \label{chiralAV1i}
    0 &=& \{ \Pi_\mu , \fop{a}^{\mu(\pm)} \} \, ,
    \\
 \label{chiralAV1ii}
    0 &=& [ \Delta_\mu , \fop{a}^{\mu(\pm)} ] \, ,
    \\
 \label{chiralAV1iii}
    0 &=& [\Delta_\nu , \fop{a}_{\mu}^{(\pm)} ] -
    [ \Delta_\mu , \fop{a}_{\nu}^{(\pm)} ]
    \pm  \varepsilon_{\alpha\beta\mu\nu}
    \{ \Pi^\alpha , \fop{a}^{\beta(\pm)}\} \, ,
 \end{eqnarray}
 \end{mathletters}%
as well as
 \begin{mathletters}
 \label{chiralFPS1}
 \begin{eqnarray}
    0 &=& [ \Delta_\mu , \fop{f}^{(\pm)} ] -
    \{ \Pi^\nu , \fop{s}_{\mu\nu}^{(\pm)} \}  \, ,
    \\
    0 &=& \{ \Pi_\mu , \fop{f}^{(\pm)} \} +
    [ \Delta^\nu , \fop{s}_{\mu\nu}^{(\pm)} ].
  \end{eqnarray}
\end{mathletters}%
This is the full set of covariant transport equations for gauge
theories in the chiral limit. $\fop{a}_{\mu}^{(\pm)}$,
$\fop{f}^{(\pm)}$ and $\fop{s}_{\mu\nu}^{(\pm)}$ generate distribution
functions for excitations with defined helicity/chirality.

In a chirally symmetric theory both vector and axial vector currents
should be conserved. Defining them by
 \begin{eqnarray}
  \cur{v}_\mu(x) &=& \int \frac{d^4p}{(2\pi)^4} \fopvc_\mu(x,p) \, ,
 \nonumber\\
  \cur{a}_\mu(x) &=& \int \frac{d^4p}{(2\pi)^4} \fopac_\mu(x,p) \, ,
 \end{eqnarray}
one finds easily\footnote{Expanding $\Delta_\mu$ in a Taylor series,
all terms containing $\partial_p$ vanish after partial integration
since $\fopac_\mu, \fopvc_\mu$ vanish at infinite momentum.}
from Eq.~(\ref{chiralAV1ii}) the conservation laws
\begin{eqnarray}
  0 &=& {[D_\mu(x), \cur{v}^\mu (x)]} \, ,
  \nonumber\\
  0 &=& {[D_\mu(x), \cur{a}^\mu (x)]} \, .
\end{eqnarray}
The color singlet components of both currents (for which $D^\mu(x)$
reduces to $\partial_x^\mu$) thus satisfy the usual conservation laws
while the color octet vector and axial vector currents are covariantly
conserved.

In the recent literature \cite{CHI} chirally invariant transport
equations were also derived in the framework of the Nambu-Jona-Lasinio
model, in the external field limit and to leading order of the
gradient expansion. The Lagrangian used there,
\begin{equation}
  \label{chirallag}
  {\mathcal L}_{\rm NJL} =
  i \overline{\psi} \gamma^\mu \partial_\mu \psi
  + G [(\overline{\psi}\psi)^2 + (\overline{\psi} i \gamma_5 \psi)^2]
  \, ,
\end{equation}
does not contain a coupling to gauge fields, but a self-coupling
of the spinors which, if sufficiently strong, leads to spontaneous
breaking of the chiral symmetry in the massless case. With the help of
the auxiliary fields $\fop{\sigma} = - 2 G \overline{\psi} \psi$ and
$\fop{\pi} = - 2G\overline{\psi} i \gamma_5 \psi$ the Lagrangian is
rewritten as
\begin{equation}
  \label{chirallag2}
  {\mathcal L}_{\rm NJL} =
  i \overline{\psi} \gamma^\mu \partial_\mu \psi
  - \fop{\sigma} \overline{\psi} \psi
  - \fop{\pi} \overline{\psi} i \gamma_5 \psi
  -\frac{\fop{\sigma}^2 + \fop{\pi}^2}{4G}.
\end{equation}
This Lagrangian is invariant under the chiral transformation
\begin{eqnarray}\label{symua}
  \psi &\rightarrow& \psi' = \exp(-i\gamma_5\frac{\chi}{2}) \psi
  \\
  \fop{\sigma} &\rightarrow& \fop{\sigma}' = \fop{\sigma} \cos \chi -
  \fop{\pi} \sin \chi
  \\
  \fop{\pi} &\rightarrow& \fop{\pi}' = \fop{\pi} \cos \chi +
  \fop{\sigma} \sin \chi,
\end{eqnarray}
resulting in a conserved axial vector current.

>From this Lagrangian the authors of \cite{CHI} derived chirally
symmetric transport equations in a vacuum state with dynamically
broken chiral symmetry ($\langle \hat\sigma^2 + \hat\pi^2 \rangle \ne
0$). In contrast to our transport equations, these equations do not
decouple $\Vc_\mu$ and $\Ac_\mu$ from $\Fc$, $\Pc$ and
$\Sc_{\mu\nu}$. The coupling between the two sectors occurs via the
vacuum expectation values $\sigma$ and $\pi$ of the auxiliary
fields. However, according to Eq.~(\ref{symua}), $\pi$ can always be
made to vanish by an appropriate chiral rotation, with a compensationg
chiral rotation on the spinor components of the Wigner density
\cite{CHI}. If one does this in the equations of Ref.~\cite{CHI},
$\sigma$ takes over the same role as the bare mass $m$ in our
equations. In other words, the dynamical mass generated by
spontaneous chiral symmetry breakdown couples the two sectors in
exactly the same way as a non-zero bare mass $m\ne 0$. In the
symmetric state, $\sigma = \pi =0$, their equations decouple like
ours.

\subsection{Classical limit}
\label{sec3.9}

Exact solutions for the kinetic equations in the free-field limit are
given in Appendix~\ref{appe}. Here we study their classical limit, by
expanding the spinor equations Eqs.~(\ref{trans}-\ref{connection}) to
leading order in the gradient expansion. We will then consider
explicitly the external field case for QED.

With the generalized mass shell operator to zeroth order $\hbar$,
\begin{equation}
  \label{massophbar0}
  (M^2)^{(0)} \fopx^{(0)} = 4(m^2 - p^2) \fopx^{(0)}
\end{equation}
which becomes just the classical mass shell operator, we find the
following equations
\begin{mathletters}\label{tothbar_0}
  \begin{eqnarray}
  \label{spinorglhbar0_mass1}
    {M^{(0)}}^{2} \fopfc^{(0)} &=& 0\\
  \label{spinorglhbar0_mass2}
    {M^{(0)}}^{2} \foppc^{(0)} &=& 0\\
  \label{spinorglhbar0_mass3}
    {M^{(0)}}^{2} \fopsc^{(0)}_{\mu\nu} &=& 0\\
  \label{spinorglhbar0_neb1}
    0 &=& p^\nu \fopsc_{\mu\nu}^{(0)} \\
  \label{spinorglhbar0_neb2}
    0 &=& p^\mu  \foppc^{(0)}\\
  \label{spinorglhbar0_con1}
    m \fopac_\mu^{(0)} &=& p^\nu \foplc_{\mu\nu}^{(0)}\\
  \label{spinorglhbar0_con2}
    m \fopvc_\mu^{(0)} &=& p_\mu \fopfc^{(0)}.
  \end{eqnarray}
\end{mathletters}%
The transport equations vanish identically, meaning that the equations
to zeroth order do not contain dynamical information. From
Eqs.~(\ref{spinorglhbar0_mass2},\ref{spinorglhbar0_neb2}) the
pseudoscalar density $\foppc^{(0)}$ proves to be zero. The equations
are still not complete: they only give information on the functional
$p$ dependence of the spinor components. Only in the semi-classical
approximation will we get transport equations for the spinor components
$\fopx^{(0)}$ and thus information on their $x$ dependence. From the
equations to first order $\hbar$ we find in addition mass shell
conditions and constraints for the first order spinor components
$\fopx^{(1)}$, that is information on their $p$ dependence.

To first order we find for the mass shell operator
\begin{eqnarray}
  \label{massop1ord}
  M^2 \fopx  &=& (M^2)^{(0)} \fopx^{(1)} +
  (M^2)^{(1)} \fopx^{(0)}\quad \mbox{with}
  \nonumber\\
  (M^2)^{(1)} \fopx^{(0)}
  &=&
  - \frac{ig}{c}p_\mu\partial_{p\nu} [\fopf^{\nu\mu}(x),\fopx^{(0)}]
\end{eqnarray}
and the generalized derivative
\begin{eqnarray}
  [\Delta_\mu,\{\Pi^\mu,\fopx\}] &\stackrel{\mbox{\small first order}}{=}&
  \{\Pi^\mu,[\Delta_\mu,\fopx]\} \nonumber\\
  & \stackrel{\mbox{\small first order}}{=} &
  2[p^\mu D_\mu(x), \fopx^{(0)}] - \frac{g}{c}p^\mu \partial_p^\nu
  \{\fopf_{\nu\mu},\fopx^{(0)}\}.
\end{eqnarray}
>From an explicit calculation we find further
 \begin{mathletters}
 \label{kommrelextfeldhbar}
 \begin{eqnarray}
 \label{k1}
    [[\Delta_\mu,\Delta_\nu],\fopx]
    &\stackrel{\mbox{\small first order}}{=}& 0
 \\
 \label{k2}
    \{ [p_\mu+\Pi_\mu,\Delta_\nu],\fopx \}
    &\stackrel{\mbox{\small first order}}{=}&
    +\frac{ig}{c}\{\fopf_{\mu\nu},\fopx^{(0)}\}
 \\
 \label{k3}
    [[p_\mu+\Pi_\mu,p_\nu+\Pi_\nu ] , \fopx ]
    &\stackrel{\mbox{\small first order}}{=}&
    +\frac{ig}{c}[\fopf_{\mu\nu},\fopx^{(0)}]
 \end{eqnarray}
\end{mathletters}
This then leads to the following spinor equations in first order of the
gradient expansion
\begin{mathletters}
  \label{spinorglhbar1_trans}
  \begin{eqnarray}
    \lefteqn{ [p_\mu D^\mu(x),\fopfc^{(0)}]  -
      \frac{g}{2c} p^\mu \partial^\nu_p
      \{ \hat F_{\nu\mu}(x), \fopfc^{(0)}\} \ =\ 0} \\
    \lefteqn{[\fopf^{\mu\nu}(x), \foplc_{\mu\nu}^{(0)}] \ =\ 0 }\\
    \lefteqn{[p^\alpha D_\alpha(x),\fopsc_{\mu\nu}^{(0)}]
      - \frac{g}{2c} p^\alpha \partial_p^\beta
      \{ \fopf_{\beta\alpha}(x),\fopsc_{\mu\nu}^{(0)}\} \ = \ }
 \nonumber\\ &&
    \frac{ig}{2c}[\fopf_{\mu\nu}(x),\fopfc^{(0)}]
    + \frac{g}{2c} \left(
      \{\fopf^\alpha_{\ \nu}(x), \fopsc_{\mu\alpha}^{(0)}\}
      - \{\fopf_\mu^{\ \alpha}(x),\fopsc_{\alpha\nu}^{(0)}\}
    \right)
  \end{eqnarray}
\end{mathletters}%
\begin{mathletters}
  \label{spinorglhbar1_mass}
  \begin{eqnarray}
    (m^2 - p^2) \fopfc^{(1)} &=&
    -\frac{ig}{4c} p^\mu \partial_p^\nu [\fopf_{\mu\nu}(x),\fopfc^{(0)}]
    +\frac{g}{4c} \{ \fopf^{\mu\nu}(x),\fopsc_{\mu\nu}^{(0)}\}
    \\
    (m^2 - p^2) \foppc^{(1)} &=&
    \frac{g}{4c} \{ \fopf^{\mu\nu}(x),\foplc_{\mu\nu}^{(0)}\}
    \\
    (m^2 - p^2) \fopsc_{\mu\nu}^{(1)} &=&
    - \frac{ig}{4c} p_\alpha \partial_{p\beta}
    [\fopf^{\alpha\beta}(x),\fopsc_{\mu\nu}^{(0)}]
    -\frac{ig}{4c} \left(
      [\fopf^\alpha_{\ \nu}(x), \fopsc_{\mu\alpha}^{(0)}]
      - [\fopf_\mu^{\ \alpha}(x),\fopsc_{\alpha\nu}^{(0)}]
    \right)
    \nonumber\\ &&
    + \frac{g}{2c}\{\fopf_{\mu\nu}(x),\fopfc^{(0)}\},
  \end{eqnarray}
\end{mathletters}%
\begin{mathletters}
  \label{spinorglhbar1_neb}
  \begin{eqnarray}
    2 p^\nu \fopsc_{\mu\nu}^{(1)} &=&
    [D_\mu(x), \fopfc^{(0)}] - \frac{g}{2c} \partial^\nu
    \{ \fopf_{\nu\mu}(x),\fopfc^{(0)}\}
    + \frac{ig}{4c} \partial_{p\alpha}
    \{ \fopf^{\alpha\nu}(x),\fopsc_{\mu\nu}^{(0)}\}\quad
    \\
    - 2 p_\mu \foppc^{(1)} &=&
    [D^\nu(x) \foplc_{\mu\nu}^{(0)}] - \frac{g}{2c}
    \partial_{p\alpha} \{ \fopf^{\alpha\nu}(x), \foplc_{\mu\nu}^{(0)}\}
  \end{eqnarray}
\end{mathletters}%
\begin{mathletters}
  \label{spinorglhbar1_con}
  \begin{eqnarray}
    2m \fopac_\mu^{(1)} &=&
    2 p^\nu \foplc_{\mu\nu}^{(1)}
    - \frac{ig}{4c} \partial_{p\alpha}
    [\fopf^{\alpha\nu}(x), \foplc_{\mu\nu}^{(0)}]
    \\
    2m \fopvc_\mu^{(1)} &=&
    2 p_\mu \fopfc^{(1)}
    - \frac{ig}{4c} \partial_p^\nu [\fopf_{\nu\mu}(x), \fopfc^{(0)}]
    \nonumber\\ &&
    + [D^\nu(x),\fopsc_{\mu\nu}^{(0)}]
    - \frac{g}{2c} \partial_{p\alpha}
    \{\fopf^{\alpha\nu}(x), \fopsc_{\mu\nu}^{(0)}\}.
  \end{eqnarray}
\end{mathletters}%
Mass shell equations and constraints couple to the spinor components
in first order, whereas the transport equations are equations for the
zeroth order spinor components. Because of the vanishing pseudoscalar
density $\foppc^{(0)}$ the second transport equation
Eq.~(\ref{spinorglhbar1_trans}ii) becomes a constraint for
$\foplc_{\mu\nu}^{(0)}$. Contracting Eq.~(\ref{spinorglhbar1_neb}i)
with $p_\mu$ and Eq.~(\ref{spinorglhbar1_neb}ii) with $p_\gamma
\varepsilon^{\mu\gamma\delta\kappa}$ and comparison with the transport
equations Eqs.~(\ref{spinorglhbar1_trans}i,iii) we find as additional
constraints
\begin{mathletters}\label{nebaddit}
  \begin{eqnarray}
    [\fopf^{\mu\nu}, \fopsc_{\mu\nu}^{(0)}] &=& 0
    \\
    {}[\fopf^{\mu\nu}, \fopfc^{(0)}] &=& 0
  \end{eqnarray}
\end{mathletters}%
telling us especially that the field strength tensor commutes with the
scalar density $\fopfc^{(0)}$.

Thus we can conclude that the leading order in the gradient expansion
is given through the mass shell and constraint equations
Eqs.~(\ref{tothbar_0}) and the transport equations
Eq.~(\ref{spinorglhbar1_trans}). (The additional constraints
Eqs.~(\ref{nebaddit}i,ii) play no role to this order of the
approximation.) To leading order the quarks (and electrons) can be
interpreted as on-shell particles obeying a transport equation with
gauge covariant drift and Vlasov terms, $[p_\mu D^\mu(x), \fopx]$ and
$g/(2c) p_\mu \partial_{p\nu}  \{ F^{\nu\mu}(x), \fopx \}$, respectively.
The pseudoscalar density will contribute only to the next higher order.

\subsubsection{Classical solutions for QED with external fields}
\label{sec3.9.1}

For QED with external fields we get for the zeroth order formally the
same equations as Eqs.~(\ref{tothbar_0}), but now the spinor
components are no longer operators. With $\Pc^{(0)}\equiv 0$ this
reads
\begin{mathletters}
  \label{spinorglqedaeu_hbar0}
  \begin{eqnarray}
    0 &=& (m^2 - p^2) \Fc^{(0)}
    \\
    0 &=& (m^2 - p^2) \Sc_{\mu\nu}^{(0)}
    \\
    0 &=& p^\nu  \Sc_{\mu\nu}^{(0)}
    \\
    2m \Ac_\mu^{(0)} &=& 2 p^\nu {{\mathcal L}}_{\mu\nu}^{(0)}\\
    2m \Vc_\mu^{(0)} &=& 2 p_\mu \Fc^{(0)}.
  \end{eqnarray}
\end{mathletters}%
To derive the first order equations we use that the commutators of the gauge
fields and field strength tensor with the spinor components vanish, whereas
the anti-commutators give a factor 2:
\begin{mathletters}
  \label{spinorglqedaeu_hbar2}
  \begin{eqnarray}
    0 \!&\!=\!&\!
    \left(
      p_\mu \partial^\mu + \frac{g}{c} F_{\mu\nu}p^\mu\partial_p^\nu
    \right) \Fc^{(0)}
    \\
    0 \!&\!=\!&\!
    \left(
      p_\alpha \partial^\alpha +
      \frac{g}{c} F_{\alpha\beta}p^\alpha\partial_p^\beta
    \right) \Sc_{\mu\nu}^{(0)}
    +\frac{g}{c}
    \left(
      F_\mu^{\ \alpha}(x) \Sc_{\alpha\nu}^{(0)} +
      F_\nu^{\ \alpha}(x) \Sc_{\mu\alpha}^{(0)}
    \right) \quad\
    \\
    (m^2\! -\! p^2) \Fc^{(1)}
    \!&\!=\!&\! \frac{g}{4c} F^{\mu\nu}(x) \Sc_{\mu\nu}^{(0)}
    \\
    (m^2\! - \!p^2) \Pc^{(1)} \!&\!=\!&\! \frac{g}{4c} F^{\mu\nu}(x)
    {{\mathcal L}}_{\mu\nu}^{(0)}
    \\
    (m^2\! -\! p^2) \Sc_{\mu\nu}^{(1)} \!&\!=\!&\! \frac{g}{2c}
      F_{\mu\nu}(x) \Fc^{(0)}
    \\
    2 p^\nu  \Sc_{\mu\nu}^{(1)} \!&\!=\!&\!
    \left(
      \partial_\mu + \frac{g}{c} F_{\mu\nu}\partial_p^\nu
    \right) \Fc^{(0)}
    \\
    2 p_\mu  \Pc^{(1)} \!&\!=\!&\!
    - \left(
      \partial^\nu + \frac{g}{c} F^{\nu\alpha}\partial_{p\alpha}
    \right) {{\mathcal L}}^{(0)}_{\mu\nu}
    \\
    2m \Ac_\mu^{(1)} \!&\!=\!&\! 2 p^\nu {{\mathcal L}}_{\mu\nu}^{(1)}
    \\
    2m \Vc_\mu^{(1)} \!&\!=\!&\! 2 p_\mu \Fc^{(1)} +
    \left(
      \partial^\nu + \frac{g}{c} F^{\nu\alpha}\partial_{p\alpha}
    \right) \Sc^{(0)}_{\mu\nu}.
  \end{eqnarray}
\end{mathletters}%
>From the systematics of the expansion one finds that the hierarchy of
equations can be solved successively for each order of the spinor components.
The equations above for instance allow for an unambiguous solution for the
classical spinor components $X^{(0)}$, without having to take into account
higher spinor components. Therefore the hierarchy of equations can be
truncated at any order for the spinor components. This remains even
true for the more general case of QCD with external fields. The reason
for this decoupling is simply that the operator $\Delta_\mu$ to zeroth
order $\hbar$ gives no contribution and therefore the equations to
$n^{\rm th}$ order contain the $n^{\rm th}$ order spinor components
only together with momenta $p_\mu$. Allowing for correlations this
remains true only if we replace multi-particle correlations by
products of one-particle Wigner functions and external fields
(Boltzmann Ansatz).

To write down the equations for $\Fc^{(0)}, \Sc_{\mu\nu}^{(0)}$ in their usual
form with electric and magnetic fields we first use the separation into
on-shell particle ($\delta(p_0 - E)$) and anti-particle ($\delta(p_0 + E)$)
contributions for $\Fc^{(0)}$ and $\Sc_{\mu\nu}^{(0)}$ according to
Eqs.~(\ref{spinorglqedaeu_hbar0}i,ii) using the following identity
\begin{equation}
  \delta(p^2 - m^2) = \frac{1}{2\vert p_0\vert}
  \left(
    \delta (p_0 - E) + \delta (p_0 + E)
  \right)
  ,\qquad E = \sqrt{{\vec p}^2 + m^2}.
\end{equation}
In the following we will restrict ourselves to particles; the
equations for anti-particles are given through $p_0 \to -p_0$. The
totally anti-symmetric field strength tensor is given in terms of
electric and magnetic fields by
{\renewcommand{\arraystretch}{1.0}
\begin{equation}
  F^{\mu\nu} =
  \left(
    \begin{array}[h]{cccc}
      0    & -E_x & -E_y & -E_z\\
      E_x  & 0    & -B_z & B_y \\
      E_y  & B_z  & 0    & -B_x\\
      E_z  & -B_y & B_x  & 0
    \end{array}
  \right).
\end{equation}
}
Similarly we define the totally anti-symmetric tensor $\Sc^{\mu\nu}$
\begin{equation}\label{Stensor}
  \Sc^{\mu\nu} =
  \left(
    \begin{array}[h]{cccc}
      0    & -S_x      & -S_y      & -S_z     \\
      S_x  & 0         & -\sigma_z & \sigma_y \\
      S_y  & \sigma_z  & 0         & -\sigma_x\\
      S_z  & -\sigma_y & \sigma_x  & 0
    \end{array}
  \right)
\end{equation}
with spin vector $\vec S$ and helicity ${\vec \sigma}$.

The  dual tensor ${\mathcal L}^{\mu\nu}$  for $\Sc^{\mu\nu}$ is then
given by ${\mathcal L}^{\mu\nu} = $ ${1\over 2}
\varepsilon^{\alpha\beta\mu\nu} \Sc_{\alpha\beta}$ in complete analogy
to the dual field strength tensor ${\mathcal F}^{\mu\nu}$
$= {1\over 2} \varepsilon^{\alpha\beta\mu\nu} F_{\alpha\beta}$ \cite{ED}.
${\mathcal L}^{\mu\nu}$ is obtained from $\Sc^{\mu\nu}$ by exchanging
${\vec S} \to {\vec \sigma}$ and ${\vec \sigma} \to -{\vec S}$.

>From  $p_\nu \Sc^{\mu\nu (0)} = 0$ we find the constraints
\begin{eqnarray}\label{bedingS}
  {\vec S}^{(0)} &=& -\vec v \times {\vec \sigma}^{(0)},
  \qquad \vec v = \frac{\vec p}{E},\\
  \vec p \cdot {\vec S}^{(0)} &=& 0.
\end{eqnarray}
The first equation connects  ${\vec S}^{(0)}$ and
${\vec \sigma}^{(0)}$, whereas the second equation is automatically
fulfilled.

Using 3-vectors we arrive after some algebra at the following
equations for the dynamics of the system as governed by the scalar density
$\Fc^{(0)}$ and helicity  ${\vec \sigma}^{(0)}$ (coupling constant $g = -e$)
\begin{mathletters}
  \label{QEDeqn2}
  \begin{eqnarray}
    \lefteqn{
      [\partial_0 + \vec v\cdot \nabla +
      \frac{e}{c} ( \vec E + \vec v \times \vec B)\cdot\nabla_p]
      \Fc^{(0)}(x,\vec p ,E) = 0
      }
    \\
    \lefteqn{
      E [\partial_0 + \vec v \cdot\nabla +
      \frac{e}{c} ( \vec E + \vec v \times \vec B)\cdot{\nabla}_p]
      {\vec \sigma}^{(0)} (x,\vec p ,E) =
      }
    \nonumber\\
    && -\frac{e}{c}[ \vec E \times (\vec v \times
    {\vec \sigma}^{(0)} (x,\vec p ,E))
    + \vec B \times {\vec \sigma}^{(0)} (x,\vec p ,E)].
  \end{eqnarray}
\end{mathletters}%
The other spinor components are dependent variables and connected to scalar
and spin density through
\begin{mathletters}
  \label{QEDeqn2b}
  \begin{eqnarray}
    {\vec S}^{(0)}(x,\vec p ,E) &=& -\vec v \times
    {\vec \sigma}^{(0)}(x,\vec p ,E)
    \\
    m \Ac_0^{(0)}(x,p ) &=& \vec p \cdot {\vec \sigma}^{(0)} (x,\vec p, E)
    \\
    m {\vec \Ac}^{(0)} (x,p) &=&
    E {\vec \sigma}^{(0)}(x,\vec p, E)
    + \vec p \times (\vec v \times {\vec \sigma}^{(0)} (x,\vec p, E))
    \\
    m \Vc_0^{(0)}(x,p ) &=& E \Fc^{(0)} (x,\vec p, E)
    \\
    m {\vec \Vc}^{(0)} (x,p) &=&
    \vec p \Fc^{(0)}(x,\vec p, E).
  \end{eqnarray}
\end{mathletters}%
Eq.~(\ref{QEDeqn2}ii) for the helicity or (with Eq.~(\ref{QEDeqn2b}i))
the spin density, is a generalization \cite{hakim82} of the
BMT equation \cite{BMT}. Eq.~(\ref{QEDeqn2}i) for the scalar density
corresponds to the Vlasov equation. The equations for the
anti-particles are given through $\vec v \to -\vec v$.

Eqs.~(\ref{QEDeqn2},\ref{QEDeqn2b}) are still not sufficient to describe the
theory, they have to be coupled to the classical Maxwell equations
\begin{eqnarray}
  \label{maxwell}
  \partial_\mu F^{\mu\nu} &=& j^\nu + j^\nu_{\rm ext}\\
  \partial_\mu {\mathcal F}^{\mu\nu} &=& 0.
\end{eqnarray}
$j^\nu_{\rm ext}$ is an external current and
\begin{equation}
  \label{strom}
  j^\nu(x) \equiv \int d^4 p \ {\rm tr} \gamma^\nu W(x,p), \quad
  j^\nu = j^\nu_{\rm particles} + j^\nu_{\rm anti particles},
\end{equation}
is the self-consistent current, coupling the field strength tensor in the
Maxwell equations to the Wigner function.
We will get contributions only from the
vector part and with the mass shell condition the current reads
\begin{equation}
  j^{\nu(0)} = \frac{2}{m}\int \frac{d^3 p}{(2 \pi)^3}
  \left(
    \begin{array}[h]{cc}
      & \Fc^{(0)}(x,\vec p, E) + \Fc^{(0)}(x,\vec p,E)\\
      \vec v &
      \left(
        \Fc^{(0)}(x,\vec p, E) - \Fc^{(0)}(x,\vec p,E)
      \right)
    \end{array}
  \right).
\end{equation}
Thus we find that $\Sc_{\mu\nu}$ couples only in next to leading order to the
Maxwell equation. Therefore to zeroth order in the gradient expansion the
self-consistent dynamics is governed by the scalar part $\Fc^{(0)}$
alone!

\subsubsection{Classical solutions for chiral QED with external fields}
\label{sec3.9.2}

Finally we study the massless mean field limit. The equations for
$\Ac_\mu{(0)}$ and $\Vc_\mu{(0)}$ are now
\begin{mathletters}\label{chiralqedAV}
  \begin{eqnarray}
    0 &=&   p_\mu \Ac^{\mu (0)} \\
    0 &=&   p_\mu \Vc^{\mu (0)} \\
    0 &=&   p_\mu \Ac_\nu^{(0)} - p_\nu \Ac_\mu^{(0)} \\
    0 &=&   p_\mu \Vc_\nu^{(0)} - p_\nu \Vc_\mu^{(0)} \\
    0 &=& \left[
      \partial_\mu - \frac{e}{c}F_{\mu\nu}\partial^\nu_p
    \right] \Ac^{\mu (0)} \\
    0 &=&  \left[
      \partial_\mu - \frac{e}{c}F_{\mu\nu}\partial^\nu_p
    \right] \Vc^{\mu (0)}.
  \end{eqnarray}
\end{mathletters}%
Multiplying Eqs.~(\ref{chiralqedAV}iii,iv) with $p_\mu$ and using
Eqs.~(\ref{chiralqedAV}i,ii) we find that  $\Ac_\mu^{(0)}$ and
$\Vc_\mu^{(0)}$ separate into a contribution from particles
$(\delta(p_0 - \vert \vec p \vert))$ and anti-particles $(\delta(p_0 -
\vert \vec p \vert))$. It further follows with
Eqs.~(\ref{chiralqedAV}iii,iv) that $\Ac_i^{(0)}$ and $\Vc_i^{(0)}$
can be expressed through $\Ac_0^{(0)}$ and $\Vc_0^{(0)}$
\begin{mathletters}
  \begin{eqnarray}
    {\vec \Ac}^{(0)} (x, \vec p ,\pm \vert\vec p \vert) &=&
    {\hat e}_p \Ac_0^{(0)} (x, \vec p ,\pm \vert\vec p \vert)\\
    {\vec \Vc}^{(0)} (x, \vec p ,\pm \vert\vec p \vert) &=&
    {\hat e}_p \Vc_0^{(0)} (x, \vec p ,\pm \vert\vec p \vert).
  \end{eqnarray}
\end{mathletters}%
${\hat e}_p$ denotes the unit vector in the direction of $\vec p$.
The transport equations in terms of the fields  $\vec E, \vec B, \vec
S, \vec \sigma$ are for particles ($\delta(p_0 - \vert \vec p \vert)$)
\begin{eqnarray}
  0  &=&
  \left[
    \partial^0 + {\hat e}_p\cdot \nabla
    + \frac{e}{c}(\vec E + {\hat e}_p \times \vec B) \cdot \nabla_p
  \right] \Ac_0^{(0)} (x, \vec p, \vert\vec p \vert)\\
  0  &=&
  \left[
    \partial^0 + {\hat e}_p \cdot \nabla
    + \frac{e}{c}(\vec E + {\hat e}_p \times \vec B)
   \cdot \nabla_p  \right] \Vc_0^{(0)} (x, \vec p, \vert\vec p \vert).
\end{eqnarray}
For anti-particles one substitutes ${\hat e}_p \to -{\hat e}_p$ and changes
$\vert \vec p \vert \to - \vert \vec p \vert$
in the arguments of $\Ac_0^{(0)}, \Vc_0^{(0)}$.
The self-consistent current is now given by
\begin{equation}
  j^{\nu(0)} (x) = \int \frac{d^3p}{(2 \pi)^3}\frac{2}{\vert \vec p\vert}
  \left(
    \begin{array}[h]{cc}
      & \Vc_0^{(0)} (x, \vec p, \vert\vec p \vert)
      + \Vc_0^{(0)} (x, \vec p, -\vert\vec p \vert)\\
      {\hat e}_p &
      \left(
          \Vc_0^{(0)} (x, \vec p, \vert\vec p \vert)
        - \Vc_0^{(0)} (x, \vec p, -\vert\vec p \vert)
      \right)
    \end{array}
  \right).
\end{equation}
Here $\Ac_\mu$ will couple only in next-to-leading order to the
self-consistent dynamics.

The equations for $\Fc,\Pc$ and $\Sc_{\mu\nu}$ decouple completely
from the equations for $\Ac_\mu$ and $\Vc_\mu$, see
Sec.~\ref{sec3.7}. Therefore they will not couple to any order in $\hbar$
to the self-consistent dynamics so we do not have to treat them here.
We should remark nevertheless that for $\Fc,\Pc$ and $\Sc_{\mu\nu}$ the
hierarchy of the gradient expansion cannot be solved successively starting with
the zeroth order components, but the equations of motion for the $n^{\rm th}$
component $\hbar$ couple to the $n + 1^{\rm st}$ component.

\section{Single-time formulation}
\label{sec4}

We are now ready to make the transition to the single-time formulation
of quantum transport theory, by taking energy moments of the covariant
transport equations derived in the preceeding Section. As shown in
Sec.\ref{sec2}, a fully equivalent single-time description requires
the calculation of {\em all} energy moments.

\subsection{Moment equations}
\label{sec4.1}

The starting point of our considerations will be again
Eqs.~(\ref{bewegkom1},\ref{bewegkom2}) for the Wigner operator.
By multiplying from the left and right with $\gamma_0$ we can isolate
the term $\{\Pi_0,\hat W\}$ on the l.h.s.\ of the equations. Adding and
subtracting then leads to
 \begin{eqnarray}
  \label{mompi0exp1}
  2 \{\Pi_0, \fopw \} &{=}& 2 m \{ \gamma^0, \fopw\}
  - [\gamma^0 \gamma^i, \{ \Pi_i, \fopw \}]
  - \{ \gamma^0\gamma^i , [i\Delta_i, \fopw] \} \\
  \label{mompi0exp2}
  0 &{=}& 2 m [\gamma^0, \fopw]
  - 2 [i\Delta_0, \fopw]
  - \{ \gamma^0\gamma^i, \{ \Pi_i, \fopw \}\}
  - [ \gamma^0\gamma^i , [i\Delta_i, \fopw]].\quad
 \end{eqnarray}
Through $\Pi_0$ Eq.~(\ref{mompi0exp1}) contains an extra factor $p_0$
on the left, but no generalized time derivative $[i\Delta_0, \fopw]$.
Eq.~(\ref{mompi0exp2}) contains the generalized time derivative but no
extra factor $p_0$. Note that the generalized derivative $\Delta$
occurs only in commutators while $\Pi$ occurs only in anticommutators
with the Wigner operator.

For the spinor component equations (\ref{spinorgl1},\ref{spinorgl2})
the analogous separation into equations with and without explicit
$p_0$ dependence yields \cite{MOM}
\begin{mathletters}\label{with-p0}
  \begin{eqnarray}
    \{\Pi_0,\fopvc^0\} &=&
    - 2 m \fopfc + \{\Pi_i,\fopvc^i\}
    \\
    \{\Pi^0,\fopac^k\} &=&
    2 m \foplc^{0k} +
    \varepsilon^{0ijk}[\Delta_i,\fopvc_j] + \{\Pi^k,\fopac^0\}
    \\
    \{\Pi_0,\fopfc\} &=& 2 m \fopvc_0 - [\Delta^i,\fopsc_{0i}]
    \\
    \{\Pi^0,\fopsc^{jk}\} &=&
    \varepsilon_{0ijk} (2 m \fopac_i + [\Delta_i,\foppc])
    - \{\Pi^k,\fopsc^{0j}\} + \{\Pi^j,\fopsc^{0k}\}
    \\
    \{\Pi_0,\fopac^0\} &=& - \{\Pi_i,\fopac^i\}
    \\
    \{\Pi_0,\fopvc_i\} &=&
    \{\Pi_i,\fopvc_0\} + \varepsilon_{0ijk} [\Delta^j,\fopac^k]
    \\
    \{\Pi^0,\fopsc_{i0}\} &=& [\Delta_i,\fopfc] - \{\Pi^j,\fopsc_{ij}\}
    \\
    \{\Pi_0,\foppc\} &=& - [\Delta^i,\foplc_{0i}]
  \end{eqnarray}
\end{mathletters}%
and
\begin{mathletters}\label{no-p0}
  \begin{eqnarray}
    2 m \foppc &=& [\Delta_\mu,\fopac^\mu]\\
    2 m \fopsc_{0i} &=& [\Delta_0,\fopvc_i] - [\Delta_i,\fopvc_0] +
    \varepsilon_{0ijk} \{\Pi^j,\fopac^k\}\\
    2 m \fopvc_i &=& \{\Pi_i,\fopfc\} + [\Delta^\nu,\fopsc_{i\nu}]\\
    2 m \fopac_0 &=& - [\Delta_0,\foppc]
    + \{\Pi^i,\foplc_{0i}\} \\
    0 &=& [\Delta_\mu,\fopvc^\mu]\\
    0 &=& \{\Pi_i,\fopvc_j\} - \{\Pi_j,\fopvc_i\} -
    \varepsilon_{0ijk} ([\Delta^0,\fopac^k] - [\Delta^k,\fopac^0]) \\
    0 &=& [\Delta_0,\fopfc] - \{\Pi^i,\fopsc_{0i}\}\\
    0 &=& \{\Pi_i,\foppc\}
    - [\Delta^0,\foplc_{0i}]
    + [\Delta^j,\foplc_{ij}].
  \end{eqnarray}
\end{mathletters}%

Before we can take $p_0$-moments of these equations we should first
analyze the dependence of the non-local operators $\Pi$ and $\Delta$
on the partial derivatives $\partial_{p_0}$. We perform a Taylor
expansion for the Schwinger string in $\partial_{p_0}$:
\begin{eqnarray}
  \label{taylor}
  {^{[x]}\fopf_{\nu\mu}}(x+is\partial_p) &=&
  \sum_{k=0}^\infty \frac{1}{k!}
  \left.
    \partial^{(k)}_{y_0}
    \ {^{[x]}\fopf_{\nu\mu}}(x+y)
  \right\vert_{{\vec y} = i s {\vec \partial_p}, y_0 = 0}
  (is{\partial_{p_0}})^k
  \quad \mbox{(QCD)} \\
  &=&  \sum_{k=0}^\infty \frac{1}{k!} \partial^{(k)}_{x_0}
  F_{\nu\mu}(x_0,\vec x +is\vec{\partial}_p) (is\partial_{p_0})^k
  \quad \mbox{(QED, external fields)}  \nonumber
\end{eqnarray}
For QED in the external field limit the higher order $p_0$ derivatives
thus couple simply to the higher order time derivatives of the
external fields. As discussed in Sec.~\ref{sec3.3}, for non-abelian
gauge fields the above expansion generates a gauge covariant temporal
gradient expansion of the transport equations.

For the calculation of the energy moments the following identity is useful:
 \begin{eqnarray}
 \label{partial-p0_1}
  \int_{-\infty}^\infty dp_0\ p_0^n \,(\partial_{p_0})^k\,
  \fopx(x,p_0,\vec p) &=&
  \left\{
    \begin{array}{lc}
      0 & n < k,\\
      (-1)^k \frac{n!}{(n-k)!}\ {^{[n-k]}\fopx}(x,\vec p) & n \ge k,
    \end{array}
  \right.\ .
\end{eqnarray}
Here $\hat X$ stands for any spinor component of the covariant Wigner
operator, and $^{[n]}\fopx$ denotes the $n^{\rm th}$ $p_0$-moment
of $\fopx$:
 \begin{equation}
 \label{defmomente}
    {^{[n]}\fopx}(x,\vec p) = \int_{-\infty}^\infty dp_0\, p_0^n\,
    \fopx(x,p).
 \end{equation}
Eq.~(\ref{partial-p0_1}) is proved by partial integration, using that
because of energy conservation we expect the Wigner operator to fall
off at large $p_0$ faster than any power.

Whenever in the (anti-)commutators in the equations of motion one of
the non-local operators $\Pi$ or $\Delta$ appears to the right of the
Wigner operator, the momentum derivatives are defined as in
(\ref{padj}) to act to the left in the sense of partial
integration. For $k$ momentum derivatives this introduces a factor
$(-1)^k$. This motivates the definition of an alternating commutator
through
 \begin{equation}
 \label{kommutator}
  [\hat A,\hat B]_k \equiv
   \hat A \hat B  - (-1)^k \hat B \hat A =
  \left\{
    \begin{array}{rl}
      [\fop{A},\fop{B}] & \qquad (k\ \mbox{even})\, ,\\
      \{\fop{A},\fop{B}\} & \qquad (k\ \mbox{odd})\, .
    \end{array}
  \right.
 \end{equation}

With these definitions one can work out the energy moments of the
required commutators resp. anticommutators of the spinor components
$\fopx$ with $\Delta$ and $\Pi$:
 \begin{eqnarray}
   \int dp_0\, &{p_0^n}& [\Delta_\mu,\fopx (x,p)]
 \nonumber\\
 \label{momentsop_qcd_1}
   &{=}& [\Delta_\mu(x,\vec p), {^{[n]}\fopx}(x,\vec p)]
   + \sum_{k=1}^n {n \choose k}
   \left[ M_{(k)\mu}(x,\vec p), {^{[n-k]}\fopx}(x,\vec p)\right]_k
 \\
 \label{momentsop_qcd_2}
   &{=}& [D_\mu(x), {^{[n]}\fopx}(x,\vec p)]
   + \sum_{k=0}^n {n \choose k}
   \left[ M_{(k)\mu}(x,\vec p), {^{[n-k]}\fopx}(x,\vec p)\right]_k
 \, ,
 \\
   \int dp_0\, &{p_0^n}& \{\Pi_0,\fopx (x,p)\}
 \nonumber\\
 \label{momentsop_qcd_4}
   &{=}& 2 \ {^{[n+1]}\fopx}(x,\vec p)
   + \sum_{k=0}^n {n \choose k}
   \left[ N_{(k)0}(x,\vec p), {^{[n-k]}\fopx}(x,\vec p)\right]_{k+1}
   \, ,
 \\
   \int dp_0\, &{p_0^n}& \{\Pi_i,\fopx (x,p)\}
 \nonumber\\
 \label{momentsop_qcd_5}
   &{=}& \{\Pi_i(x,\vec p), {^{[n]}\fopx}(x,\vec p)\}
   + \sum_{k=1}^n {n \choose k} \left[N_{(k) i}(x,\vec p),
   {^{[n-k]}\fopx} (x,\vec p)\right]_{k+1}
 \\
 \label{momentsop_qcd_6}
   &{=}& 2 p_i \ {^{[n]}\fopx}(x,\vec p)
   + \sum_{k=0}^n {n \choose k}
   \left[ N_{(k)i}(x,\vec p), {^{[n-k]}\fopx}(x,\vec p)\right]_{k+1}
   \, .
\end{eqnarray}
Here we defined the following non-local operators in 3-dimensional
momentum space:
 \begin{eqnarray}
 \label{3dpi}
   \Pi_\mu(x,\vec p) &{=}& p_\mu + {2g\over c}\int_{-1/2}^0 ds\,
       is\hbar \partial_p^i\
       {^{[x]}\fopf_{i\mu}}(x+is\hbar\partial_p) \, ,
 \\
 \label{3ddelta}
   \Delta_\mu(x,\vec p) &{=}& \hbar D_\mu(x) +
   {ig\over c}\int_{-1/2}^0 ds\, i\hbar\partial_p^i\
   {^{[x]}\fopf_{i\mu}}(x+is\partial_p) \, ,
 \\
 \label{momentsop_qcd2}
   M_{(k)\mu}(x,\vec p)
   &{=}& {g\hbar\over c} \int_{-1/2}^0 ds\, (-i\hbar s)^k
   \left[
    \left.\partial^{(k)}_{y_0}
      \ {^{[x]}\fopf_{\mu i}}(x_0+y_0,\vec x + i\hbar s \vec \partial_p)
    \right\vert_{y_0 = 0}\partial_p^i
   \right.
 \nonumber\\ &&
   \left. \qquad\qquad - \frac{ik}{\hbar s}
    \left.\partial^{(k-1)}_{y_0}
      \ {^{[x]}\fopf_{\mu 0}}(x_0+y_0,\vec x + i\hbar s \vec \partial_p)
    \right\vert_{y_0 = 0}
   \right] \, ,
 \\
   N_{(k)\mu}(x,\vec p)
   &{=}& {2g\over c}\int_{-1/2}^0 ds\,(-i\hbar s)^{k+1}
   \left[
    \left.\partial^{(k)}_{y_0}
      \ {^{[x]}\fopf_{\mu i}}(x_0+y_0,\vec x + i\hbar s \vec \partial_p)
    \right\vert_{y_0 = 0}\partial_p^i
   \right.
 \nonumber\\ &&
   \left. \qquad\qquad - \frac{ik}{\hbar s}
    \left.\partial^{(k-1)}_{y_0}
      \ {^{[x]}\fopf_{\mu 0}}(x_0+y_0,\vec x + i\hbar s \vec \partial_p)
    \right\vert_{y_0 = 0}
  \right] \, .
 \end{eqnarray}
Note that we have used for the 3-dimensional momentum space versions
of the operators $\Pi$ and $\Delta$ the same symbols as for the
4-dimensional ones; which one is meant in a particular equation should
be obvious from the context.

For a given power $n$ of $p_0$, the term $\Pi_0$ gives rise to moments
which are one order higher than those from all other terms. This term
thus indeed couples different orders of the moment hierarchy so that
it will not trivially truncate.

The moment equations are now given by
 \begin{eqnarray}
  \label{momglexp1}
   4 {^{[n+1]} \fopw} \!&\!=\!&\!
  2 m \{ \gamma^0 , {^{[n]}\fopw} \}
  - 2 \{ N_{(0)0} , {^{[n]}\fopw} \}
  - [\gamma^0 \gamma^i , \{ \Pi_i,{^{[n]}\fopw} \} ]
  - \{ \gamma^0 \gamma^i , [i\Delta_i , {^{[n]}\fopw} ] \}
  \nonumber\\  &&
  - \sum_{k=1}^n {n \choose k}
  \Bigl(
  2 [ N_{(k)0} , {^{[n-k]}\fopw} ]_{k+1}
  + [\gamma^0 \gamma^i , [N_{(k)i},{^{[n-k]}\fopw}]_{k+1}]
  \nonumber\\ &&
  \qquad \qquad \quad
  + \{ \gamma^0 \gamma^i , [iM_{(k)i} , {^{[n-k]}\fopw}]_k \}
  \Bigr)
  \nonumber\\
  &&  \mbox{(constraints)}\\
  && \nonumber\\
  \label{momglexp2}
  0 \!&\!=\!&\!
  2 m [\gamma^0 , {^{[n]}\fopw} ]
  - 2 [i\Delta_0 , {^{[n]}\fopw}]
  - \{ \gamma^0 \gamma^i , \{ \Pi_i,{^{[n]}\fopw} \} \}
  - [\gamma^0 \gamma^i , [i\Delta_i , {^{[n]}\fopw} ]]
  \nonumber\\
  && - \sum_{k=1}^n {n \choose k}
  \Bigl(
  2 [iM_{(k)0} , {^{[n-k]}\fopw}]_k
  + \{ \gamma^0 \gamma^i , [N_{(k)i}, {^{[n-k]}\fopw}]_{k+1} \}
  \nonumber\\ &&
  \qquad \qquad \quad
  + [\gamma^0 \gamma^i , [iM_{(k)i} , {^{[n-k]}\fopw}]_k]
  \Bigr).
  \nonumber\\
  && \mbox{(dynamics)}
 \end{eqnarray}
The first of the two sets of equations, which couples the $n$ lowest
moments to the $n+1^{\rm st}$ one, contains only derivatives with
respect to the spatial components of $x$ and no time derivatives. We
can formally write it as
 \begin{equation}
  \label{momform1}
  {^{[n+1]}{\hat W}}(\vec x,\vec p,x_0)
    = \sum_{k=0}^n {\mathcal O}_{(k)}^n
    (\vec x, \vec p, x_0; \vec{\partial}_x,\vec{\partial}_p) \
    {^{[k]}{\hat W}}(\vec x,\vec p,x_0)\, ,
 \end{equation}
which shows that it forms a set of constraints.

The second set of equations determines the time derivative of the
$n^{\rm th}$ moment in terms of all lower moments including the
$n^{\rm th}$ one, which we write formally as
 \begin{eqnarray}
 \label{momform2}
   \partial_0 \, {^{[n]}{\hat W}}(\vec x,\vec p,x_0)
   &{+}& {\mathcal Q}_{(n)}^n (\vec x, \vec p, x_0; \vec{\partial}_x,
     \vec{\partial}_p) \, {^{[n]}{\hat W}}(\vec x, \vec p, x_0)
 \nonumber\\
   &{=}& -\sum_{k=0}^{n-1} {\mathcal Q}_{(k)}^n
     (\vec x, \vec p, x_0; \vec{\partial}_x,\vec{\partial}_p) \,
     {^{[k]}{\hat W}}(\vec x, \vec p, x_0).
 \end{eqnarray}
This is a time-dependent partial differential equation for the $n^{\rm
  th}$ moment, with a source term from the lower moments, which we
interpret as a single-time transport equation. The equation for $n=0$
is special because in this case the source term on the r.h.s. of
(\ref{momform2}) vanishes. The operators ${\mathcal O}_{(k)}^n,
{\mathcal Q}_{(k)}^n$ are functionals of the gauge fields and contain
arbitrary powers of the spatial coordinate and momentum derivatives.

The spinor decomposition of the moment equations is given in
Appendix~\ref{appf}.

\subsection{BGR equations}
\label{sec4.2}

Limiting ourselves to external fields and considering only the lowest
(zeroth) moments we can reduce the dynamical equations
(\ref{engyno-p01}-\ref{engyno-p08}) for the spinor components to the
BGR equations \cite{BGR1,MOM}. For QED with external fields we find
for the non-local operators
\begin{eqnarray}
  \int dp_0\, p_0^n\, [\Delta_\mu, \fopx (x,p)]
  &\!\!\stackrel{{\mbox{\scriptsize external}\atop\mbox{\scriptsize fields}}}
  {\longrightarrow}\!\!&
  {\mathcal D}_\mu \ {^{[n]}X}(x,\vec p)
  + \sum_{k=1}^n {n \choose k} K_{(k)\mu}
  \ {^{[n-k]}X} (x,\vec p)
  \\
  \int dp_0\, p_0^n\, \{ \Pi_0,\fopx (x,p)\}
  &\!\!\stackrel{{\mbox{\scriptsize external}\atop\mbox{\scriptsize fields}}}
  {\longrightarrow}\!\!&
  2{^{[n+1]}X} (x,\vec p)
  + 2\sum_{k=0}^n {n \choose k} L_{(k)0} \ {^{[n-k]}X} (x,\vec p)
  \\
  \int dp_0\, p_0^n\, \{ \Pi_i,\fopx (x,p)\}
  &\!\!\stackrel{{\mbox{\scriptsize external}\atop\mbox{\scriptsize fields}}}
  {\longrightarrow}\!\!&
  2{\mathcal P}_i \ {^{[n]}X}(x,\vec p) + 2\sum_{k=1}^n {n \choose k} L_{(k)i}
  \ {^{[n-k]}X} (x,\vec p)
\end{eqnarray}
with the definitions (\ref{qedopa},\ref{qedopb}) for ${\mathcal
  P}_\mu$, ${\mathcal D}_\mu$ and the following operators $K_{(k)\mu}$ and
$L_{(k)\mu}$:
 \begin{eqnarray}
 \label{momentsop_qed2}
  K_{(k)\mu} \!&\!=\!&\! g\!\!\!\int\limits_{-1/2}^{1/2}\!\!\! ds (-is)^k
  [\partial^{(k)}_{x_0} \fopf_{\mu j}(\vec x +is\vec{\partial}_p,x_0)
  \partial_p^j + \frac{ik}{s}
  \partial^{(k-1)}_{x_0}
  \fopf_{0 \mu}(\vec x +is\vec{\partial}_p,x_0)]\, ,
 \\
  L_{(k)\mu} \!&\!=\!&\! g\!\!\!\int\limits_{-1/2}^{1/2}\!\!\! ds (-is)^{k+1}
  [\partial^{(k)}_{x_0}
  \fopf_{\mu j}(\vec x +is \vec{\partial}_p,x_0)
   \partial_p^j + \frac{ik}{s}
  \partial^{(k-1)}_{x_0}\fopf_{0 \mu}(\vec x +is\vec{\partial}_p,x_0)].
 \end{eqnarray}
The BGR equations are thus given as follows
\begin{equation}
  \label{BGRallgemein}
  0 =  2 m [\gamma^0 , {^{[0]}W} ]
  - 2 i{\mathcal D}_0 \ {^{[0]}W}
  - 2 \{ \gamma^0 \gamma^i , {\mathcal P}_i\ {^{[0]}W} \}
  - [\gamma^0 \gamma^i , i{\mathcal D}_i \ {^{[0]}W} ].
\end{equation}
In spinor decomposition they are
\begin{mathletters}
  \label{BGRspinor}
  \begin{eqnarray}
    2 m \ {^{[0]} \Pc} \!&\!=\!&\! {\mathcal D}_\mu \ {^{[0]}\Ac^\mu}
    \\
    2 m \ {^{[0]}\Sc_{0i}} \!&\!=\!&\! {\mathcal D}_0 \ {^{[0]}\Vc_i} -
    {\mathcal D}_i \ {^{[0]}\Vc_0} +
    2 \varepsilon_{0ijk} {\mathcal P}^j \ {^{[0]}\Ac^k}
    \\
    2 m \ {^{[0]}\Vc_i} \!&\!=\!&\! 2 {\mathcal P}_i \ {^{[0]}\Fc} +
    {\mathcal D}^\nu\ {^{[0]}\Sc_{i\nu}}
    \\
    2 m \ {^{[0]}\Ac_0} \!&\!=\!&\! - {\mathcal D}_0 \ {^{[0]}\Pc}
    + 2 {\mathcal P}^i \ {^{[0]}\Lc_{0i}}
    \\
    0 \!&\!=\!&\! {\mathcal D}_\mu \ {^{[0]}\Vc^\mu}
    \\
    0 \!&\!=\!&\! 2 {\mathcal P}_i \ {^{[0]}\Vc_j} -
    2 {\mathcal P}_j \ {^{[0]}\Vc_i} -
    \varepsilon_{0ijk} ({\mathcal D}^0 \ {^{[0]}\Ac^k} -
    {\mathcal D}^k \ {^{[0]}\Ac^0})
    \\
    0 \!&\!=\!&\! {\mathcal D}_0 \ {^{[0]}\Fc} -
    2 {\mathcal P}^i \ {^{[0]}\Sc_{0i}}
    \\
    0 \!&\!=\!&\! 2 {\mathcal P}_i \ {^{[0]}\Pc} -
    {\mathcal D}^0 \ {^{[0]}\Lc_{0i}} +
    {\mathcal D}^j \ {^{[0]}\Lc_{ij}}
  \end{eqnarray}
\end{mathletters}%
The BGR equations therefore are just the dynamical equations for the
zeroth moments for the case of QED with external fields. The
connection with the spinor components $f_i, \vec{g}_i$, $i = 0,1,2,3$,
as introduced by BGR in \cite{BGR1} is
\begin{equation}
  f_0 = {^{[0]}\Vc_0},\quad f_1 = -{^{[0]}\Ac_0},\quad f_2 = {^{[0]}\Pc},
  \quad f_3 = {^{[0]}\Fc},\quad
\end{equation}
\begin{equation}
  \vec{g}_0 = -{^{[0]}\vec{\Ac}},\quad \vec{g}_1 = -{^{[0]}\vec{\Vc}},
  \quad \vec{g}_2 = {^{[0]}\vec{S}},
  \quad \vec{g}_3 = -{^{[0]}\vec{\sigma}}.\quad
\end{equation}

The hierarchy of moment equations Eqs.~(\ref{momglexp1},\ref{momglexp2})
will in general not truncate. The initial problem, namely that the
covariant Wigner operator involves the fermion fields at all times
and thus cannot be properly initialized at $t=-\infty$, has now been
reformulated into the task of solving simultaneously infinitely many
equations as initial value problems. On the other hand the BGR
equations (\ref{BGRallgemein},\ref{BGRspinor}) can not describe the
theory completely since they lack information on all higher moments.
How we can truncate the hierarchy naturally and what role the BGR
equations play with respect to the infinite hierarchy of moment
equations will become clear in the next subsection where we discuss
the moment hierarchy in the limit of external fields.

\subsection{Moment hierarchy for external fields}
\label{sec4.3}

Since the Dirac equation and its adjoint for (time dependent) external
fields are inhomogeneous first order differential equations for the wave
functions, we need to specify as initial conditions only the fields
$\psi, \psi^\dag$ for all $\vec x$ at some initial time $x_0=t_i$.
In terms of moments it should be sufficient to specify ${^{[0]}W}$ at
some initial time $t_i$. From the structure of Eqs.~(\ref{momform1})
it is obvious that there exists a recursion relation which allows to
express all higher order moments ${^{[n]}W}(\vec{x},\vec{p},x_0)$,
$n\geq 1$, through the lowest moment ${^{[0]}W}(\vec{x},\vec{p},x_0)$.
Let us write it schematically as
 \begin{equation}
 \label{recur}
    {^{[n]}W}(\vec{x},\vec{p},x_0) =
    {\mathcal P}_n(\vec{x},\vec{p},x_0;\vec{\partial}_x,\vec{\partial}_p)
    \,{^{[0]}W}(\vec{x},\vec{p},x_0)\, ,\quad n\geq 1,
 \end{equation}
where ${\mathcal P}_n$ is a combination of products of the operators
${\mathcal O}^n_{(k)}$ and thus contains derivatives with respect to
$\vec{x}$ and $\vec{p}$ but no time derivatives. With the help of
(\ref{recur}) one can rewrite (\ref{momform2}) for $n \geq 1$ as
 \begin{equation}
 \label{recur1}
    \partial_0 \left( {\mathcal P}_n \, {^{[0]}W} \right) =
    - \sum_{k=0}^n {\mathcal Q}^n_{(k)} {\mathcal P}_k\,{^{[0]}W}\, ,
    \quad, n\geq 1,
 \end{equation}
or
 \begin{equation}
 \label{recur2}
    {\mathcal P}_n \left(\partial_0\, {^{[0]}W} \right) =
    - \left[\left(\partial_0 {\mathcal P}_n\right)
            + \sum_{k=0}^n {\mathcal Q}^n_{(k)} {\mathcal P}_k
      \right]\,{^{[0]}W}\, , \quad, n\geq 1.
 \end{equation}
Eliminating on the l.h.s. of (\ref{recur2}) \ $\partial_0\, {^{[0]}W}$
with the help of Eq.~(\ref{momform2}) for $n=0$ we obtain
 \begin{equation}
 \label{recur3}
    {\mathcal P}_n\,{\mathcal Q}^0_{(0)}\, {^{[0]}W}  =
    - \left[\left(\partial_0 {\mathcal P}_n\right)
            + \sum_{k=0}^n {\mathcal Q}^n_{(k)} {\mathcal P}_k
      \right]\,{^{[0]}W}\, , \quad, n\geq 1.
 \end{equation}
This is, for each $n\geq 1$, a new constraint equation for
${^{[0]}W}$.

We have thus replaced the set (\ref{momform2}) of single-time
transport equations by a single transport equation for ${^{[0]}W}$
(the case $n=0$ in (\ref{momform2})) plus an infinite set of new
constraint equations (\ref{recur3}) for $n\geq 1$. The single
remaining transport equation reads explicitly
 \begin{equation}
 \label{bewegmom_0}
  2 [i\Delta_0 , {^{[0]}W}] =
  2 m [\gamma^0 , {^{[0]}W}]
  - \{ \gamma^0 \gamma^i , \{ \Pi_i , {^{[0]}W} \} \}
  - [\gamma^0 \gamma^i , [i\Delta_i , {^{[0]}W} ]]\, .
 \end{equation}
This is the non-abelian generalization of the BGR equation
(\ref{BGRallgemein}) for QED. Eq.~(\ref{bewegmom_0}) is a first order
differential equation in $x_0$ for ${^{[0]}W}$ which can be solved
uniquely by specifying ${^{[0]}W}(\vec x,\vec p,x_0)$ at $x_0=t_i$.
But what is the meaning of the additional constraints (\ref{recur3})?

We will now show that they are redundant, i.e. turn into identities
after using the constraints (\ref{recur}). A brute force proof
of this statement was given in \cite{HIE} for the case of a simple
scalar field theory. For QED the statement was proven only for the
lowest non-trivial values of $n$. The reason for this was that for
higher $n$ the explicit calculation of the operators ${\mathcal P}_n$
in (\ref{recur}) becomes extremely cumbersome. We will here give a
more elegant general proof which is also directly applicable to the
non-Abelian case which exploits the ``equivalence'' (with respect to
taking higher $p_0$-moments) of the two {\em covariant} equations from
which the single-time transport and constraint equations were derived.

The proof is made more presentable by introducing the following
notation for multiple commutators:
 \begin{equation}
   A_- X = [A,X], \quad A_+ X = \{A,X\}, \quad A_- B_- X = [A,[B,X]]
   \qquad \mbox{etc.},\quad
 \end{equation}
where the commutators or anti-commutators extend to everything to the
right of the operator $A_\pm$. For the spinor structure we will use in
addition
 \begin{equation}
   g^0 \equiv \gamma^0, \qquad g^i \equiv \gamma^0 \gamma^i
 \end{equation}

Let us now look at the integrands under the moment integral $\int
dp_0\, p_0^n$ for the two equations which gave rise to
Eqs.~(\ref{momform1},\ref{momform2}):
 \begin{eqnarray}
 \label{bewegna2}
   2 p_{0+} W &=& \left[ 2mg^0_+ - 2 N_0^+ - g^i_- \Pi_{i+}
                       - g^i_+ i \Delta_{i-}\right] W\, ,
 \\
 \label{bewegna1}
   0 &=& \left[ 2m g_-^0 - 2i \Delta_{0-} - g^i_+ \Pi_{i+}
              - g^i_- i \Delta_{i-}\right] W \, .
 \end{eqnarray}
We will now consider in addition Eq.~(\ref{bewegna1}) multiplied with
$2 p_0$ (which can be written as $p_{0+}[\mbox{Eq.~(\ref{bewegna1})}]$~)
 \begin{equation}
 \label{bewegnb}
  p_{0+} \left[ 2m g_-^0 - 2i  \Delta_{0-}
              - g^i_+ \Pi_{i+} - i\,g^i_- \Delta_{i-}
         \right] W = 0.
 \end{equation}
After taking moments $\int dp_0\, p_0^n$,
Eqs.~(\ref{bewegna1},\ref{bewegnb}) give dynamical equations for
the moments $n,n+1$ while Eq.~(\ref{bewegna2}) gives constraints which
tell us how to eliminate the moment $n+1$ through the moments
$k=0,\dots,n$.

The idea of the proof is to push in Eq.~(\ref{bewegnb}) the operator
$p_{0+}$ to the right until it stands in front of $W$, then to use
Eq.~(\ref{bewegna2}) to eliminate $p_{0+}W$ and afterwards exploit
Eq.~(\ref{bewegna1}) to eliminate the dynamical part $\Delta_{0-} W$.
The result will be an identity. On the level of moments this is
equivalent to the elimination procedure discussed at the beginning of
this subsection.

We first push $p_{0+}$ in Eq.~(\ref{bewegnb}) to the right and insert
Eq.~(\ref{bewegna2}). Noting that $g_\pm^\mu$ commutes with $p_{0+}$,
$\Delta_{\mu-}$, $\Pi_{\mu+}$ we find
 \begin{eqnarray}
  \label{bewegnc}
  0 &=& \Bigl[
    m g^0_- - i \Delta_{0-} - \frac{1}{2} g^i_+ \Pi_{i+}
    - {\textstyle{1\over 2}} g^i_- \Delta_{i-} \Bigr (2 p_{0+} W)
 \nonumber\\
  && + \Bigl[
    2( i\Delta_{0-} p_{0+} - p_{0+} i\Delta_{0-})
    + g^i_+ ( \Pi_{i+} p_{0+} - p_{0+} \Pi_{i+})
 \nonumber\\
  && \ \ + g^i_- ( \Delta_{i-} p_{0+} - p_{0+} \Delta_{i-})
     \Bigr] W
 \nonumber\\
  &=& \Bigl[ m g^0_- - i \Delta_{0-}
             - {\textstyle{1\over 2}} g^i_+ \Pi_{i+}
             - {\textstyle{1\over 2}} g^i_- \Delta_{i-} \Bigr]
      \Bigl[ 2 m g^0_+ - 2 N_{0+} -  g^i_- \Pi_{i+}
             - g^i_+ \Delta_{i-} \Bigr] W
 \nonumber\\
  && + \Bigl[
    2( i\Delta_{0-} p_{0+} - p_{0+} i\Delta_{0-})
    + g^i_+ ( \Pi_{i+} p_{0+} - p_{0+} \Pi_{i+})
 \nonumber\\
  &&\ \  + g^i_- ( \Delta_{i-} p_{0+} - p_{0+} \Delta_{i-}) \Bigr] W.
 \end{eqnarray}
Now (i.e. after having inserted (\ref{bewegna2})!) we push
$\Delta_{0-}$ again back to the right and use Eq.~(\ref{bewegna1}) for
$\Delta_{0-} W$:
 \begin{eqnarray}
  \label{bewegnd}
  0 &=&\ \Bigl[ m g^0_- - {\textstyle{1\over 2}} g^i_+ \Pi_{i+}
             - {\textstyle{1\over 2}} g^i_- \Delta_{i-} \Bigr]
        \Bigl[ 2 m g^0_+ - 2 N_{0+} -  g^i_- \Pi_{i+}
             - g^i_+ \Delta_{i-} \Bigr] W
 \nonumber\\
  && - \Bigl[ 2 m g^0_+ - 2 N_{0+} -  g^i_- \Pi_{i+}
            - g^i_+ \Delta_{i-} \Bigr]
       \Bigl[ m g^0_- - {\textstyle{1\over 2}} g^i_+ \Pi_{i+}
            - {\textstyle{1\over 2}} g^i_- \Delta_{i-} \Bigr] W
 \nonumber\\
  && + 2 \Bigl( i\Delta_{0-} p_{0+} - p_{0+} i\Delta_{0-}
     + i\Delta_{0-} N_{0+} - N_{0+} i\Delta_{0-}\Bigr) W
 \nonumber\\
  && + g^i_+ \Bigl( \Pi_{i+} p_{0+} - p_{0+} \Pi_{i+}
     + i \Delta_{0-} i\Delta_{i-} - i \Delta_{i-} i \Delta_{0-}
  \Bigr) W
 \nonumber\\
  && + g^i_- \Bigl( i\Delta_{i-} p_{0+} - p_{0+} i\Delta_{i-}
     + i\Delta_{0-} \Pi_{i+} - \Pi_{i+} i\Delta_{0-} \Bigr)  W.
\end{eqnarray}
Expanding and combining terms with the same spinor structure leads to
\begin{eqnarray}
  \label{bewegne}
  0 &=& \biggl[
    2m^2 ( g^0_- g^0_+ - g^0_+ g^0_-) - 2m N_{0+}(g^0_- - g^0_-)
    \nonumber\\&&
    - m \Pi_{i+} (g^0_- g^i_- + g^i_+ g^0_+ - g^i_- g^0_- - g^0_+ g^i_+ )
    \nonumber\\ &&
    - m i\Delta_{i-} ( g^0_- g^i_+ + g^i_- g^0_+ - g^0_+ g^i_- - g^i_+ g^0_- )
    \nonumber\\ &&
    + g^i_+ [\Pi_{i+} \Pi_{0+} - \Pi_{0+} \Pi_{i+} + i\Delta_{0-} i\Delta_{i-}
    - i\Delta_{i-}  i\Delta_{0-} ]
    \nonumber\\ &&
    + g^i_- [i\Delta_{i-} \Pi_{0+} - \Pi_{0+} i\Delta_{i-} + i\Delta_{0-}
    \Pi_{i+} - \Pi_{i+} i\Delta_{0-} ]
    \nonumber\\ &&
    + {\textstyle{1\over 2}} (g^i_- g^j_+ - g^i_+ g^j_-)
      [i\Delta_{i-} i\Delta_{j-} - \Pi_{i+}\Pi_{j+}]
    \nonumber\\ &&
    + {\textstyle{1\over 2}} (g^i_+ g^j_+ - g^i_- g^j_-)
      [\Pi_{i+}i\Delta_{j-} - i\Delta_{i-} \Pi_{j+}]
  \biggr] W.
\end{eqnarray}
As can be easily checked the following relations for the spinor
matrices hold:
\begin{eqnarray}
  \label{spinorkomrel}
  (g_-^0 g_+^0 - g_+^0 g_-^0)W &=& 0\\
  (g_-^0 g_+^i -  g_+^0 g_-^i + g_-^i g_+^0 - g_+^i g_-^0)W &=& 0\\
  (g_-^0 g_-^i -  g_+^0 g_+^i + g_+^i g_+^0 - g_-^i g_-^0)W &=& 0\\
  (g_+^i g_+^j - g_-^i g_-^j)W &=& 2(g^i W g^j + g^j W g^i)\\
  (g_-^i g_+^j - g_+^i g_-^j)W &=& 2(g^i W g^j - g^j W g^i).
\end{eqnarray}
Therefore the terms containing the mass $m$ will vanish. The rest can
be combined, using the explicit form of the $\gamma$ matrices, to give
\begin{equation}
  \label{bewegnf}
  0 = \left[
    (\Pi_{\mu+} + i \Delta_{\mu-})(\Pi_{\nu+} - i \Delta_{\nu-}) -
    (\Pi_{\nu+} - i \Delta_{\nu-})(\Pi_{\mu+} + i \Delta_{\mu-})
  \right] \gamma^0 \gamma^\mu W \gamma^\nu \gamma^0.
\end{equation}
With the help of Eqs.~(\ref{kommrelextfeld}) we finally arrive at
\begin{eqnarray}
  \label{bewegng}
  0 &=& [p_\mu + \Pi_\mu + i \Delta_\mu, p_\nu + \Pi_\nu - i \Delta_\nu]
  \gamma^0 \gamma^\mu W \gamma^\nu \gamma^0
  \nonumber\\ &&
  - \gamma^0 \gamma^\nu W \gamma^\mu \gamma^0
  [ p_\nu + \Pi_\nu - i \Delta_\nu, p_\mu + \Pi_\mu + i \Delta_\mu].
\end{eqnarray}
In Appendix~\ref{appg} we show that the commutator vanishes:
 \begin{equation}
 \label{comvan}
    [ p_\mu + \Pi_\mu + i \Delta_\mu, p_\nu + \Pi_\nu - i \Delta_\nu]
    \equiv 0 \, .
 \end{equation}
Eq.~(\ref{bewegng}) is therefore an identity.

In the case of external fields we thus have a natural truncation of the
transport hierarchy on the level of ${^{[0]}W}$. Its time evolution
is uniquely determined by Eq.~(\ref{bewegmom_0}) once $W(\vec x,\vec
p, t)\vert_{t = t_0}$ has been specified. All higher moments can be
found through successive insertion of the solution into
Eq.~(\ref{momglexp1}). These ``constraints'' are differential equations
in $\vec x$ and $\vec p$ without time evolution. For the external
field problem the ``initial value problem'' for the quark Wigner
operator has thus been solved, and we have also given an explicit
prescription how to find the higher moments. The latter was missing
in the work of Refs.~\cite{BGR1,BGR2,BGR3}, although the BGR equations
Eqs.~(\ref{BGRallgemein}) properly describe the dynamics of the
zeroth moment. Higher moments do, however, contain important physical
information even for external fields. For example, the energy momentum
tensor is given by the first moments (for explicit expressions see
\cite{HIE}) and can therefore be calculated only with the help of the
constraint Eq.~(\ref{momglexp1}) once the solution for the zeroth
moment has been obtained.

\subsection{QED and external fields}
\label{sec4.4}

In this subsection we make the results obtained above explicit for the
case of an {\em abelian} external gauge field, i.e. for QED with
external fields. We will show the explicit connection between the
covariant theory and the moment equations in leading order of the
gradient expansion and demonstrate that the correct Vlasov (including
spin precession terms) limit is recovered from the single-time
transport equations.

To zeroth order in the gradient expansion the equations for the moments are
 \begin{mathletters}    \label{momhbar0VF}
  \begin{eqnarray}
    {^{[n+1]}\Vc^{0(0)}} &=& m \ {^{[n]}\Fc^{(0)}} - p_i \ {^{[n]}\Vc^{i(0)}}
    \\
    {^{[n+1]}\Fc^{(0)}} &=& m \ {^{[n]}\Vc^{0(0)}}
    \\
    {^{[n+1]}\Vc^{i(0)}} &=& p^i \ {^{[n]}\Vc^{0(0)}}
    \\
    m {^{[n]}\Vc^{i(0)}} &=& p^i \ {^{[n]}\Fc^{(0)}}
    \\
    p^i \ {^{[n]}\Vc^{j(0)}} &=& p^j \ {^{[n]}\Vc^{i(0)}}
    \\
    {^{[n+1]}\Pc^{(0)}} &=& 0
    \\
    {^{[n]}\Pc^{(0)}} &=& 0
    \\
    p^i \ {^{[n]}\Pc^{(0)}} &=& 0
  \end{eqnarray}
\end{mathletters}%
and
\begin{mathletters}\label{momhbar0AS}
  \begin{eqnarray}
    {^{[n+1]}\Ac^{k(0)}} &=& -m \ {^{[n]}\Lc^{0k(0)}} + p^k
    \ {^{[n]}\Ac^{0(0)}}
    \\
    {^{[n+1]}\Sc^{jk(0)}} &=& m \varepsilon^{0ijk} \ {^{[n]}\Ac^{(0)}_i}
    - p^k \ {^{[n]}\Sc^{0j(0)}} + p^j \ {^{[n]}\Sc^{0k(0)}}
    \\
    {^{[n+1]}\Ac^{0(0)}} &=& - p_i \ {^{[n]}\Ac^{i(0)}}
    \\
    {^{[n+1]}\Sc^{0i(0)}} &=& p_j \ {^{[n]}\Sc^{ij(0)}}
    \\
    m \ {^{[n]}\Sc^{0i(0)}} &=& \varepsilon^{0ijk} p_j \ {^{[n]}\Ac^{(0)}}_k
    \\
    m \ {^{[n]}\Ac^{0(0)}} &=& p_i \ {^{[n]}\Lc^{0i(0)}}
    \\
    0 &=& p_i \ {^{[n]}\Sc^{0i(0)}}.
  \end{eqnarray}
\end{mathletters}%
Eqs.~(\ref{momhbar0VF}vi,vii,viii) show that ${^{[n]}\Pc^{(0)}}$ vanishes for
all $n$ identically. In addition we get that
$\Vc_\mu^{(0)}$ can be expressed by $\Fc^{(0)}$ and $\Ac_\mu^{(0)}$ by
$\Sc_{\mu\nu}^{(0)}$.
Eqs.~(\ref{momhbar0VF}) can be solved now for ${^{[n]}\Fc^{(0)}}$ and
${^{[n]}\Vc_\mu^{(0)}}$ by specifying the lowest moments
${^{[0]}\Fc^{(0)}}$ and ${^{[0]}\Vc^{0(0)}}$
($E = \sqrt{{\vec p}^{\ 2} + m^2}> 0, m>0$):
{\renewcommand{\arraystretch}{2.0}
  \begin{equation}
      \begin{array}[h]{rclcrcl}
        \D {^{[2n]}\Vc^{0(0)}} &=& \D E^{2n} \ {^{[0]}\Vc^{0(0)}} &\qquad&
        \D{^{[2n+1]}\Vc^{0(0)}} &=&\D E^{2n+1}\frac{E}{m} \
        {^{[0]}\Fc^{(0)}} \\
        \D{^{[2n]}{\vec \Vc}^{(0)}} &=&\D E^{2n}
        \frac{\vec p}{m}
        \ {^{[0]}\Fc^{(0)}} &\qquad&
        \D{^{[2n+1]}{\vec \Vc}^{(0)}} &=&
        \D E^{2n+1}\frac{\vec p}{E} \ {^{[0]}\Vc^{0(0)}} \\
        \D{^{[2n]}\Fc^{(0)}} &=&
        E^{2n} \ {^{[0]}\Fc^{(0)}} &\qquad&
        \D{^{[2n+1]}\Fc^{(0)}} &=& \D E^{2n+1}\frac{m}{E} \
        {^{[0]}\Vc^{0(0)}}.
      \end{array}
  \end{equation}
}%
Using
\begin{equation}
  m \ {^{[0]}\Vc^{0(0)}} = {^{[1]}\Fc^{(0)}}
\end{equation}
we can also write
{\renewcommand{\arraystretch}{2.0}
  \begin{equation}
    \begin{array}{rclcrcl}
      \D {^{[2n]}\Vc^{0(0)}} &=&
      \D E^{2n} \frac{1}{m}\ {^{[1]}\Fc^{(0)}}
      &\qquad&
      \D {^{[2n+1]}\Vc^{0(0)}} &=& \D E^{2n+1}\frac{E}{m} \
      {^{[0]}\Fc^{(0)}} \\
      \D {^{[2n]}{\vec \Vc}^{(0)}} &=&
      \D E^{2n}\frac{\vec p}{m}
      \ {^{[0]}\Fc^{(0)}}
      &\qquad&
      \D {^{[2n+1]}{\vec \Vc}^{(0)}} &=&
      \D E^{2n+1}\frac{\vec p}{Em} \ {^{[1]}\Fc^{(0)}} \\
      \D {^{[2n]}\Fc^{(0)}} &=&
      \D E^{2n} \ {^{[0]}\Fc^{(0)}}
      &\qquad&
      \D {^{[2n+1]}\Fc^{(0)}} &=& \D E^{2n+1}\frac{1}{E} \
      {^{[1]}\Fc^{(0)}},
    \end{array}
  \end{equation}
}%
thus specifying rather ${^{[0]}\Fc^{(0)}}$ and ${^{[1]}\Fc^{(0)}}$ instead of
${^{[0]}\Fc^{(0)}}$ and ${^{[0]}\Vc^{0(0)}}$.
The connection with particle (upper sign) and anti-particle (lower sign)
distribution functions from the Lorentz-covariant theory is found to be
\begin{equation}\label{verteiqed}
  \Fc(x, \vec p , \pm E) = E \ {^{[0]}\Fc^{(0)}} \pm m \ {^{[0]}\Vc^{0(0)}} =
  E \ {^{[0]}\Fc^{(0)}} \pm {^{[1]}\Fc^{(0)}}.
\end{equation}
Thus the particle and anti-particle contributions of the covariant theory
$\Fc(x,\vec p , \pm E)$ translate into the sum and difference respectively of
${^{[0]}\Fc}$ and ${^{[1]}\Fc}$ in single-time formulation.
This is immediately obvious from the fact that ${^{[0]}\Fc}$ is derived from
$\Fc$ as an integral
over all energies, that is particle and anti-particle contributions
\begin{equation}
{^{[0]}\Fc(x,\vec p)} = \int\limits_{-\infty}^{\infty} dp_0 \Fc(x,p),
\end{equation}
whereas ${^{[1]}\Fc}$ has an additional weight $p_0$, leading to the difference
between particles with positive and negative energies.

The transport equations for the zeroth order moments in the semi-classical
approximation are now given by
(see Appendix~\ref{apph})
\begin{eqnarray}  \label{qedmombeweg1}
  0 \!&\!=\!&\! E^n \left[
    E(\partial^0 + \frac{e}{c} \nabla_p\cdot \vec E) {^{[1]}\Fc^{(0)}}
    + (\vec p \cdot \nabla + \frac{e}{c} (\vec p \times \vec B )
   \cdot\nabla_p) (E {^{[0]}\Fc^{(0)}})
  \right]
  \\  \label{qedmombeweg1b}
  0 \!&\!=\!&\! E^n \left[
    E(\partial^0 + \frac{e}{c} \nabla_p\cdot \vec E) (E{^{[0]}\Fc^{(0)}})
    + (\vec p \cdot \nabla + \frac{e}{c} (\vec p \times \vec B )
    \cdot \nabla_p) {^{[1]}\Fc^{(0)}}
  \right].
\end{eqnarray}
Adding and subtracting gives us back exactly the transport equations we get
from putting the distribution function of the covariant theory on the mass
shell
\begin{displaymath}
\Fc(x,p) = \Fc(x, \vec p ,\pm E) \delta(p_0 \mp E),
\end{displaymath}
inserting into the transport equations of the covariant theory
Eqs.~(\ref{spinorglqedaeu_hbar2}i,\ref{QEDeqn2}i) and integrating over
$\int\!dp_0\ p_0^n$.
This shows by example the equivalence between covariant and single-time
formulation.

Similar we get for ${^{[n]}\Ac^{(0)}}$ and ${^{[n]}\Sc_{\mu\nu}^{(0)}}$  the
following equations, if we specify the lowest moments of ${^{[n]}\Sc^{ij(0)}}$,
that is ${^{[0]}{\vec \sigma}^{i(0)}}$ and
${^{[1]}{\vec \sigma}^{i(0)}}$
\begin{mathletters}
  \begin{eqnarray}
    {^{[2n]}\Ac^{0(0)}} &=&
    E^{2n} \frac{1}{m} \vec p \cdot {^{[0]}{\vec \sigma}^{(0)}}
    \\
    {^{[2n]}{\vec \Ac}^{(0)}} &=&
    E^{2n} \frac{1}{m}
    \left(
      {^{[1]}{\vec \sigma}^{(0)}} +
      \vec p \times (\vec p \times \frac{{^{[1]}{\vec \sigma}^{(0)}}}{E^2})
    \right)
    \\
    {^{[2n]}{\vec S}^{(0)}} &=&
    - E^{2n} \vec p \times \frac{{^{[1]}{\vec \sigma}^{(0)}}}{E^2}
    \\
    {^{[2n]}{\vec \sigma}}^{(0)} &=&
    E^{2n} \ {^{[0]}{\vec \sigma}^{(0)}}
    \\
    {^{[2n+1]}\Ac^{0(0)}} &=&
    E^{2n+1} \frac{1}{m}
    \vec p \cdot \frac{{^{[0]}{\vec \sigma}^{(0)}}}{E}
    \\
    {^{[2n+1]}{\vec \Ac}^{(0)}} &=&
    E^{2n+1} \frac{1}{m}
    \left(
      \frac{{^{[0]}{\vec \sigma}^{(0)}}}{E} +
      \vec p \times (\vec p \times \frac{{^{[0]}{\vec \sigma}^{(0)}}}{E})
    \right)
    \\
    {^{[2n+1]}{\vec S}^{(0)}} &=&
    - E^{2n+1} \vec p \times \frac{{^{[0]}{\vec \sigma}^{(0)}}}{E}
    \\
    {^{[2n+1]}{\vec \sigma}}^{(0)} &=&
    E^{2n+1} \frac{{^{[1]}{\vec \sigma}^{(0)}}}{E}.
  \end{eqnarray}
\end{mathletters}%
As before we have for the connection with the particle and anti-particle
distribution of the covariant theory
\begin{equation}\label{verteiqed2}
  \vec \sigma (x, \vec p , \pm E) = E \ {^{[0]}{\vec \sigma}^{(0)}}
  \pm  {^{[1]}{\vec \sigma}^{0(0)}}.
\end{equation}
The transport equations as derived in Appendix~\ref{apph} are
\begin{mathletters}    \label{bewegmomsigma}
  \begin{eqnarray}
    0 &=&
      E(\partial^0 + \frac{e}{c} \nabla_p\cdot \vec E )
      \frac{{^{[1]}{\vec \sigma}^{(0)}}}{E}
      + (\vec p \cdot \nabla + \frac{e}{c} (\vec p \times \vec B )
      \cdot \nabla_p) {^{[0]}{\vec \sigma}^{(0)}}
      \nonumber\\ &&
      \qquad + \frac{e}{c}
      \left(
        \vec p \cdot ( \vec E \cdot
        \frac{{^{[1]}{\vec \sigma}^{(0)}}}{E^2})
        + \vec B \times {^{[0]}{\vec \sigma}^{(0)}} )
      \right)
    \\
    0 &=&
    \biggl[
      E (\partial^0 + \frac{e}{c} \nabla_p\cdot \vec E )
      {^{[0]}{\vec \sigma}^{(0)}}
      + (\vec p \cdot \nabla + \frac{e}{c} (\vec p \times \vec B )
      \cdot \nabla_p) \frac{{^{[1]}{\vec \sigma}^{(0)}}}{E}
      \nonumber\\ &&
      \qquad\qquad+ \frac{e}{c} \left(
        \vec p \cdot ( \vec E \cdot \frac{{^{[0]}{\vec \sigma}^{(0)}}}{E})
        + \vec B \times
      \frac{{^{[1]}{\vec \sigma}^{(0)}}}{E}
    \right)
    \biggr].
  \end{eqnarray}
\end{mathletters}%
Adding and subtracting gives us again the transport equations for the
particle and anti-particle distributions of the covariant theory
Eqs.~(\ref{spinorglqedaeu_hbar2}ii,\ref{QEDeqn2}ii) after integrating over
$\int\! dp_0\ p_0^n$ and employing the mass shell condition $\vec \sigma$
$= \vec \sigma (x, \vec p ,\pm E) \delta(p_0 \mp E)$.
Eqs.~(\ref{bewegmomsigma}) are therefore again the generalized BMT-equations
Eqs.~(\ref{QEDeqn2}), but now formulated with moments.

In the case of QED with external fields we thus showed the explicit connection
between the formulation with moments and the covariant theory. To this end we
had to solve the moment hierarchy exactly in the classical limit. This
solution proved then to correspond to the classical mass shell condition of
the covariant theory.

\section{Conclusions}
\label{sec5}

In this paper we have established the connection between the covariant
and the so-called single-time formulations of the quantum kinetic
theory, which is based on the dynamics of the quantum mechanical
Wigner operator. We discussed the role of the time parameter in the
density matrix, and we could show that a formulation with a single
time parameter necessarily leads to a hierarchy of equations for the
energy moments for the Wigner operator. For these moments we found a
direct connection with the time derivatives of the wave functions.

A covariant field theory is equivalent to the complete set of its
$p_0$-moments. The impossibility to formulate the covariant transport
theory as an initial value problem, which we discussed in
Sec.~\ref{sec2}, thus translates into the impossible task of solving
an infinite hierarchy of moment equations. The hierarchy structure
lends itself, however, to a systematic approximation scheme via
successive truncation. For fermion dynamics in external classical
gauge fields the truncation problem can be solved {\rm exactly}, in
the case of Dirac fields at the level of the lowest $p_0$-moment, the
single-time Wigner function $W_3(\vec x, \vec p, t)$. This reflects
the fact that for external gauge fields the fermionic theory is a
single-particle theory whose dynamics can be solved uniquely once
initial conditions for the fermion wave functions $\psi(\vec x,x_0)$
at $x_0=t_i$ have been specified.

To find the equations of motion for the moments we first reviewed the
covariant transport theory, extending some of the results obtained
previously in \cite{LINK}, \cite{QTT}. We introduced a compact
notation for generalized non-local momentum and derivative operators
which allowed us to cast the equations of motion for the Wigner
operator into a form which permitted further manipulation. We then
performed a gradient expansion followed by a color decomposition and
established the connection between the (local) momentum operator and
covariant derivative with their non-local generalizations.

To find the proper mass shell and transport equations for the  Wigner
operator we had also to perform a spinor decomposition. Thereby we
could eliminate half of the initial degrees of freedom to end up with
a system of equations for only eight independent spinor components,
whereas the other eight components could be expressed as dependent
variables. This improves on previous covariant treatments were the
system of mass shell and transport equations always contained
redundant information. We derived this system of mass shell and
transport equations plus constraints on the one hand for scalar,
pseudo-scalar and tensor components, which are a suitable basis of
spinor components for massive quarks; alternatively we also derived
mass shell, transport and constraint equations for vector and axial
vector components, which are more suitable as independent components
for the case of vanishing bare quark masses (chiral limit).

That our equations have the proper classical limit was proven by an
explicit gradient expansion. We could also find the expected vacuum
solutions where the pseudo-scalar and spin density did vanish
identically. For chiral theory we found the expected conservation of
vector and axial vector current. We also compared our results with
recent investigations of quantum transport equations for the
Nambu-Jona-Lasinio theory. We showed that the failure of those
equations to decouple scalar, pseudoscalar and tensor densities from
the vector and axial vector ones was related to the spontaneous
breaking of chiral symmetry in that theory, with the dynamically
generated mass acting like a non-zero bare mass in our case, thereby
coupling the two sectors.

After the discussion of the covariant theory we considered the
expansion of our equations in energy moments. This moment expansion
was performed with the help of a partial gradient expansion in
$\partial_{p_0}$ and resulted in an infinite hierarchy of dynamical
equations for the moments, which are coupled by constraints. The
equivalence of both formulations was given by construction. For
external Abelian fields to leading order of the gradient expansion we
showed the explicit connection of the particle and anti-particle
distributions of covariant theory with the zeroth and first moments in
the equal time formulation. Finally we were able to show that for the
external field case we need only the dynamical equations for the
zeroth moments. All higher moments can then be successively calculated
from the lowest moments via the constraint equations. This leads for
the external field case to a natural truncation condition for the
moment hierarchy where the minimal truncation, that is ignoring all
but the lowest moments, just corresponds to the BGR equations.

The question now arises whether now the equations of motion for the
single-time Wigner operators can be finally applied for the (numerical)
description of a QGP or electron-ion plasma. For the transport
equations with the full operator structure the answer is certainly no
in practice. Further work is required to analyze correlations and
collision terms in this case which should be obtained by taking an
ensemble average and using appropriate factorization of expectation
values. For external fields, however, where the equations reduce to
transport equations without a collision term, quantum transport theory
has now been formulated in the single-time approach in a way which
allows for immediate numerical implementation, for both QED and QCD.
For each color and spin component {\em one} transport equation must be
solved, together (but not simultaneously!) with as many
time-independent constraints as given by the highest order of
$p_0$-moments which one wants to know.

Quantum transport theory beyond the external field limit and the
extraction of suitable collision terms remains a great challenge.
For gauge theories the Wigner formalism, through the occurrence of
Schwinger strings, strongly suggests to use the radial gauge
\cite{RAD} for which perturbation theory has only recently become
viable after a set of consistent Feynman-rules was established in
Ref.~\cite{RAD2}. Infrared divergences and the need for resummation in
non-abelian gauge theories with massless degrees of freedom have,
however, so far hampered practical progress.

\appendix
\section{Color matrices}
\label{appa}

The generators of $SU(3)$ in the fundamental representation,
$t_a={\hbar\over 2}\lambda_a$ (where $\lambda_a$ are the Gell-Mann
matrices), satisfy the following identities
 \begin{eqnarray}
 \label{A_farbid1}
  (t^a)^\dag &=& t^a\\
  \label{A_farbid2}
  {[ t^a , t^b ]} &=& i \hbar f^{abc} t^c \\
  \label{A_farbid3}
  \{ t^a,t^b\} &=& \frac{\hbar^2}{3}\delta^{ab} {\bf 1}
  +  \hbar d^{abc} t^c\\
  \label{A_farbid4}
  t^a t^b &=& \frac{1}{2}
  \left( \frac{\hbar^2}{3} \delta^{ab} {\bf 1} + \hbar
    \left(
      d^{abc} + i f^{abc}
    \right) t^c
  \right)
  \nonumber\\
  &\equiv& \frac{1}{2} \left(
    \frac{\hbar^2}{3}\delta^{ab} {\bf 1} + \hbar h^{abc}t^c
  \right).
\end{eqnarray}
$f^{abc}$ and $d^{abc}$ are the totally anti-symmetric and totally
symmetric structure constants of $SU(3)$.

The traces of the generators are given by
\begin{eqnarray}
  \label{A_farbsp}
  {\rm tr \ } t^a &=& 0\\
  {\rm tr \ }t^at^b &=& \frac{\hbar^2}{2} \delta^{ab}\\
  {\rm tr \ }t^a t^b t^c &=& \frac{\hbar^3}{4}\left( d^{abc} + if^{abc}\right)
  \equiv \frac{\hbar^3}{4} h^{abc}\\
  {\rm tr \ }t^a t^b t^a t^c &=& -\frac{\hbar^4}{12} \delta^{bc}.
\end{eqnarray}
One also has
\begin{eqnarray}
  \label{A_farbst}
  f^{abc} &=& -i \frac{2}{\hbar^3}\ {\rm tr \ }([t^a,t^b]t^c)\\
  d^{abc} &=& \frac{2}{\hbar^3}\ {\rm tr \ }(\{t^a,t^b\}t^c)\\
  f^{abb} &=& 0\\
  d^{abb} &=& 0\\
  f^{abc}f^{bcd} &=& 3 \delta^{ab}.
\end{eqnarray}

\section{Gauge transformations}
\label{appb}

Let $S(x)$ be a gauge transformation in $U(1)$ or $SU(3)$:
 \begin{equation}
 \label{eichtrafo0}
  S(x) = e^{i\theta_a(x) t_a}, \qquad [\theta_a(x) t_a, S(x)] = 0.
 \end{equation}
(For $U(1)$ there is only one generator, $t_0=1$.) Then the following
transformation rules apply:
 \begin{eqnarray}
 \label{eichtrafo1}
  \psi(x) &\rightarrow& S(x)\psi(x)\\
  \psi^\dag(x) &\rightarrow& \psi^\dag(x) S^{-1}(x)\\
  \fopa_\mu(x) &\rightarrow& S(x)\left[ \fopa_\mu(x) +
  \frac{1}{g} (\partial_\mu \theta_a(x)) t_a\right] S^{-1}(x)\\
  D_\mu(x) &\rightarrow& S(x) D_\mu(x) S^{-1}(x)\\
  U(x,y) &\rightarrow& S(x) U(x,y) S^{-1}(y)\\
  \fop{\varrho}(x+y/2,x-y/2) &\rightarrow&
  S(x)\fop{\varrho}(x+y/2,x-y/2) S^{-1}(x)\label{eichtrafo5}\\
  \fopw(x,p) &\rightarrow& S(x) \fopw(x,p) S^{-1}(x)\\
  {^{[x]}\fopf_{\mu\nu}}(z(s)) &\rightarrow&
  S(x) {^{[x]}\fopf_{\mu\nu}}(z(s)) S^{-1}(x)\\{}
  \label{eichtrafo1x}
  [D_\mu (x) , \fopw(x,p)] &\rightarrow&
  S(x) [D_\mu (x) , \fopw(x,p)] S^{-1}(x)\\
  \Pi_\mu &\rightarrow& S(x) \Pi_\mu S^{-1}(x)\\
  \{ \Pi_\mu , \fopw(x,p) \} &\rightarrow&
  S(x) \{ \Pi_\mu , \fopw(x,p) \} S^{-1}(x)\\
  \Delta_\mu & \rightarrow& S(x) \Delta_\mu S^{-1}(x)\\{}
  [\Delta_\mu, \fopw(x,p)] &\rightarrow&
  S(x) [\Delta_\mu, \fopw(x,p)] S^{-1}(x).
 \end{eqnarray}

\section{Link operators}
\label{appc}

We recall the formula for the variation of link operators with respect
to their end points \cite{LINK}:
 \begin{eqnarray}
    \delta U(b,a) &=&  \delta b\, ig \fopa(b) U(b,a)
                         - ig U(b,a) \fopa(a) \, \delta a\\
  && - ig \int_0^1 ds\, U(b,z(s))\, F_{\mu\nu}(z(s)) \,U(z(s),a)
     \,(b-a)^\mu \,  (\delta a + s(\delta b - \delta a))^\nu,
 \nonumber\\
  && z(s) = a + (b-a)s
 \nonumber,
 \end{eqnarray}
This yields the following explicit expressions:
 \begin{eqnarray}
 \label{A_linkabl}
   \partial_{+\mu}U(x_+,x)
   \!\!&{=}&\!\! - \frac{1}{2}igU(x_+,x)\Bigl[\fopa_\mu(x) -
   2U(x,x_+)\fopa_\mu(x_+)U(x_+,x)
 \nonumber\\
   &&  \qquad\quad\quad + \int_0^{1/2}ds\,(1+2s)\,
   {^{[x]}\fopf_{\nu\mu}}(x+sy)y^\nu
   \Bigr]  ,
 \\
   \partial_{+\mu}U(x,x_-) \!\!&=&\!\! -\frac{1}{2}ig\Bigl[-\fopa_\mu(x) +
   \int_{-1/2}^{0} \!\!\!\!ds\,(1+2s)\,
   {^{[x]}\fopf_{\nu\mu}}(x+sy)y^\nu
   \Bigr]U(x,x_-)  ,
 \\
   \partial_{-\mu}U(x_+,x) \!\!&=&\!\!
   -\frac{1}{2}igU(x_+,x)\Bigl[\fopa_\mu(x) +
   \int_{0}^{1/2}ds\,(1-2s)
   {^{[x]}\fopf_{\nu\mu}}(x+sy)y^\nu
   \Bigr] ,
 \\
   \partial_{-\mu}U(x,x_-) \!\!&=&\!\! -\frac{1}{2}ig\Bigl[-\fopa_\mu(x) +
   2U(x,x_-)\fopa_\mu(x_-)U(x_-,x)
 \nonumber\\
   && \quad \qquad +
   \int_{-1/2}^{0}ds\,(1-2s)
   {^{[x]}\fopf_{\nu\mu}}(x+sy)y^\nu
  \Bigr]U(x,x_-).
\end{eqnarray}
For the derivative of the Schwinger string ${^{[x]}\fopf}_{\nu\mu}(x +
sy)$ (see Eq.~(\ref{schwingerstringdef})) with respect to $y$ one finds
 \begin{eqnarray}
 \label{schwingabl}
   \partial_{y\alpha} {^{[x]}F_{\mu\nu}}(x+y) &=&
   U(x,x+y)\left[
     \partial_{y\alpha} - ig A_\alpha(x+y) , F_{\mu\nu}(x+y)
   \right] U(x+y,x)
 \nonumber\\
   && - ig \int_0^1 ds \, s\, y^\beta
  \left[
    {^{[x]}F_{\alpha\beta}}(x+sy), {^{[x]}F_{\mu\nu}}(x+y)
  \right].
\end{eqnarray}
Its covariant derivative with respect to $x$ is
 \begin{eqnarray}\label{schwingerderivative}
  \tilde{\mathcal D}_\mu(x) {^{[x]}F_{\alpha\beta}}(x+y) &{\equiv}&
    [D_\mu(x), {^{[x]}F_{\alpha\beta}}(x+y) ]
 \nonumber\\
  &{=}& U(x,x+y)\left(
    \tilde{\mathcal D}_\mu(x+y) F_{\alpha\beta}(x+y)
  \right)U(x+y,x)
 \nonumber\\
  && - ig \int_0^1 ds\, y^\nu
  \left[
    {^{[x]}F_{\mu\nu}}(x+sy) , {^{[x]}F_{\alpha\beta}}(x+y)
  \right].
 \end{eqnarray}
Combining the last two equations one finds
 \begin{eqnarray}
   \tilde{\mathcal D}_\mu(x) {^{[x]}F_{\alpha\beta}}(x+y)
   &{=}&
 \nonumber\\
   \partial_{y\mu} {^{[x]}F_{\alpha\beta}}(x+y)
   &{-}& ig \int_0^1 ds\, (1-s)\, y^\nu
  \left[
    {^{[x]}F_{\mu\nu}}(x+sy) , {^{[x]}F_{\alpha\beta}}(x+y)
  \right].
\end{eqnarray}


\section{Spinor equations}
\label{appd}

In this Appendix we combine the spinor equations
(\ref{spinorgl1},\ref{spinorgl2}) into generalized mass shell and
transport equations for the spinor components. Multiplying the first
three of equations in (\ref{spinorgl1}) and (\ref{spinorgl2}) each by
$2m$ and inserting the last two equations
(\ref{spinorgl1}) yields
 \begin{eqnarray}
 \label{A_spinormass1}
  4m^2\fopfc &=& \{\Pi_\mu,\{\Pi^\mu,\fopfc\}\} +
  \{\Pi_\mu,[\Delta_\nu,\fopsc^{\mu\nu}]\} \\
  \label{A_spinormass2}
  4m^2\foppc &=& - [\Delta_\mu,[\Delta^\mu,\foppc]]
  +
  \frac{1}{2}\varepsilon^{\mu\nu\alpha\beta}
  [\Delta_\mu,\{\Pi_\nu,\fopsc_{\alpha\beta}\}] \\
  \label{A_spinormass3}
  4 m^2 \fopsc_{\mu\nu} &=& [\Delta_\mu,\{\Pi_\nu,\fopfc\}] -
  [\Delta_\nu,\{\Pi_\mu,\fopfc\}] +
  [\Delta_\mu,[\Delta^\alpha,\fopsc_{\nu\alpha}]] -
  [\Delta_\nu,[\Delta^\alpha,\fopsc_{\mu\alpha}]] -
  \nonumber\\
  &&
  \varepsilon_{\alpha\beta\mu\nu} \{\Pi^\alpha,[\Delta^\beta,\foppc]\} +
  \frac{1}{2}\varepsilon_{\alpha\beta\mu\nu}
  \varepsilon^{\beta\gamma\delta\kappa}
  \{\Pi^\alpha,\{\Pi_\gamma,\fopsc_{\delta\kappa}\}\}\
 \end{eqnarray}
as well as
 \begin{eqnarray}
  \label{A_spinortrans1}
  0 &=& [\Delta_\mu,\{\Pi^\mu,\fopfc\}]
  + [\Delta_\mu,[\Delta_\nu,\fopsc^{\mu\nu}]]\\
  \label{A_spinortrans2}
  0 &=& \{\Pi_\mu,[\Delta^\mu,\foppc]\} - \frac{1}{2}
  \varepsilon^{\mu\nu\alpha\beta}
  \{\Pi_\mu,\{\Pi_\nu,\fopsc_{\alpha\beta}\}\}\\
  \label{A_spinortrans3}
  0 &=& \{\Pi_\mu,\{\Pi_\nu,\fopfc\}\}
  - \{\Pi_\nu,\{\Pi_\mu,\fopfc\}\} +
  \{\Pi_\mu,[\Delta^\alpha,\fopsc_{\nu\alpha}]\} -
  \{\Pi_\nu,[\Delta^\alpha,\fopsc_{\mu\alpha}]\} +
  \nonumber\\
  &&
  \varepsilon_{\alpha\beta\mu\nu}
  [\Delta^\alpha,[\Delta^\beta,\foppc]] - \frac{1}{2}
  \varepsilon_{\alpha\beta\mu\nu}
  \varepsilon^{\beta\gamma\delta\kappa}
  [\Delta^\alpha,\{\Pi_\gamma,\fopsc_{\delta\kappa}\}].
 \end{eqnarray}
With some further manipulations the first three equations will give
rise to generalized mass shell constraints, the last three to
generalized transport equations for $\fopfc$, $\foppc$, and
$\fopsc_{\mu\nu}$. In terms of the solutions to these equations
the remaining spinor components $\fopac_\mu$ and $\fopvc_\mu$
are given by the last two equations (\ref{spinorgl1}).

For the further manipulations the following identities will be useful:
 \begin{mathletters}
 \label{kommrelextfeld}
 \begin{eqnarray}
  \label{kommrel1}
    [\Delta_{\mu},[\Delta_{\nu},\hat X]]
    - [\Delta_{\nu},[\Delta_{\mu},\hat X]]
    &=& [[ \Delta_\mu,  \Delta_\nu], \hat X]
  \\
  \label{kommrel2}
    \{\Pi_{\mu},[\Delta_{\nu},\hat X]\}
    - [\Delta_{\nu},\{\Pi_{\mu},\hat X\}]
    &=& \{ [p_\mu+\Pi_\mu , \Delta_\nu], \hat X\}
  \\
  \label{kommrel3}
    \{\Pi_{\mu},\{\Pi_{\nu},\hat X\}\}
    - \{\Pi_{\nu},\{\Pi_{\mu},\hat X\}\}
    &=& [[p_\mu+\Pi_\mu, p_\mu + \Pi_\nu], \hat X].
  \end{eqnarray}
 \end{mathletters}%
The additional factors $\sim p_\mu$ on the r.h.s. arise from the fact
that the commutators in these equations refer to the color and Fock
space structure of the operators only and do not consistently include
also the momentum derivative operators which, according to
(\ref{padj}), were defined to act only to the left when standing
to the right of a Wigner function. Splitting the generalized momentum
operator (\ref{allgop}) according to $\Pi_\mu = p_\mu + \pi_\mu$ where
the second term $\pi_\mu$ (like $\Delta_\mu$ in (\ref{allgop2}))
contains only momentum {\em derivatives}, but no momentum
{\em factors}, Eq.~(\ref{kommrel2}), for example, therefore has to be
evaluated as follows:
 \begin{eqnarray}
 \label{explain}
   &&\{p_\mu+\pi_\mu,[\Delta_\nu,\hat X]\}
   - [\Delta_\nu,\{p_\mu+\pi_\mu,\hat X\}]
 \nonumber\\
   &{=}&  + p_\mu (\Delta_\nu \hat X) - p_\mu (\hat X \Delta_\nu)
        + (\Delta_\nu \hat X) p_\mu  - (\hat X \Delta_\nu) p_\mu
 \nonumber\\
   && - \Delta_\nu (p_\mu \hat X) - \Delta_\nu (\hat X p_\mu)
      + (p_\mu  \hat X) \Delta_\nu + (\hat X p_\mu) \Delta_\nu
 \nonumber\\
   && + \{\pi_\mu,[\Delta_\nu, \hat X]\}
      - [\Delta_\nu, \{\pi_\mu, \hat X\}]
 \nonumber\\
   &{=}& 2 [p_\mu, \Delta_\nu] \hat X - 2 \hat X [\Delta_\nu, p_\mu]
         + [\pi_\mu, \Delta_\nu] \hat X - \hat X [\Delta_\nu, \pi_\mu]
 \nonumber\\
   &{=}& [p_\mu + \Pi_\mu, \Delta_\nu] \hat X
       - \hat X [\Delta_\nu, p_\mu + \Pi_\mu]
\nonumber\\
    &{=}& [[p_\mu + \Pi_\mu, \Delta_\nu], \hat X],
 \end{eqnarray}
and similarly for Eq.~(\ref{kommrel3}). In the intermediate
expressions the round brackets indicate, where ambiguous, how the
momentum derivatives act via (\ref{padj}); since the momentum
derivatives in $\pi_\mu$ and $\Delta_\nu$ commute the remaining terms
are unambiguous.

We will now use the generalized quadratic mass shell operator
 \begin{equation}
 \label{A_massop}
  M^2 \fopx = 4 m^2 \fopx - \{\Pi_\mu,\{\Pi^\mu,\fopx\}\} +
  [\Delta_\mu,[\Delta^\mu,\fopx]]
\end{equation}
to combine equal spinor components on the left hand sides of the
equations.

\subsection{Mass shell equations}
\label{appd1}

We now derive the mass shell constraints (\ref{mass}). To get the mass
shell condition for $\fopfc$ with the operator $M^2$ from above we
bring the term $\{\Pi_\mu,\{\Pi^\mu,\fopfc\}\}$ in
Eq.~(\ref{A_massop}) to the left and add on both sides
$[\Delta_\mu,[\Delta^\mu,\fopfc]]$. Furthermore we employ the
constraint Eq.~(\ref{spinorgl2}iv) together with the anti-symmetry of
$\fopsc_{\mu\nu}$ in the form
\begin{equation}
  [\Delta_\mu,[\Delta^\mu,\fopfc]] + \{\Pi_\mu,[\Delta_\nu,\fopsc^{\mu\nu}]\} =
  [\Delta_\mu,\{\Pi_\nu,\fopsc^{\mu\nu}\}]
  - \{\Pi_\nu,[\Delta_\mu,\fopsc^{\mu\nu}]\}\, ,
\end{equation}
and then use (\ref{kommrel2}).

The mass shell condition for $\foppc$ we obtain by subtracting
$\{\Pi_\mu,\{\Pi^\mu,P\}\}$ from Eq.~(\ref{A_spinormass2}) and using
\begin{equation}
  [\Delta_\mu, \{\Pi_\nu , \foplc^{\mu\nu}\}]
  - \{\Pi_\mu,\{\Pi^\mu,P\}\}
  = [\Delta_\mu,\{\Pi_\nu,\foplc^{\mu\nu}\}]
  - \{\Pi_\nu,[\Delta_\mu,\foplc^{\mu\nu}]\}\, ,
\end{equation}
which follows from Eq.~(\ref{spinorgl2}v). We then again use
(\ref{kommrel2}).

For the mass shell condition for $\fopsc_{\mu\nu}$, finally, we rewrite
$\varepsilon_{\alpha\beta\mu\nu}\{\Pi^\alpha,[\Delta^\beta,\foppc]\}$ as
\begin{equation}
  \varepsilon_{\alpha\beta\mu\nu}\{\Pi^\alpha,[\Delta^\beta,\foppc]\}
  = \varepsilon_{\alpha\beta\mu\nu}([\Delta^\beta,\{\Pi^\alpha,\foppc \}]
  + \{\Pi^\alpha,[\Delta^\beta,\foppc]\} -
  [\Delta^\beta,\{\Pi^\alpha,\foppc\}]),
\end{equation}
and use the constraint (\ref{spinorgl2}v):
 \begin{eqnarray}
 \label{A_sneb1}
  \varepsilon_{\alpha\beta\mu\nu}\{\Pi^\alpha,[\Delta^\beta,\foppc]\}
  \!&\!=\!&\! \varepsilon_{\alpha\beta\mu\nu}
  \left(
    \{\Pi^\alpha,[\Delta^\beta,\foppc ]\}
    - [\Delta^\beta,\{\Pi^\alpha,\foppc\}]
  \right)
  \nonumber\\ &&\!
  - [\Delta^\beta,[\Delta_\nu,\fopsc_{\mu\beta}]]
  - [\Delta^\beta,[\Delta_\mu,\fopsc_{\beta\nu}]]
  + [\Delta^\beta,[\Delta_\beta,\fopsc_{\mu\nu}]].\qquad
\end{eqnarray}
One further employs
\begin{eqnarray}
  \label{A_sneb2}
  [\Delta_\mu,\{\Pi_\nu,\fopfc\}]
  - [\Delta_\nu,\{\Pi_\mu,\fopfc\}]
  &=&
  [\Delta_\mu,\{\Pi_\nu,\fopfc\}] - \{\Pi_\nu,[\Delta_\mu,\fopfc]\}
  \nonumber\\ &&
  - \left(
    [\Delta_\nu,\{\Pi_\mu,\fopfc\}] - \{\Pi_\mu,[\Delta_\nu,\fopfc]\}
  \right)
  \nonumber\\ &&
  + \{\Pi_\nu,\{\Pi^\alpha,\fopsc_{\mu\alpha}\}\}
  + \{\Pi_\mu,\{\Pi^\alpha,\fopsc_{\alpha\nu}\}\}\quad
\end{eqnarray}
where the constraint (\ref{spinorgl2}iv) was used in the last line.
Inserting now Eqs.~(\ref{A_sneb1},\ref{A_sneb2}) into
(\ref{A_spinormass3}) and using Eqs.~(\ref{kommrelextfeld}) leads to
the mass shell condition for $\fopsc_{\mu\nu}$.

\subsection{Transport equations}
\label{appd2}

In the following we want to express the transport equations with the
generalization
\begin{equation}
  \label{A_transab}
  [\Delta_\mu,\{\Pi^\mu,\fopx\}]
\end{equation}
of the Lorentz covariant (proper time) derivative $m \frac{d}{d\tau}
\fopx = p_\mu \partial^\mu \fopx$ ($\tau$: proper time, $x_0$:
global time). For $\fopfc$ Eq.~(\ref{A_spinortrans1}) is already in
the desired form. The transport equation for $\foppc$ follows by
acting with $\Delta^\mu$ on Eq.~(\ref{spinorgl1}v). For
$\fopsc_{\mu\nu}$ we use the following identity
\begin{eqnarray}
  \label{A_sneb5}
  \{\Pi_\mu,[\Delta^\alpha,\fopsc_{\alpha\nu}]\}
  + \{\Pi_\nu,[\Delta^\alpha,\fopsc_{\mu\alpha}]\} &=&
  \{\Pi_\mu,[\Delta^\alpha,\fopsc_{\alpha\nu}]\} -
  [\Delta^\alpha,\{\Pi_\mu,\fopsc_{\alpha\nu}]\}
  \nonumber\\ &&
  \{\Pi_\nu,[\Delta^\alpha,\fopsc_{\mu\alpha}]\} -
  [\Delta^\alpha,\{\Pi_\nu,\fopsc_{\mu\alpha}]\}.\quad
\end{eqnarray}

\subsection{Equations for the vector and axial vector densities}
\label{appd3}

Alternatively one can derive transport equations, mass shell
conditions, and constraints for the eight vector and axial vector
components $\fopvc_\mu$ as independent functions and obtain the eight
other components $\fopfc, \foppc$ and $\fopsc_{\mu\nu}$ as dependent
functions. The results are:

\noindent (a) transport equations:
 \begin{eqnarray}
 \label{A_transalt1}
   [\Delta^\nu,\{\Pi_\nu,\fopac_\mu\}] &=&
   \{[\Delta_\nu,p_\mu+\Pi_\mu],\fopac^\nu\}
   - \varepsilon_{\alpha\beta\mu\nu}
     [\Delta^\alpha \Delta^\beta,\fopvc^\nu]
 \\
 \label{A_transalt2}
   [\Delta^\nu,\{\Pi_\nu,\fopvc_\mu\}] &=&
   \{[\Delta_\nu,p_\mu+\Pi_\mu],\fopvc^\nu\}
   - \varepsilon_{\alpha\beta\mu\nu}
     [\Delta^\alpha \Delta^\beta,\fopac^\nu]
 \end{eqnarray}
\noindent (b) mass shell equations:
 \begin{eqnarray}
 \label{A_massalt1}
   M^2\fopac_\mu &=& [[p_\mu+\Pi_\mu,p_\nu+\Pi_\nu],\fopac^\nu]
             - [[\Delta_\mu,\Delta_\nu],\fopac^\nu]
 \nonumber\\
   &&+ \varepsilon_{\alpha\beta\mu\nu}
     \{[p^\alpha+\Pi^\alpha,\Delta^\beta],\fopvc^\nu\}
 \\
 \label{A_massalt2}
   M^2\fopvc_\mu &=& [[p_\mu+\Pi_\mu,p_\nu+\Pi_\nu],\fopvc^\nu]
             - [[\Delta_\mu,\Delta_\nu],\fopvc^\nu]
 \nonumber\\
   &&+ \varepsilon_{\alpha\beta\mu\nu}
     \{[p^\alpha+\Pi^\alpha,\Delta^\beta],\fopac^\nu\}
 \end{eqnarray}
\noindent (c) constraints:
 \begin{eqnarray}
  \label{A_constraintalt1}
  0 &=& [\Delta_\mu,\fopvc^\mu]\\
  \label{A_constraintalt2}
  0 &=& \{\Pi_\mu,\fopac^\mu\}\\
  \label{A_constraintalt3}
  0 &=& \{\Pi_\mu,\fopvc_\nu\} - \{\Pi_\nu,\fopvc_\mu\} -
  \varepsilon_{\alpha\beta\mu\nu} [\Delta^\alpha,\fopac^\beta]
 \end{eqnarray}
\noindent (d) equations defining the dependent functions:
 \begin{eqnarray}
 \label{A_connectionalt1}
  2m\fopfc &=& \{\Pi_\mu,\fopvc^\mu\}\\
 \label{A_connectionalt2}
  2m\foppc &=& [\Delta_\mu,\fopac^\mu]\\
 \label{A_connectionalt3}
  2m\fopsc_{\mu\nu} &=& [\Delta_\mu,\fopvc_\nu] -
  [\Delta_\nu,\fopvc_\mu] + \varepsilon_{\alpha\beta\mu\nu}
  \{\Pi^\alpha,\fopac^\beta\}.
\end{eqnarray}

\section{Vacuum solutions}
\label{appe}
\subsection{Massive quarks}
\label{appe1}

In the field-free limit the spinor equations can be solved
exactly. Setting $\fopf_{\mu\nu}(x) \equiv 0$ and choosing a gauge
in which then also $\fopa_\mu(x) \equiv 0$, the spinor equations
reduce to
\begin{mathletters}    \label{spvacuum}
  \begin{eqnarray}
    0 &=&  p_\mu \partial^\mu \fopfc
    \\
    0 &=&  p_\mu \partial^\mu \foppc
    \\
    0 &=&  p_\mu \partial^\mu \fopsc_{\alpha\beta}
    \\
    0 &=&  \left[4(E^2 - p_0^2) + \Box \right]\fopfc
    \\
    0 &=&  \left[4(E^2 - p_0^2) + \Box \right]\foppc
    \\
    0 &=&  \left[4(E^2 - p_0^2) + \Box \right]\fopsc_{\mu\nu}
    \\
    0 &=&  \frac{1}{2}\partial_\mu \fopfc - p^\nu \fopsc_{\mu\nu}
    \\
     0 &=&  p_\mu \foppc +
     \frac{1}{4}\varepsilon_{\alpha\beta\mu\nu}\partial^\nu
    \fopsc^{\alpha\beta}
    \\
    2m \fopvc_\mu &=& 2 p_\mu \fopfc + \partial^\nu \fopsc_{\mu\nu}\\
    2 m\fopac_\mu &=& -\partial_\mu \foppc + \varepsilon_{\alpha\beta\mu\nu}
    p^\nu  \fopsc^{\alpha\beta}.
  \end{eqnarray}
\end{mathletters}%
Here $E^2 = {\vec p}^{\, 2} + m^2$, and the operator $\Box$ is defined
through $\Box = \partial_\mu \partial^\mu$.

One can see easily that the transport equations
(\ref{spvacuum}i,ii,iii) are already contained in the constraints
(\ref{spvacuum}vii,viii): For $\fopfc$ and $\foppc$ this follows
simply by contraction with $p_\mu$ and $\partial_\mu$ respectively and
exploiting the anti-symmetry of  $\fopsc_{\mu\nu}= -
\fopsc_{\nu\mu}$. Eq.~(\ref{spvacuum}iii) follows from
Eq.~(\ref{spvacuum}viii) by contraction with
$\varepsilon^{\alpha'\beta'\mu\nu'}p_{\nu'}$.

Spinor components which vanish at $t=-\infty$ remain zero for all
times: Formal integration of the transport equation $p_\mu
\partial^\mu \fopx(x,p) = 0$ for a spinor component $\fopx$ leads to
\begin{equation}
  p_0 \fopx (t + \Delta t) = p_0 \fopx (t) - p_i \partial^i \fopx (t) \Delta t.
\end{equation}
With the initial condition $\fopx(t \to -\infty) = 0$ we find
$\fopx(t) \equiv 0$ unless $p_0=0$. But $p_0 = 0$ inserted into
$[4(E^2 - p_0^2) + \Box ] \fopx = 0$ gives after formal integration
\begin{equation}
  \partial_0 \fopx(t+\Delta t) =
  \partial_0 \fopx(t) + (\nabla^2 - 4 E^2)\Delta t \fopx(t),
\end{equation}
so that with the initial condition $\fopx(t \to -\infty) = 0$ we have
\begin{equation}
    \partial_0 \fopx(t+\Delta t) =
  \partial_0 \fopx(t).
\end{equation}
Another formal integration then again yields $\fopx (t) \equiv 0$.
Imposing for the spin and pseudoscalar densities vacuum initial
conditions at $t=-\infty$, $\fopsc_{\mu\nu}(t\to -\infty) = \foppc (t
\to -\infty) = 0$, we see that they remain zero for all times, and
with them the axial vector density $\fopac_\mu$ (see
Eq.~(\ref{spvacuum}x)).

The scalar density $\fopfc$ evolves according to
\begin{eqnarray}
  \partial_\mu \fopfc (x,p) &=& 0\\{}
  [\Box + 4(E^2 - p_0^2)] \fopfc(x,p) &=& 0
\end{eqnarray}
>From the first equation we get $\fopfc (x,p) = \fopfc(p)$ while the second
gives $(E^2 - p_0^2)\fopfc(p) = 0$. The solution is stationary and
homogeneous and can be written as a sum of particle and antiparticle
contributions,
\begin{equation}
  \fopfc(x,p) = \frac{1}{2E} \left[ \delta(p_0 - E) + \delta(p_0 + E)\right]
  \fopfc(p) ,
\end{equation}
with an arbitrary momentum spectrum $\fopfc(p)$. The vector density is
obtained through Eq.~(\ref{spvacuum}ix) as
\begin{equation}
  \fopvc_\mu(x,p) = {p_\mu\over m} \fopfc(x,p)
  = \frac{p_\mu}{2mE}\left[\delta(E-p_0) +
    \delta(E+p_0)\right] \fopfc(p).
\end{equation}

\subsection{Massless quarks}
\label{appe2}

In the chiral limit the equations now read
\begin{mathletters}
  \begin{eqnarray}
    \label{vacchirAV}
    0 &=& p_\mu \fop{a}^{\mu(\pm)} \\
    0 &=& \partial_\mu \fop{a}^{\mu(\pm)} \\
    0 &=& \partial_\mu \fop{a}_\nu^{(\pm)} - \partial_\nu \fop{a}_\mu^{(\pm)}
    \pm 2 \varepsilon_{\alpha\beta\mu\nu} p^\alpha \fop{a}^{\beta(\pm)}
  \end{eqnarray}
\end{mathletters}%
for $\fop{a}_\mu^{(\pm)}$ and
\begin{mathletters}
  \begin{eqnarray}
    \label{vacchirFPS}
    0 &=& 2 p_\mu \fop{f}^{(\pm)} - \partial^\nu \fop{s}_{\nu\mu}^{(\pm)} \\
    0 &=& \partial_\mu \fop{f}^{(\pm)} + 2 p^\nu \fop{s}_{\nu\mu}^{(\pm)}
  \end{eqnarray}
\end{mathletters}%
for $\fop{f}^{(\pm)}, \fop{s}_{\mu\nu}^{(\pm)}$.

With the help of the -- linearly dependent -- ``mass shell'' equations
we know that the solutions have to fulfill
\begin{eqnarray}
  (4 p^2 - \Box) \fop{a}_\mu^{(\pm)} &=& 0 \\
  (4 p^2 - \Box) \fop{f}^{(\pm)} &=& 0 \\
  (4 p^2 - \Box) \fop{s}_{\mu\nu}^{(\pm)} &=& 0.
\end{eqnarray}
The solutions for the $x$ dependence would therefore have the form
$\exp(\pm p_\mu x^\mu)$, which does not allow for reasonable boundary
conditions at $x_\mu \to \pm \infty$. Therefore we use the following Ansatz
\begin{eqnarray}
  \fop{f}^{(\pm)}(x,p) &=& \fop{f}^{(\pm)}(\vec p) \delta(p^2)\\
  \fop{a}_\mu^{(\pm)}(x,p) &=& \fop{a}_\mu^{(\pm)}(\vec p) \delta(p^2)\\
  \fop{s}_{\mu\nu}^{(\pm)}(x,p) &=&
  \fop{s}_{\mu\nu}^{(\pm)}(\vec p) \delta(p^2).
\end{eqnarray}
Using Eqs.~(\ref{vacchirAV},\ref{vacchirFPS}) we find the additional
restriction
\begin{eqnarray}
  \fop{f}^{(\pm)}(x,p) &=& \fop{f}^{(\pm)}_0 \delta(p_\mu), \quad
  \fop{f}^{(\pm)}_0 = \mbox{const.}\\
  \fop{a}_\mu^{(\pm)}(x,p) &=& p_\mu \fop{a}_0^{(\pm)}(\vec p) \delta(p^2)
  \nonumber\\
  &=& \frac{1}{2}\fop{a}_0^{(\pm)}(\vec p)
  { \delta(p_0 - \vert \vec p \vert) -
    \delta(p_0 + \vert \vec p \vert) \choose - {\hat e}_p
    \delta(p_0 - \vert \vec p \vert) - {\hat e}_p
    \delta(p_0 + \vert \vec p \vert)
    },
  \qquad {\hat e}_p  =
  \frac{\vec p}{\vert \vec p \vert}.
\end{eqnarray}
Assuming now that the spin density vanishes for $x_0 \to -\infty$ we
find that $\fop{s}^{(\pm)}_{\mu\nu}$ has to vanish identically since
$\fop{s}^{(\pm)}_{\mu\nu}$ is independent of $x_\mu$.
The contribution from $\fop{f}^{(\pm)}$ is restricted to $p_\mu
= 0$ and constant in space and therefore irrelevant. Thus we are left with
$\fop{a}_\mu^{(\pm)}$, giving contributions to vector and axial vector
components as follows
\begin{eqnarray}
  \fopvc_\mu(x,p) &=& \frac{1}{4}
  (\fop{a}_0^{(+)}(\vec p) + \fop{a}_0^{(-)}(\vec p))
  { \delta(p_0 - \vert \vec p \vert) -
    \delta(p_0 + \vert \vec p \vert) \choose - {\hat e}_p
    \delta(p_0 - \vert \vec p \vert) - {\hat e}_p
    \delta(p_0 + \vert \vec p \vert)
    }\\
  \fopac_\mu(x,p) &=& \frac{1}{4}
  (\fop{a}_0^{(+)}(\vec p) - \fop{a}_0^{(-)}(\vec p))
  { \delta(p_0 - \vert \vec p \vert) -
    \delta(p_0 + \vert \vec p \vert) \choose - {\hat e}_p
    \delta(p_0 - \vert \vec p \vert) - {\hat e}_p
    \delta(p_0 + \vert \vec p \vert)
    }
\end{eqnarray}
which have no longer a space and time dependence.

\section{Moment equations in spinor decomposition}
\label{appf}

The corresponding moment equations to Eqs.~(\ref{with-p0},\ref{no-p0}) in
spinor decomposition are for the constraints (Eq.~(\ref{with-p0})), which
couple the $n+1^{\rm st}$ to all the lower moments,
\begin{eqnarray}
  \label{engywith-p01}
  2\ {^{[n+1]}\fopvc^0} \!&\!=\!&\! 2 m\ {^{[n]}\fopfc}
  - \{ N_{(0)\ 0},{^{[n]}\fopvc^0} \} -
  \{\Pi_i,{^{[n]}\fopvc^i} \} -
  \nonumber\\ &&
  \sum_{k=1}^n
  {n \choose k} [ N_{(k) \mu} ,{^{[n-k]}\fopvc^\mu}]_{k+1} \\
  \label{engywith-p02}
  2\ {^{[n+1]}\fopac^k} \!&\!=\!&\!
  - 2 m \ {^{[n]}\foplc^{0k}}
  +  \varepsilon^{0ijk} [\Delta_i,{^{[n]}\fopvc_j}]
  -  \{ N_{(0)}^0, {^{[n]}\fopac^k}\} + \{\Pi^k,{^{[n]}\fopac^0}\}
  + \\
  &&  \sum_{k=1}^n {n \choose k}
  \left[\varepsilon^{0ijk}
    [ M_{(k) i},{^{[n-k]}\fopvc_j}]_k
    - [N_{(k)}^0,{^{[n-k]}\fopac^k}]_{k+1}
    + [N_{(k)}^k,{^{[n-k]}\fopac^0}]_{k+1}
  \right]\nonumber\\
  \label{engywith-p03}
  2 \ {^{[n+1]}\fopfc}  \!&\!=\!&\! 2 m \ {^{[n]}\fopvc_0} - \{ N_{(0)0}\
  ,{^{[n]}\fopfc} \} - [\Delta^i,
  {^{[n]}\fopsc_{0i}}] - \nonumber\\
  && \sum_{k=1}^n {n \choose k}
  \left(
    [N_{(k) 0},{^{[n-k]}\fopfc}]_{k+1}
    + [M_{(k)}^i,{^{[n-k]}\fopsc_{0i}}]_k
  \right)\\
  \label{engywith-p04}
  2{^{[n+1]}\fopsc^{jk}} \!&\!=\!&\! \varepsilon^{0ijk}
  \left(
    2 m \  {^{[n]}\fopac_i} + [\Delta_i,{^{[n]}\foppc}]
  \right)
  - \{N_{(0) 0},{^{[n]}\fopsc^{jk}}\}
  \nonumber\\ &&
  - \{\Pi^k, {^{[n]}\fopsc^{0j}} \}
  +  \{\Pi^j,{^{[n]}\fopsc^{0k}} \} +
  \nonumber\\
  &&  \sum_{k=1}^n {n \choose k}
  \left[\varepsilon^{0ijk}
    [M_{(k) i},{^{[n-k]}\foppc}]_k
    - [N_{(k)}^0,{^{[n-k]}\fopsc^{jk}}]_{k+1}
  \right.
  \nonumber\\&&    \left.
    \qquad  - [N_{(k)}^k,{^{[n-k]}\fopsc^{0j}}]_{k+1}
    + [N_{(k)}^j,{^{[n-k]}\fopsc^{0k}}]_{k+1}
  \right]
  \\
  \label{engywith-p05}
  2\ {^{[n+1]}\fopac^0} \!&\!=\!&\!
  -\{N_{(0) 0},{^{[n]}\fopac^0}\} -
  \{ \Pi_i,{^{[n]}\fopac^i}\} -
  \nonumber\\ &&
  \sum_{k=1}^n {n \choose k}
  [N_{(k) \mu},{^{[n-k]}\fopac^\mu}]_{k+1}
  \\
  \label{engywith-p06}
  2\ {^{[n+1]}\fopvc_i} \!&\!=\!&\! -
  \{N_{(0) 0},{^{[n]}\fopvc_i}\} +
  \{ \Pi_i,{^{[n]}\fopvc_0}\} +
  \varepsilon_{0ijk}[\Delta^j,{^{[n]}\fopac^k}] -
  \\ &&
  \sum_{k=1}^n {n \choose k}
  \left[
    [N_{(k) 0},{^{[n-k]}\fopvc_i}]_{k+1}
    - [N_{(k) i},{^{[n-k]}\fopvc_0}]_{k+1}
    - \varepsilon_{0ijk}
    [M_{(k)}^j,{^{[n-k]}\fopac^k}]_k
  \right]
  \nonumber\\
  \label{engywith-p07}
  2\ {^{[n+1]}\fopsc_{i0}} \!&\!=\!&\!
  - \{N_{(0) 0},{^{[n]}\fopsc_{i0}}\} -
  \{\Pi^j,{^{[n]}\fopsc_{ij}}\} +
  [\Delta_i,{^{[n]}\fopfc}] + \nonumber\\
  && \sum_{k=1}^n {n \choose k}
  \left[
    [M_{(k) i},{^{[n-k]}\fopfc}]_k
    - [N_{(k)}^\nu,{^{[n-k]}\fopsc_{i\nu}}]_{k+1}
  \right]
  \\
  \label{engywith-p08}
  2\ {^{[n+1]}\foppc} \!&\!=\!&\!  -
  \{N_{(0) 0},{^{[n]}\foppc}\} -
  [\Delta^i,{^{[n]}\foplc_{0i}}] -
  \nonumber\\
  && \sum_{k=1}^n {n \choose k}
  \left[
    [N_{(k) 0},{^{[n-k]}\foppc}]_{k+1} +
    [M_{(k)}^i,{^{[n-k]}\foplc_{0i}}]_k
  \right]
\end{eqnarray}
and for the dynamical equations (Eq.~(\ref{no-p0})) containing explicit
$\partial_0$ derivatives
\begin{eqnarray}
  \label{engyno-p01}
  2 m {^{[n]}\foppc} \!&\!=\!&\! [\Delta_\mu,{^{[n]}\fopac^\mu}] +
  \nonumber\\
  && \sum_{k=1}^n {n \choose k}
  \left[ M_{(k)\mu},{^{[n-k]}\fopac^\mu}\right]_k
  \\
  \label{engyno-p02}
  2 m {^{[n]}\fopsc_{0i}} \!&\!=\!&\! [\Delta_0,{^{[n]}\fopvc_i}] -
  [\Delta_i,{^{[n]}\fopvc_0}] +
  \varepsilon_{0ijk} \{\Pi^j,{^{[n]}\fopac^k}\} +
  \nonumber\\
  && \sum_{k=1}^n {n \choose k}
  \left(
    \left[M_{(k)0},{^{[n-k]}\fopvc_i}\right]_k -
    \left[M_{(k)i},{^{[n-k]}\fopvc_0}\right]_k
  \right) \\
  \label{engyno-p03}
  2 m {^{[n]}\fopvc_i} \!&\!=\!&\! \{\Pi_i,{^{[n]}\fopfc}\} +
  [\Delta^\nu,{^{[n]}\fopsc_{i\nu}}] +
  \nonumber\\
  && \sum_{k=1}^n {n \choose k}
  \left(
    \left[N_{(k)i},{^{[n-k]}\fopfc}\right]_{k+1} +
    \left[M_{(k)}^\nu,{^{[n-k]}\fopsc_{i\nu}}\right]_k
  \right)\ \\
  \label{engyno-p04}
  2 m {^{[n]}\fopac_0} \!&\!=\!&\! - [\Delta_0,{^{[n]}\foppc}]
  + \{\Pi^i,{^{[n]}\foplc_{0i}}\} -
  \nonumber\\ && \sum_{k=1}^n {n \choose k}
  \left(
    \left[M_{(k)0},{^{[n-k]}\foppc}\right]_k -
    \left[ N_{(k)}^i,{^{[n-k]}\foplc_{0i}}\right]_{k+1}
  \right) \\
  \label{engyno-p05}
  0 \!&\!=\!&\! [\Delta_\mu,{^{[n]}\fopvc^\mu}] +
  \sum_{k=1}^n {n \choose k}
  \left[ M_{(k)\mu},{^{[n-k]}\fopvc^\mu}\right]_k    \\
  \label{engyno-p06}
  0 \!&\!=\!&\! \{\Pi_i,{^{[n]}\fopvc_j}\} - \{\Pi_j,{^{[n]}\fopvc_i}\} -
  \varepsilon_{0ijk} ([\Delta^0,{^{[n]}\fopac^k}] -
  [\Delta^k,{^{[n]}\fopac^0}]) +
  \nonumber\\
  && \sum_{k=1}^n {n \choose k}
  \left(
    \left[N_{(k)i},{^{[n-k]}\fopvc_j}\right]_{k+1} -
    \left[N_{(k)j},{^{[n-k]}\fopvc_i}\right]_{k+1}
  \right.
  \nonumber\\ &&
  \left.
    \qquad \qquad- \varepsilon_{0ijk}
    \left(
      \left[M_{(k)}^0,{^{[n-k]}\fopac^k}\right]_k -
      \left[M_{(k)}^k,{^{[n-k]}\fopac^0}\right]_k
    \right)
  \right)
  \\
  \label{engyno-p07}
  0 \!&\!=\!&\! [\Delta_0,{^{[n]}\fopfc}] -
  \{\Pi^i,{^{[n]}\fopsc_{0i}}\} +
  \nonumber\\
  && \sum_{k=1}^n {n \choose k}
  \left(
    \left[M_{(k)0},{^{[n-k]}\fopfc}\right]_k -
    \left[N_{(k)}^i,{^{[n-k]}\fopsc_{0i}}\right]_{k+1}
  \right)
  \\
  \label{engyno-p08}
  0 \!&\!=\!&\! \{\Pi_i,{^{[n]}\foppc}\} -
  [\Delta^0,{^{[n]}\foplc_{0i}}] +
   [\Delta^j,{^{[n]}\foplc_{ij}}] +
  \\
  && \sum_{k=1}^n {n \choose k}
  \left(
    \left[N_{(k)i},{^{[n-k]}\foppc}\right]_{k+1} -
    \left[M_{(k)}^0,{^{[n-k]}\foplc_{0i}}\right]_k +
    \left[ M_{(k)}^j,{^{[n-k]}\foplc_{ij}}\right]_k
  \right).\nonumber
\end{eqnarray}
We used again the notations
Eq.~(\ref{momentsop_qcd_1}) for the non-local operators $M_{k\mu}$ and
$N_{k\mu}$.

\section{The commutator Eq.~{\lowercase{(\ref{comvan})}}}
\label{appg}

We show that for classical external fields the commutator
Eq.~(\ref{comvan}) vanishes identically. Written out explicitly it
reads
 \begin{eqnarray}
 \label{bewegnh}
   0 &\stackrel{!}{=}& \Bigl[2 p_\mu +  i D_\mu (x)
   - ig \int_{-1/2}^0 ds(1-2s)\,
   {^{[x]}F_{\alpha\mu}(x + is  \partial_p)} \partial_p^\alpha  ,
 \nonumber\\
   && \ 2 p_\nu - i D_\nu (x) + ig \int_{-1/2}^0 ds(1+2s) \,
  {^{[x]}F_{\alpha\mu}(x + is  \partial_p)} \partial_p^\alpha \Bigr].
 \end{eqnarray}
To evaluate the commutator we first perform a Taylor expansion for the
Schwinger string
\begin{eqnarray}
  \label{sstrtaylor}
  {^{[x]}F_{\mu\nu}(x + is \partial_p)} &=&
  \sum_{n=0}^\infty \frac{(is)^n}{n!}
  \lbrack\!\lbrack
  (\partial_p^\alpha D_\alpha)^n,
  F_{\mu\nu}(x)
  \rbrack\!\rbrack
  \nonumber\\ &=&
  \sum_{n=0}^\infty \frac{(is)^n}{n!}
  (\partial_p^\alpha\tilde{{\mathcal D}}_\alpha)^n
  F_{\mu\nu}(x),
\end{eqnarray}
with the definitions Eq.~(\ref{sstrdef1},\ref{deriv}). We will then
need the two identities
 \begin{eqnarray}
 \label{indida1}
  \tilde{{\mathcal D}}_\mu (\partial_p \cdot  \tilde{{\mathcal D}})^n
  F_{\beta\nu} &=&  (\partial_p \cdot  \tilde{{\mathcal D}})^n
  \tilde{{\mathcal D}}_\mu F_{\beta\nu}
  \\ &&
  - ig \sum_{m=0}^{n-1} {n \choose m+1}
  [(\partial_p \cdot  \tilde{{\mathcal D}})^m F_{\mu\alpha},
  (\partial_p \cdot  \tilde{{\mathcal D}})^{n-m-1} F_{\beta\nu}]
  \partial_p^\alpha
  \nonumber\\ \label{indida2}
  {}[p_\mu, (\partial_p \cdot  \tilde{{\mathcal D}})^n F_{\beta\nu}] &=&
  -n (\partial_p \cdot  \tilde{{\mathcal D}})^{n-1}
  \tilde{{\mathcal D}}_\mu F_{\beta\nu}
  \\ &&
  + ig \sum_{m=0}^{n-1} m {n \choose m+1}
  [(\partial_p \cdot  \tilde{{\mathcal D}})^{m-1} F_{\mu\alpha},
  (\partial_p \cdot  \tilde{{\mathcal D}})^{n-m-1} F_{\beta\nu}]
  \partial_p^\alpha
  \nonumber.
 \end{eqnarray}
The proof goes by induction as follows:\\
The validity for $n=0$ is obvious. To make in Eq.~(\ref{indida1}) the
step from $n$ to $n+1$ we need the relations
  \begin{eqnarray}\label{idindh1}
    \tilde{{\mathcal D}}_\mu \tilde{{\mathcal D}}_\nu X
    - \tilde{{\mathcal D}}_\nu \tilde{{\mathcal D}}_\mu X
    &=&  [[ D_\mu,D_\nu], X]
    \ =\ -ig [F_{\mu\nu}, X]
    \\ \label{idindh2}
    \partial_p \cdot  \tilde{{\mathcal D}} [A(x),B(x)] &=&
    [ (\partial_p \cdot  \tilde{{\mathcal D}}) A(x),B(x)]
    + [ A(x), (\partial_p \cdot  \tilde{{\mathcal D}}) B(x)]
  \end{eqnarray}
Thus we get
\begin{eqnarray}
  \label{indid1}
  \lefteqn{ \tilde{{\mathcal D}}_\mu (\partial_p \cdot
    \tilde{{\mathcal D}})^{n+1} F_{\beta\nu} \ =
    }\nonumber\\
  &\!\!\!\stackrel{\mbox{\scriptsize Eq.~(\ref{idindh1})}}{=}\!\!\!&
  \!\!\!-ig [ F_{\mu\alpha}, (\partial_p \cdot
  \tilde{{\mathcal D}})^n F_{\beta\nu}]
  \partial_p^\alpha
  + (\partial_p \cdot  \tilde{{\mathcal D}})
  \tilde{{\mathcal D}}_\mu
  (\partial_p \cdot  \tilde{{\mathcal D}})^n F_{\beta\nu}
  \nonumber\\
  &\!\!\!\stackrel{\mbox{\scriptsize Eq.~(\ref{indida1})}}{=}\!\!\!&
  \!\!\!-ig [ F_{\mu\alpha}, (\partial_p \cdot
  \tilde{{\mathcal D}})^n F_{\beta\nu}]
  \partial_p^\alpha
  + (\partial_p \cdot  \tilde{{\mathcal D}})^{n+1}
  \tilde{{\mathcal D}}_\mu F_{\beta\nu}
  \nonumber\\ &&
  \!\!\!- ig (\partial_p \cdot  \tilde{{\mathcal D}})
  \sum_{m=0}^{n-1} {n \choose m+1}
  [(\partial_p \cdot  \tilde{{\mathcal D}})^m F_{\mu\alpha},
  (\partial_p \cdot  \tilde{{\mathcal D}})^{n-m-1} F_{\beta\nu}]
  \partial_p^\alpha
  \nonumber\\
  &\!\!\!\stackrel{ \mbox{\scriptsize Eq.~(\ref{idindh2})}}{=}\!\!\!&
  \!\!\!-ig [ F_{\mu\alpha},
  (\partial_p \cdot  \tilde{{\mathcal D}})^n F_{\beta\nu}]\partial_p^\alpha
  + (\partial_p \cdot  \tilde{{\mathcal D}})^{n+1}
  \tilde{{\mathcal D}}_\mu F_{\beta\nu}
  \nonumber\\&&
  \!\!\!- ig \sum_{m=0}^{n-1} {n \choose m+1}
  \biggl(
  [(\partial_p \cdot  \tilde{{\mathcal D}})^{m+1} F_{\mu\alpha},
  (\partial_p \cdot  \tilde{{\mathcal D}})^{n-m-1} F_{\beta\nu}]
  \nonumber\\ && \qquad \qquad \qquad
  + [(\partial_p \cdot  \tilde{{\mathcal D}})^m F_{\mu\alpha},
  (\partial_p \cdot  \tilde{{\mathcal D}})^{n-m} F_{\beta\nu}]
  \biggr)    \partial_p^\alpha
  \nonumber\\ &=&
  \!\!\!-ig [ F_{\mu\alpha},
  (\partial_p \cdot  \tilde{{\mathcal D}})^n F_{\beta\nu}]\partial_p^\alpha
  + (\partial_p \cdot  \tilde{{\mathcal D}})^{n+1}
  \tilde{{\mathcal D}}_\mu F_{\beta\nu}
  \nonumber\\ &&
  \!\!\!- ig \sum_{m=1}^{n} {n \choose m+1}
  [(\partial_p \cdot  \tilde{{\mathcal D}})^{m} F_{\mu\alpha},
  (\partial_p \cdot  \tilde{{\mathcal D}})^{n-m} F_{\beta\nu}]
  \partial_p^\alpha
  \nonumber\\ &&
  \!\!\!- ig \sum_{m=0}^{n-1} {n \choose m+1}
  [(\partial_p \cdot  \tilde{{\mathcal D}})^m F_{\mu\alpha},
  (\partial_p \cdot  \tilde{{\mathcal D}})^{n-m} F_{\beta\nu}]
  \partial_p^\alpha
  \nonumber\\ &=&
  \!\!\!-ig (n+1) [ F_{\mu\alpha},
  (\partial_p \cdot  \tilde{{\mathcal D}})^n F_{\beta\nu}]\partial_p^\alpha
  \nonumber\\ &&
  \!\!\!-ig  [ (\partial_p \cdot  \tilde{{\mathcal D}})^n F_{\mu\alpha},
  F_{\beta\nu}]\partial_p^\alpha
  + (\partial_p \cdot  \tilde{{\mathcal D}})^{n+1}
  \tilde{{\mathcal D}}_\mu F_{\beta\nu}
  \nonumber\\ &&
  \!\!\!- ig \sum_{m=1}^{n}
  \underbrace{\left[{n \choose m} + {n \choose m+1} \right]}_{\D
    {n+1 \choose  m+1}}
  [(\partial_p \cdot  \tilde{{\mathcal D}})^{m} F_{\mu\alpha},
  (\partial_p \cdot  \tilde{{\mathcal D}})^{n-m} F_{\beta\nu}]
  \nonumber\\ &=&
  (\partial_p \cdot  \tilde{{\mathcal D}})^{n+1}
  \tilde{{\mathcal D}}_\mu F_{\beta\nu}
  - ig \sum_{m=0}^{n} {n+1 \choose m+1}
  [(\partial_p \cdot  \tilde{{\mathcal D}})^m F_{\mu\alpha},
  (\partial_p \cdot  \tilde{{\mathcal D}})^{n-m} F_{\beta\nu}]
  \partial_p^\alpha .
\end{eqnarray}
This proves Eq.~(\ref{indida1}). For Eq.~(\ref{indida2}) we find for $n=1$
 \begin{equation}
  [p_\mu, \partial_p^\alpha \tilde{{\mathcal D}}_\alpha F_{\beta\nu}]
  = - g_\mu^\alpha \tilde{{\mathcal D}}_\alpha F_{\beta\nu}.
  = - \tilde{{\mathcal D}}_\mu F_{\beta\nu}\, ,
 \end{equation}
i.e. the equation is correct. The step $n \rightarrow n+1$ is given by
\begin{eqnarray}
  \label{indid3}
  \lefteqn{ [p_\mu, (\partial_p \cdot
    \tilde{{\mathcal D}})^{n+1} F_{\beta\nu}] \ =
    }\nonumber\\ &=&
  [p_\mu, (\partial_p \cdot  \tilde{{\mathcal D}})]
  (\partial_p \cdot  \tilde{{\mathcal D}})^n F_{\beta\nu}
  + (\partial_p \cdot  \tilde{{\mathcal D}})
  [p_\mu, (\partial_p \cdot  \tilde{{\mathcal D}})^n F_{\beta\nu}]
  \nonumber\\
  &\!\!\! \stackrel{\mbox{\scriptsize Eqs.~(\ref{indida2},\ref{idindh2})}}
  {=}\!\!\!&
  - \tilde{{\mathcal D}}_\mu
  (\partial_p \cdot  \tilde{{\mathcal D}})^{n} F_{\beta\nu}
  - n (\partial_p \cdot  \tilde{{\mathcal D}})^{n}
  \tilde{{\mathcal D}}_\mu F_{\beta\nu}
  \nonumber\\ &&
  + ig \sum_{m=0}^{n-1} m {n \choose m+1}
  [(\partial_p \cdot  \tilde{{\mathcal D}})^{m} F_{\mu\alpha},
  (\partial_p \cdot  \tilde{{\mathcal D}})^{n-m-1} F_{\beta\nu}]
  \partial_p^\alpha
  \nonumber\\ &&
  + ig \sum_{m=0}^{n-1} m {n \choose m+1}
  [(\partial_p \cdot  \tilde{{\mathcal D}})^{m-1} F_{\mu\alpha},
  (\partial_p \cdot  \tilde{{\mathcal D}})^{n-m} F_{\beta\nu}]
  \partial_p^\alpha
  \nonumber\\
  &\!\!\!\stackrel{\mbox{\scriptsize Eq.~(\ref{indida1})}}{=}\!\!\!&
  - (n+1) (\partial_p \cdot  \tilde{{\mathcal D}})^{n}
  \tilde{{\mathcal D}}_\mu F_{\beta\nu}
  \nonumber\\ &&
  + ig \sum_{m=0}^{n-1} {n \choose m+1}
  [(\partial_p \cdot  \tilde{{\mathcal D}})^{m} F_{\mu\alpha},
  (\partial_p \cdot  \tilde{{\mathcal D}})^{n-m-1} F_{\beta\nu}]
  \partial_p^\alpha
  \nonumber\\ &&
  + ig \sum_{m=0}^{n-1} m {n \choose m+1}
  [(\partial_p \cdot  \tilde{{\mathcal D}})^{m} F_{\mu\alpha},
  (\partial_p \cdot  \tilde{{\mathcal D}})^{n-m-1} F_{\beta\nu}]
  \partial_p^\alpha
  \nonumber\\ &&
  + ig \sum_{m=0}^{n-1} m {n \choose m+1}
  [(\partial_p \cdot  \tilde{{\mathcal D}})^{m-1} F_{\mu\alpha},
  (\partial_p \cdot  \tilde{{\mathcal D}})^{n-m} F_{\beta\nu}]
  \partial_p^\alpha
  \nonumber\\&=&
  - (n+1) (\partial_p \cdot  \tilde{{\mathcal D}})^{n}
  \tilde{{\mathcal D}}_\mu F_{\beta\nu}
  \nonumber\\ &&
  + ig \sum_{m=1}^{n} {n \choose m}
  [(\partial_p \cdot  \tilde{{\mathcal D}})^{m-1} F_{\mu\alpha},
  (\partial_p \cdot  \tilde{{\mathcal D}})^{n-m} F_{\beta\nu}]
  \partial_p^\alpha
  \nonumber\\ &&
  + ig \sum_{m=1}^{n} (m-1) {n \choose m}
  [(\partial_p \cdot  \tilde{{\mathcal D}})^{m-1} F_{\mu\alpha},
  (\partial_p \cdot  \tilde{{\mathcal D}})^{n-m} F_{\beta\nu}]
  \partial_p^\alpha
  \nonumber\\ &&
  + ig \sum_{m=0}^{n-1} m {n \choose m+1}
  [(\partial_p \cdot  \tilde{{\mathcal D}})^{m-1} F_{\mu\alpha},
  (\partial_p \cdot  \tilde{{\mathcal D}})^{n-m} F_{\beta\nu}]
  \partial_p^\alpha
  \nonumber\\ &=&
  - (n+1) (\partial_p \cdot  \tilde{{\mathcal D}})^{n}
  \tilde{{\mathcal D}}_\mu F_{\beta\nu}
  \nonumber\\ &&
  + ig n [(\partial_p \cdot  \tilde{{\mathcal D}})^{n-1} F_{\mu\alpha},
  F_{\beta\nu}] \partial_p^\alpha
  \nonumber\\ &&
  + ig \sum_{m=1}^{n-1}
  [(\partial_p \cdot  \tilde{{\mathcal D}})^{m-1} F_{\mu\alpha},
  (\partial_p \cdot  \tilde{{\mathcal D}})^{n-m} F_{\beta\nu}]
  \partial_p^\alpha
  \nonumber\\ && \qquad
  \underbrace{ \left[
    {n \choose m} + (m-1) {n \choose m} + m {n \choose m+1}
  \right]}_{ \D m {n+1 \choose m+1}}
  \nonumber\\ &=&
  - (n+1) (\partial_p \cdot  \tilde{{\mathcal D}})^{n}
  \tilde{{\mathcal D}}_\mu F_{\beta\nu}
  \nonumber\\ &&
  + ig \sum_{m=0}^{n}{n+1 \choose m+1}
  [(\partial_p \cdot  \tilde{{\mathcal D}})^{m-1} F_{\mu\alpha},
  (\partial_p \cdot  \tilde{{\mathcal D}})^{n-m} F_{\beta\nu}]
  \partial_p^\alpha\quad \Box.
\end{eqnarray}

For the Schwinger string we get with the definitions of
Eq.~(\ref{sstrtaylor}) the following commutator relations
\begin{eqnarray}
  \label{finalkom1}
  \lefteqn{\tilde{{\mathcal D}}_\mu
    {^{[x]}F_{\beta\nu}(x + is\partial_p)}\partial_p^\beta = }
  \nonumber\\
  && \sum_{n=0}^\infty \frac{(is)^n}{n!}
  (\partial_p \cdot  \tilde{{\mathcal D}})^{n}
  \tilde{{\mathcal D}}_\mu
  F_{\beta\nu}(x)\partial_p^\beta
  \nonumber\\ &&
  + g \sum_{n=0}^\infty \frac{i^n s^{n+1}}{n!}
  \sum_{m=0}^n \frac{1}{m+1}{n \choose m}
  [(\partial_p \cdot  \tilde{{\mathcal D}})^{m} F_{\mu\alpha},
  (\partial_p \cdot  \tilde{{\mathcal D}})^{n-m} F_{\beta\nu}]
  \partial_p^\alpha\partial_p^\beta
\end{eqnarray}
and
\begin{eqnarray}
  \label{finalkom2}
  \lefteqn{[p_\mu ,{^{[x]}F_{\beta\nu}(x + is\partial_p)}\partial_p^\beta ]
    \ =\ } \nonumber\\ &&
  - F_{\mu\nu}(x)
  - i \sum_{n=0}^\infty \frac{i^n s^{n+1}}{n!}
  (\partial_p \cdot  \tilde{{\mathcal D}})^{n}
  \left[
    \frac{1}{n+1} \tilde{{\mathcal D}}_\alpha F_{\mu\nu}(x)
    + \tilde{{\mathcal D}}_\mu
    F_{\alpha\nu}(x)
  \right]\partial_p^\alpha
  \nonumber\\ &&
  - ig \sum_{n=0}^\infty \frac{i^n s^{n+2}}{n!}
  \sum_{m=0}^n \frac{1}{m+2}{n \choose m}
  [(\partial_p \cdot  \tilde{{\mathcal D}})^{m} F_{\mu\alpha},
  (\partial_p \cdot  \tilde{{\mathcal D}})^{n-m} F_{\beta\nu}]
  \partial_p^\alpha\partial_p^\beta.\quad
\end{eqnarray}
In addition we find for the commutator of the Schwinger string
\begin{eqnarray}\label{finalkom3}
  \lefteqn{  [{^{[x]}F_{\alpha\mu}(x + is\partial_p)}\partial_p^\alpha ,
    {^{[x]}F_{\beta\nu}(x + it\partial_p)}\partial_p^\beta ] = }
  \\ &&
  - \sum_{n=0}^\infty \frac{i^n}{n!}
  \sum_{m=0}^n {n \choose m} s^m t^{n-m}
  [(\partial_p \cdot  \tilde{{\mathcal D}})^{m} F_{\mu\alpha},
  (\partial_p \cdot  \tilde{{\mathcal D}})^{n-m} F_{\beta\nu}]
  \partial_p^\alpha\partial_p^\beta.
  \nonumber
\end{eqnarray}

Plugging in Eqs.~(\ref{finalkom1},\ref{finalkom2},\ref{finalkom3})
into the commutator Eq.~(\ref{bewegnh}) and performing the $s$
integration we get the following equation:
\begin{eqnarray}
  \label{finaleq}
  \lefteqn{  [ p_\mu + \Pi_\mu + i \Delta_\mu,
               p_\nu + \Pi_\nu - i \Delta_\nu] =}
  \nonumber\\ &&
  - 2 ig (F_{\mu\nu}(x) + F_{\nu\mu}(x))
  \nonumber\\ &&
  + g \sum_{n = 0}^\infty \frac{(-i/2)^n}{n!}
  (\partial_p \cdot  \tilde{{\mathcal D}})^{n} \frac{1}{(n+1)(n+3)}
  \left[
     \tilde{{\mathcal D}}_\alpha F_{\mu\nu}(x)
     +  \tilde{{\mathcal D}}_\mu F_{\nu\alpha}(x)
     +  \tilde{{\mathcal D}}_\nu F_{\alpha\mu}(x)
   \right]\partial_p^\alpha
   \nonumber\\ &&
   + g^2 \sum_{n = 0}^\infty \frac{(-i/2)^n}{4n!}
   \sum_{m=0}^n {n \choose m} f(n,m)
    [(\partial_p \cdot  \tilde{{\mathcal D}})^{m} F_{\mu\alpha},
    (\partial_p \cdot  \tilde{{\mathcal D}})^{n-m} F_{\beta\nu}]
    \partial_p^\alpha\partial_p^\beta \, ,
\end{eqnarray}
with
\begin{eqnarray}
  \label{fnm}
  f(n,m) &=& \frac{1}{(n+2)(n+3)}
  \left(\frac{1}{m+1} + \frac{2n+5}{n-m+1}\right)
  \nonumber\\ &&
  + \frac{1}{(n+3)(n+4)}
  \left(\frac{1}{m+2} - \frac{2n+7}{n-m+2}\right)
  \nonumber\\ &&
  - \frac{2m+3}{(m+1)(m+2)(n-m+1)(n-m+2)}.
\end{eqnarray}
The first and second line on the r.h.s of Eq.~(\ref{finaleq}) vanish
due to the antisymmetry of the field strength tensor and the Jacobi
identity (\ref{jacobi1}). In the third line one shows easily
that for each combination of the indices $n\ge 0$, $0
\le m \le n$ the function $f(n,m)$ vanishes identically. Thus the
commutator (\ref{comvan}) vanishes.

\section{QED moment equations in the classical limit}
\label{apph}

To derive the transport equations for the moments in leading order in the
gradient expansion we need the moment equations to order  $\hbar$.
These are with coupling constant $g=-e$ inserted:
\begin{eqnarray}
  \label{momqedapp1i}
  2 {^{[n+1]}\Vc^{0(1)}} \!&\!=\!&\!
  2 m {^{[n]}\Fc^{(1)}} - 2 p_i {^{[n]}\Vc^{i(1)}}
  \\
  \label{momqedapp1ii}
  2 {^{[n+1]}\Ac^{k(1)}} \!&\!=\!&\!
  -2 m {^{[n]}\Lc^{0k(1)}} + 2 p^k {^{[n]}\Ac^{0(1)}}
  \nonumber\\ &&
  + \varepsilon^{0ijk}
  \left(
    (\partial_i -\frac{e}{c} \partial_p^j F_{ij}(x)) {^{[n]}\Vc_j^{(0)}}
    + \frac{ne}{c} F_{i0}(x) {^{[n-1]}\Vc_j^{(0)}}
  \right)
  \\
  \label{momqedapp1iii}
  2 {^{[n+1]}\Fc^{(1)}} \!&\!=\!&\!
  2 m {^{[n]}\Vc^{0(1)}}
  - (\partial_i -\frac{e}{c} \partial_p^j F_{ij}(x)) {^{[n]}\Sc^{0i(0)}}
  - \frac{ne}{c} F_{i0}(x) {^{[n-1]}\Sc^{0i(0)}}\
  \\
  \label{momqedapp1iv}
  2 {^{[n+1]}\Sc^{jk(1)}} \!&\!=\!&\!
  2m \varepsilon^{0ijk} {^{[n]}\Ac_i^{(1)}}
  - 2 p^k {^{[n]}\Sc^{0j(1)}}
  + 2 p^j {^{[n]}\Sc^{0k(1)}}
  \\
  \label{momqedapp1v}
  2 {^{[n+1]}\Ac^{0(1)}} \!&\!=\!&\!
  -2 p_i {^{[n]}\Ac^{i(1)}}
  \\
  \label{momqedapp1vi}
  2 {^{[n+1]}\Vc^{i(1)}} \!&\!=\!&\!
  2 p^i {^{[n]}\Vc^{0(1)}}
  \nonumber\\ &&
  + \varepsilon^{0ijk}
  \left(
    (\partial_j -\frac{e}{c} \partial_p^l F_{jl}(x)){^{[n]}\Ac_k^{(0)}}
    + \frac{ne}{c} F_{j0}(x) {^{[n-1]}\Ac_k^{(0)}}
  \right)
  \\
  \label{momqedapp1vii}
  2 {^{[n+1]}\Sc^{i0(1)}} \!&\!=\!&\!
  - 2 p_j {^{[n]}\Sc^{ij(1)}}
  \nonumber\\ &&
  + (\partial_i -\frac{e}{c} \partial_p^j F_{ij}(x)) {^{[n]}\Fc^{(0)}}
  + \frac{ne}{c} F_{i0}(x) {^{[n-1]}\Fc^{(0)}}
  \\
  \label{momqedapp1viii}
  2 {^{[n+1]}\Pc^{(1)}} \!&\!=\!&\!
  - (\partial_i -\frac{e}{c} \partial_p^j F_{ij}(x)) {^{[n]}\Lc^{0i(0)}}
  - \frac{ne}{c} F_{i0}(x) {^{[n-1]}\Lc^{0i(0)}}
\end{eqnarray}
and
\begin{eqnarray}
  \label{momqedapp2i}
  2 m {^{[n]}\Pc^{(1)}} \!&\!=\!&\!
  (\partial_\mu -\frac{e}{c} \partial_p^j F_{\mu j}(x)) {^{[n]}\Ac^{\mu(0)}}
  + \frac{ne}{c} F_{\mu0}(x) {^{[n-1]}\Ac^{\mu(0)}}
  \\
  \label{momqedapp2ii}
  2 m {^{[n]}\Sc^{0i(1)}} \!&\!=\!&\!
  2 \varepsilon^{0ijk} p_j {^{[n]}\Ac_k^{(1)}}
  + (\partial^0 -\frac{e}{c} \partial_{pj} F^{0j}(x)) {^{[n]}\Vc^{i(0)}}
  \nonumber\\ &&
  - (\partial^i -\frac{e}{c} \partial_{pj} F^{ij}(x)) {^{[n]}\Vc^{0(0)}}
  - \frac{ne}{c} F_{i0}(x) {^{[n-1]}\Vc^{0(0)}}
  \\
  \label{momqedapp2iii}
  2 m {^{[n]}\Vc^{i(1)}} \!&\!=\!&\!
  2 p_i {^{[n]}\Fc^{(1)}}
  + (\partial_\nu -\frac{e}{c} \partial_p^j F_{\nu j}(x)) {^{[n]}\Sc^{i\nu(0)}}
  + \frac{ne}{c} F_{\nu0}(x) {^{[n-1]}\Sc^{i\nu(0)}}\quad
  \\
  \label{momqedapp2iv}
  2 m {^{[n]}\Ac^{0(1)}} \!&\!=\!&\!  2 p_i {^{[n]}\Lc^{0i(1)}}
  \\
  \label{momqedapp2v}
  0 \!&\!=\!&\!
  (\partial_\mu -\frac{e}{c} \partial_p^j F_{\mu j}(x)) {^{[n]}\Vc^{\mu(0)}}
  + \frac{ne}{c} F_{\mu0}(x) {^{[n-1]}\Vc^{\mu(0)}}
  \\
  \label{momqedapp2vi}
  0 \!&\!=\!&\! 2 p^i {^{[n]}\Vc^{j(1)}} - 2 p^j {^{[n]}\Vc^{i(1)}}
  - \varepsilon^{0ijk}
  (\partial_0 -\frac{e}{c} \partial_p^j F_{0j}(x)) {^{[n]}\Ac_k^{(0)}}
  \nonumber\\ &&
  +\varepsilon^{0ijk}
  \left(
    (\partial_k -\frac{e}{c} \partial_p^l F_{kl}(x)) {^{[n]}\Ac^{0(0)}}
    \frac{ne}{c} F_{k0}(x){^{[n-1]}\Ac^{0(0)}}
  \right)
  \\
  \label{momqedapp2vii}
  0 \!&\!=\!&\! -2 p_i {^{[n]}\Sc^{0i(1)}}
  + (\partial_0 -\frac{e}{c} \partial_p^j F_{0j}(x)) {^{[n]}\Fc^{(0)}}
  \\
  \label{momqedapp2viii}
  0 \!&\!=\!&\! 2 p^i {^{[n]}\Pc^{(1)}}
  -  (\partial_0 -\frac{e}{c} \partial_p^j F_{0j}(x))
  {^{[n]}\Lc^{0i(0)}}
  \nonumber\\ &&
  +  (\partial_j -\frac{e}{c} \partial_p^l F_{jl}(x))
  {^{[n]}\Lc^{ij(0)}}
  +  \frac{ne}{c} F_{j0}(x) {^{[n-1]}\Lc^{ij(0)}}.
\end{eqnarray}
To find the equations for the moments to zeroth order $\hbar$ we
eliminate in Eqs.~(\ref{momqedapp2i}-\ref{momqedapp2viii})
with $n \to n+1$ the components of order $\hbar$ with the help of
Eqs.~(\ref{momqedapp1i}-\ref{momqedapp1viii}).
Inserting Eqs.~(\ref{momqedapp1ii},\ref{momqedapp1vii}) into
Eq.~(\ref{momqedapp2ii})
and Eqs.~(\ref{momqedapp1vii}) into Eq.~(\ref{momqedapp2vii}) gives us (with
the
unchanged Eq.~(\ref{momqedapp2v})) the dynamical equations for the components
${^{[n]}\Vc^{\mu(0)}}$, ${^{[n]}\Fc^{(0)}}$
\begin{eqnarray}
  \label{bewegmomqedapp1i}
  0 \!&\!=\!&\!
  (\partial_0 - \frac{e}{c} \partial_p^j F_{0j}(x)) {^{[n+1]}\Vc^{i(0)}}
  + p^j (\partial_j - \frac{e}{c} \partial_p^k F_{jk}(x)) {^{[n]}\Vc^{i(0)}}
  + \frac{ne}{c} F_{j0}(x) p^j {^{[n-1]}\Vc^{i(0)}}
  \nonumber\\ &&
  - p_j (\partial^i - \frac{e}{c} \partial_{pk} F^{ik}(x))
  {^{[n]}\Vc^{j(0)}}
  - \frac{ne}{c} F_{i0}(x) p_j {^{[n-1]}\Vc^{j(0)}}
  \nonumber\\ &&
  - (\partial^i - \frac{e}{c} \partial_{pj} F^{ij}(x)) {^{[n+1]}\Vc^{0(0)}}
  - \frac{(n+1) e}{c} F^{i0}(x) {^{[n]}\Vc^{0(0)}}
  \nonumber\\ &&
  + m (\partial^i - \frac{e}{c} \partial_{pj} F^{ij}(x)) {^{[n]}\Fc^{(0)}}
  + m \frac{ne}{c} F^{i0}(x) {^{[n-1]}\Fc^{(0)}}
  \\
  \label{bewegmomqedapp1ii}
  0 \!&\!=\!&\! (\partial_\mu - \frac{e}{c} \partial_p^j F_{\mu j}(x))
  {^{[n]}\Vc^{\mu(0)}}
  + \frac{ne}{c} F_{\mu 0}(x)  {^{[n-1]}\Vc^{\mu(0)}}
  \\
  \label{bewegmomqedapp1iii}
  0 \!&\!=\!&\! (\partial_0 - \frac{e}{c} \partial_p^j F_{0j}(x))
{^{[n+1]}\Fc^{(0)}}
  \nonumber\\ &&
  + p^i (\partial_i - \frac{e}{c} \partial_p^j F_{ij}(x)) {^{[n]}\Fc^{(0)}}
  + \frac{ne}{c} F_{i0}(x) p^i {^{[n-1]}\Fc^{(0)}} .
\end{eqnarray}
To apply now the constraints for the moments to zeroth order $\hbar$
{
  \renewcommand{\arraystretch}{2.0}
  \begin{equation}
    \label{momhbar0Fapp}
    \begin{array}{rclcrcl}
      \D {^{[2n]}\Vc}^{0(0)} \!&\!=\!&\!
      \D E^{2n} \frac{1}{m}{^{[1]}\Fc}^{(0)}
      &\qquad&
      \D {^{[2n+1]}\Vc}^{0(0)} \!&\!=\!&\! \D E^{2n+1}\frac{E}{m}
{^{[0]}\Fc}^{(0)} \\
      \D {^{[2n]}{\vec \Vc}}^{(0)} \!&\!=\!&\!
      \D E^{2n}\frac{\vec p}{m}
      {^{[0]}\Fc}^{(0)}
      &\qquad&
      \D {^{[2n+1]}{\vec \Vc}}^{(0)} \!&\!=\!&\!
      \D E^{2n+1}\frac{\vec p}{Em} {^{[1]}\Fc}^{(0)} \\
      \D {^{[2n]}\Fc}^{(0)} \!&\!=\!&\!
      \D E^{2n} {^{[0]}\Fc}^{(0)}
      &\qquad&
      \D {^{[2n+1]}\Fc}^{(0)} \!&\!=\!&\! \D E^{2n+1}\frac{1}{E}
{^{[1]}\Fc}^{(0)}
    \end{array}
  \end{equation}
}%
we have to distinguish between even and odd $n$.
For  $n$ even we find with
\begin{equation}
  \partial_p^j p^i = p^i \partial^j_p + g^{ij},\quad
  \partial_p^j E = E \partial_p^j - \frac{p^j}{E}
\end{equation}
from Eq.~(\ref{bewegmomqedapp1ii}) the following transport equation for the
zeroth moment ${^{[0]}\Fc^{(0)}}$
\begin{equation}
  \label{bewegendqed1}
  0 = E^n \left[
    E(\partial_0 - \frac{e}{c} \partial_p^j F_{0j}(x)) (E {^{[0]}\Fc^{(0)}})
    + p_j (\partial^j - \frac{e}{c} \partial_p^k F_{jk}(x)) {^{[1]}\Fc^{(0)}}
  \right].
\end{equation}
Similar we find for the odd moments with the help of
$E {^{[n]}\Fc^{(0)}} \leftrightarrow {^{[n]}\Fc^{(1)}}$ the transport equation
for the first moment ${^{[1]}\Fc^{(0)}}$
\begin{equation}
  \label{bewegendqed2}
  0 = E^n \left[
    E(\partial_0 - \frac{e}{c} \partial_p^j F_{0j}(x)) {^{[1]}\Fc^{(0)}}
    + p_j (\partial^j - \frac{e}{c} \partial_p^k F_{jk}(x))(E
{^{[0]}\Fc^{(0)}})
  \right].
\end{equation}
Expressed with fields $\vec E$, $\vec B$ we finally get
\begin{eqnarray}  \label{bewegendqed2a}
  0 \!&\!=\!&\! E^n \left[
    E(\partial^0 + \frac{e}{c} \nabla_p\cdot \vec E) {^{[1]}\Fc^{(0)}}
    + (\vec p \cdot \nabla + \frac{e}{c} (\vec p \times \vec B )
    \cdot \nabla_p) (E {^{[0]}\Fc^{(0)}})
  \right]
  \\  \label{bewegendqed2b}
  0 \!&\!=\!&\! E^n \left[
    E(\partial^0 + \frac{e}{c} \nabla_p \vec E) (E{^{[0]}\Fc^{(0)}})
    + (\vec p \cdot \nabla + \frac{e}{c} (\vec p \times \vec B )
    \cdot \nabla_p) {^{[1]}\Fc^{(0)}}
  \right].
\end{eqnarray}
The remaining dynamical equations
Eq.~(\ref{bewegmomqedapp1i},\ref{bewegmomqedapp1iii}) prove to contain
no new information: Eq.~(\ref{bewegmomqedapp1i}) can be derived from
$p_i \cdot$ Eq.~(\ref{bewegendqed1}) or $p_i \cdot$
Eq.~(\ref{bewegendqed2a}), Eq.~(\ref{bewegmomqedapp1iii}) leads just to
Eq.~(\ref{bewegendqed1}) and Eq.~(\ref{bewegendqed2}) with $n$
replaced by $n-1$.

The transport equations for the other spinor components
${^{[n]}\Sc_{\mu\nu}^{(0)}}$, ${^{[n]}\Ac_\mu^{(0)}}$ we find from
Eqs.~(\ref{momqedapp2i},\ref{momqedapp2iii},\ref{momqedapp2vi},\ref{momqedapp2viii})
for the $n+1^{\rm st}$~moments by eliminating the components to first
order $\hbar$ with the help of Eqs.~(\ref{momqedapp1i}-\ref{momqedapp1viii}).
The transport equations are thus
\begin{eqnarray}
  \label{bewegmomqedapp3i}
  0 \!&\!=\!&\!
  (\partial_0 -\frac{e}{c}\partial_p^j F_{0j}(x)){^{[n+1]}\Ac^{0(0)}}
  \nonumber\\ &&
  + (\partial_i -\frac{e}{c}\partial_p^j F_{ij}(x)) {^{[n+1]}\Ac^{i(0)}}
  + \frac{(n+1)e}{c} F_{i0}(x) {^{[n]}\Ac^{i(0)}}
  \nonumber\\ &&
  + m (\partial_i -\frac{e}{c}\partial_p^j F_{ij}(x))
  {^{[n]}\Lc^{0i(0)}}
  + m \frac{ne}{c} F_{i0}(x) {^{[n-1]}\Lc^{0i(0)}}
  \\
  \label{bewegmomqedapp3ii}
  0 \!&\!=\!&\!
  (\partial_0 -\frac{e}{c}\partial_p^j F_{0j}(x)) {^{[n+1]}\Sc^{i0(0)}}
  \nonumber\\ &&
  + (\partial_j -\frac{e}{c}\partial_p^k F_{jk}(x))
  {^{[n+1]}\Sc^{ij(0)}}
  + \frac{(n+1)e}{c} F_{j0}(x) {^{[n]}\Sc^{ij(0)}}
  \nonumber\\ &&
  - p^i (\partial_j -\frac{e}{c}\partial_p^k F_{jk}(x))
  {^{[n]}\Sc^{0j(0)}}
  - \frac{ne}{c} p^i F_{j0}(x) {^{[n-1]}\Sc^{0j(0)}}
  \nonumber\\ &&
  - m \varepsilon^{0ijk}
  \left[
    (\partial_j -\frac{e}{c}\partial_p^l F_{jl}(x)) {^{[n]}\Ac_k^{(0)}}
    + \frac{ne}{c} F_{j0}(x) {^{[n-1]}\Ac_k^{(0)}}
  \right]
  \\
  \label{bewegmomqedapp3iii}
  0 \!&\!=\!&\!
  (\partial_0 \!-\!\frac{e}{c}\partial_p^j F_{0j}(x)) {^{[n+1]\!}\Ac^{i(0)}}
  \!+\! p_j (\partial^j \!-\!\frac{e}{c}\partial_{pk} F^{jk}(x))
  {^{[n]\!}\Ac^{i(0)}}
  \!+\! \frac{ne}{c} p_j F^{j0}(x) {^{[n-1]\!}\Ac^{i(0)}}
  \nonumber\\ &&
  - (\partial^i -\frac{e}{c}\partial_{pk} F^{ik}(x)){^{[n]}\Ac^{j(0)}}
  - \frac{ne}{c} p_j F^{i0}(x) {^{[n-1]}\Ac^{j(0)}}
  \nonumber\\ &&
  - (\partial^i -\frac{e}{c}\partial_{pk} F^{ik}(x)) {^{[n+1]}\Ac^{0(0)}}
  - \frac{(n+1)e}{c} F^{i0}(x) {^{[n]}\Ac^{0(0)}}
  \\
  \label{bewegmomqedapp3iv}
  0 \!&\!=\!&\!
  (\partial_0 -\frac{e}{c}\partial_p^j F_{0j}(x))
  {^{[n+1]}\Lc^{0i(0)}}
  \nonumber\\ &&
  + p^i (\partial_j - \frac{e}{c}\partial_p^k F_{jk}(x))
  {^{[n]}\Lc^{0j(0)}}
  + \frac{ne}{c} p^i F_{j0}(x) {^{[n-1]}\Lc^{0j(0)}}
  \nonumber\\ &&
  - (\partial_j -\frac{e}{c}\partial_p^k F_{jk}(x))
  {^{[n+1]}\Lc^{ij(0)}}
  - \frac{(n+1)e}{c} F_{j0}(x) {^{[n]}\Lc^{ij(0)}}.
\end{eqnarray}
Inserting the constraints for the moments to zeroth order  $\hbar$
\begin{eqnarray}
  \label{momhbar0ASappi}
  {^{[2n]}\Ac^{0(0)}} \!&\!=\!&\!
  E^{2n} \frac{1}{m} \vec p \cdot {^{[0]}{\vec \sigma}^{(0)}}
  \\
  \label{momhbar0ASappii}
  {^{[2n]}{\vec \Ac}^{(0)}} \!&\!=\!&\!
  E^{2n} \frac{1}{m}
  \left(
    {^{[1]}{\vec \sigma}^{(0)}} +
    \vec p \times (\vec p \times
    \frac{{^{[1]}{\vec \sigma}^{(0)}}}{E^2})
  \right)
  \\
  \label{momhbar0ASappiii}
  {^{[2n]}{\vec S}^{(0)}} \!&\!=\!&\!
  - E^{2n} \vec p \times \frac{{^{[1]}{\vec \sigma}^{(0)}}}{E^2}
  \\
  \label{momhbar0ASappiv}
  {^{[2n]}{\vec \sigma}^{(0)}} \!&\!=\!&\!
  E^{2n} {^{[0]}{\vec \sigma}^{(0)}}
  \\
  \label{momhbar0ASappv}
  {^{[2n+1]}\Ac^{0(0)}} \!&\!=\!&\!
  E^{2n+1} \frac{1}{m}
  \vec p \cdot \frac{{^{[0]}{\vec \sigma}^{(0)}}}{E}
  \\
  \label{momhbar0ASappvi}
  {^{[2n+1]}{\vec \Ac}^{(0)}} \!&\!=\!&\!
  E^{2n+1} \frac{1}{m}
  \left(
    \frac{{^{[0]}{\vec \sigma}^{(0)}}}{E} +
    \vec p \times (\vec p \times \frac{{^{[0]}{\vec \sigma}^{(0)}}}{E})
  \right)
  \\
  \label{momhbar0ASappvii}
  {^{[2n+1]}{\vec S}^{(0)}} \!&\!=\!&\!
  - E^{2n+1} \vec p \times \frac{{^{[0]}{\vec \sigma}^{(0)}}}{E}
  \\
  \label{momhbar0ASappviii}
  {^{[2n+1]}{\vec \sigma}^{(0)}} \!&\!=\!&\!
  E^{2n+1} \frac{{^{[1]}{\vec \sigma}^{(0)}}}{E} ,
\end{eqnarray}
we get for instance from Eq.~(\ref{bewegmomqedapp3iv}) for even and odd moments
transport equations for the first two moments of ${\vec \sigma}^{(0)}$.
Written with fields $\vec E$ and $\vec B$ these are
\begin{eqnarray}\label{bmtgl1}
  0 \!&\!=\!&\! E^{2n}
  \biggl[
  E\left(
    \partial^0 +\frac{e}{c} \nabla_p \cdot \vec E
  \right)
  \frac{{^{[1]}{\vec \sigma}^{(0)}}}{E}
  + \left(
    \vec p \cdot \nabla + \frac{e}{c} (\vec p \times \vec B )
  \cdot \nabla_p  \right)
  {^{[0]}{\vec \sigma}^{(0)}}
  \nonumber\\ &&
  + \frac{e}{c}
  \left(
    \vec p
    \left(
      \vec E \cdot \frac{{^{[1]}{\vec \sigma}^{(0)}}}{E^2}
    \right)
    + \vec B \times {^{[0]}{\vec \sigma}^{(0)}}
  \right)
  \biggr]
  \\
  \label{bmtgl2}
  0 \!&\!=\!&\! E^{2n+1} \biggl[
  E(\partial^0 +\frac{e}{c} \nabla_p \cdot \vec E )
  {^{[0]}{\vec \sigma}^{(0)}}
  + (\vec p \cdot \nabla + \frac{e}{c} (\vec p \times \vec B )
  \cdot \nabla_p )
  \frac{{^{[1]}{\vec \sigma}^{(0)}}}{E}
  \nonumber\\ &&
  + \frac{e}{c}\left(
    \vec p \left(\vec E \cdot \frac{{^{[0]}{\vec \sigma}^{(0)}}}{E}\right)
    + \vec B \times \frac{{^{[1]}{\vec \sigma}^{(0)}}}{E}
  \right)
  \biggr].
\end{eqnarray}
The other dynamical equations do again contain no new information,
leading to Eqs.~(\ref{bmtgl1},\ref{bmtgl2}) or $\vec p \times$
Eqs.~(\ref{bmtgl1},\ref{bmtgl2}).

\newpage

\newpage

%
%

\setcounter{totalnumber}{3}

\newpage

{\renewcommand{\arraystretch}{2.0}
\begin{table}[htb]
  \begin{center}
    \leavevmode
    \begin{tabular}{|r||l|l|}
      \hline
      & $\Pi_\mu$ & $\Delta_\mu$\\
      \hline \hline
      $0^{\rm th}$ order & $p_\mu$ & $0$\\
      \hline
      $1^{\rm st}$ order & $-\frac{ig}{4c} \fopf_{\nu\mu}(x)\partial_p^\nu
      \hbar$ & $
      \left[D_{\mu}(x) - \frac{g}{2c} \fopf_{\nu\mu}(x)\partial_p^\nu
      \right] \hbar$\\
      \hline
    \end{tabular}
    \vspace*{1.0cm}
    \caption{The first two orders of the $\hbar$-expansion for the
      non-local operators $\Pi_\mu$ and $\Delta_\mu$.}
    \label{tabhbarord}
  \end{center}
\end{table}
}

{\renewcommand{\arraystretch}{2.0}
\begin{table}[htb]
  \begin{center}
    \leavevmode
    \begin{tabular}{|r||l|l|}
      \hline
      & ${\mathcal P}_\mu$ & ${\mathcal D}_\mu$\\
      \hline \hline
      $0^{\rm th}$ order & $p_\mu$ & 0\\
      \hline
      $1^{\rm st}$ order & 0 & $
        \left[
          \partial_\mu - \frac{g}{c} F_{\nu\mu}(x)
          \partial_p^\nu
        \right] \hbar$\\
        \hline
    \end{tabular}
    \vspace*{1.0cm}
    \caption[$\hbar$-expansion for ${\mathcal P}_\mu$, ${\mathcal
      D}_\mu$]{The first two orders for the gradient expansion for the
      non-local operators  ${\mathcal P}_\mu$ and ${\mathcal D}_\mu$ for QED
      with external fields.}\label{tabhbarordQED}
  \end{center}
\end{table}
}

\end{document}